\pdfoutput=1
\documentclass[aps,prb,twocolumn,floatfix,amsmath,amssymb,footinbib,showpacs]{revtex4}

\usepackage{graphicx}
\usepackage{dsfont}

\newcommand{\ix}[1]{\ensuremath{\mathrm{#1}}}

\newcommand{\inv}{\ix{inv}}

\newcommand{\R}{\ix{R}}

\usepackage{amsopn}
\DeclareMathOperator{\sgn}{sgn}
\DeclareMathOperator{\tr}{tr}
\DeclareMathOperator{\Tr}{Tr}

\begin{document}

\title{Renormalization group flow of the Luttinger-Ward functional:
  conserving approximations and application to the Anderson impurity model}

\author{J.F. Rentrop, V. Meden and S.G. Jakobs}

\affiliation{\small Institut f\"ur Theorie der Statistischen Physik, RWTH Aachen
  University and\\ \small JARA - Fundamentals of Future Information Technology,
  52062 Aachen, Germany}


\begin{abstract} 
  We study the renormalization group flow of the Luttinger-Ward
  functional and of its two-particle irreducible vertex functions,
  given a cut-off in the two-particle interaction. We derive a
  conserving approximation to the flow and relate it to the
  fluctuation exchange approximation as well as to non-conserving
  approximations introduced in an earlier publication ctuation exchange approximation as well as to nonconserving approximations introducen [J. F. Rentrop, S. G. Jakobs, and V. Meden, J. Phys. A: Math. Theor. 48, 145002 (2015)]. We apply the
  different approximate flow equations to the single impurity Anderson
  model in thermal equilibrium at vanishing temperature. Numerical results
  for the effective mass, the spin susceptibility, the charge
  susceptibility, and the linear conductance reflect the similarity of
  the methods to the fluctuation exchange approximation.  We find the
  majority of the approximations to deviate stronger from the exact
  results than one-particle irreducible functional renormalization
  group schemes.  However, we identify a simple static two-particle
  irreducible flow scheme which performs remarkably well and produces
  an exponential Kondo-like scale in the renormalized level position.
\end{abstract} 

\maketitle


\section{Introduction}

Baym and Kadanoff start the abstract of their seminal paper on how to
construct conserving approximations to many-particle Green functions
with the words ``in describing transport phenomena, it is vital to
build the conservation laws of number, energy, momentum, and angular
momentum into the structure of the approximation''.\cite{Bay61} For
the following decades it was indeed a paradigm that approximate
solutions to quantum many-body problems ought to be conserving.
However, for low-dimensional systems known conserving approximation
schemes suffer from severe artifacts.  The conserving self-consistent
Hartree-Fock approximation, for example, predicts an unphysical
breaking of spin symmetry for the single impurity Anderson model at
moderate interactions;\cite{And61} and it wrongly predicts the
formation of a charge density wave in one-dimensional quantum wires
with weak repulsive interaction.\cite{Sha94}  As another example,
two-particle Green functions computed with conserving approximations
as proposed by Baym and Kadanoff~\cite{Bay61, Bay62} violate the Pauli
principle in form of the crossing symmetry relation.\cite{Sta85,
  Bic91a, Bic91b}

Maintaining conservation laws is usually not in the focus of
renormalization group (RG) approaches to quantum many-body problems.
Typical RG-based approximations are non-conserving, for instance
standard truncations of the ``functional'' (or ``exact'') RG (fRG) in
the one-particle irreducible (1PI) vertex expansion.\cite{Kat04,
  Met12}  The relation between fRG approximations and conservation
laws was repeatedly under investigation.  In particular the connection
of the fRG to Ward identities attracted interest; Ward identities are
relations between many-body correlation functions which encode the
respective conservation laws.  Katanin showed how the deviation of
1PI fRG results from Ward identities can be reduced by modifications
in the truncation procedure.\cite{Kat04}  Enss found the commonly
employed fRG truncation schemes to be in principle incompatible with
the Ward-identities typically used in the condensed-matter
literature.\cite{Ens05}  Kopietz and coworkers proceeded reversely
and used Ward identities to create new truncations of the hierarchy of
fRG flow equations.\cite{sue05, Bar09, Str13}

Another topic that raised attention in this context is the relation of
the fRG to the conserving approximations proposed by Baym and
Kadanoff.\cite{Bay61, Bay62}  These are often called
``$\Phi$-derivable'' in reference to their construction.  First an
approximation to the Luttinger-Ward functional~\cite{Lut60a} $\Phi$ is
devised which is invariant under the symmetry transformations
associated with the conservation law.  Then correlation functions are
computed from the two-particle irreducible (2PI) vertex functions of
$\Phi$, the physical value being determined by a self-consistency
equation for the self-energy.  It was shown for a scalar field theory
that the physical self-energy of any given $\Phi$-derivable
approximation can be obtained from a 1PI fRG flow;\cite{Bla11} for
that purpose one expresses the 1PI four-point function that enters the
flow equation via the 2PI four-point function that corresponds to the
given approximation.  This finding highlights the renormalizability of
the vertex functions in $\Phi$-derivable approximations, which was
studied intensively already before, see Ref.~\onlinecite{Ber05} and references
therein. In Ref.~\onlinecite{Car15}, it is discussed how the
$\Phi$-derivable approximation based on the second order approximation
to $\Phi$ can be obtained from a 2PI fRG flow. References~\onlinecite{Bla11} and~\onlinecite{Car15} thus show how a given $\Phi$-derivable approximation can
be reconstructed by the fRG.  In this paper we address the opposite
question.  Can the fRG be used to construct new $\Phi$-derivable
approximations?  So far, the required invariant approximate functional
$\Phi$ is usually given by some subset of (skeleton) diagrams from the
expansion of $\Phi$ in powers of the interaction.\cite{Bay62}  The
fRG could be used to construct completely new, non-diagrammatic
invariant approximations to $\Phi$.  The precise form of the
functional might even not be required if the fRG describes the flow of
the corresponding physical values of the vertex functions.

A natural starting point for our investigation is the fRG flow of the
Luttinger-Ward functional and of its vertex functions as described
in Refs.~\onlinecite{Dup05,Dup14} and~\onlinecite{Ren15}.  In Ref.~\onlinecite{Ren15}, we used the notions of
``$C$-flow'' and ``$U$-flow'' in order to distinguish whether the flow
parameter is introduced into the free propagator $C$ or into the
two-particle interaction~$U$. We showed that
the hierarchy of $C$-flow equations for the physical vertex functions
truncated straightforwardly at level $2m$ is solved by $m$-th order
self-consistent perturbation theory.  This generalizes the result
of Ref.~\onlinecite{Car15} to arbitrary order (however only for condensed-matter
problems without ultra-violet divergencies).  Truncated $C$-flow is
completely equivalent to the well known $\Phi$-derivable
self-consistent perturbation theory.  In particular, the result does
not depend in any form on the choice of the flow parameter and its
possibly regularizing properties.  For models with infrared
divergencies in perturbation theory, the straightforward application
of $C$-flow RG is only possible if self-consistency has a regularizing
effect.

Concerning the $U$-flow, we did not discuss in Ref.~\onlinecite{Ren15} how the
truncation schemes relate to conserving approximations; we do so in
this paper.  Here, we show how $\Phi$-derivable approximations
to the $U$-flow can be constructed.  We carry out the relevant steps
in a nontrivial truncation and find an approximation which is closely
related to the fluctuation exchange (FLEX) approximation.\cite{Bic89a}  The corresponding
invariant approximations to $\Phi$ are diagrammatically equivalent
except for prefactors.  Furthermore we study the $U$-flow of the
physical vertex functions and identify the $U$-flow approximations
from Ref.~\onlinecite{Ren15} as non-$\Phi$-derivable approximations to our
$\Phi$-derivable one.  Additionally, in the lowest order truncation we
find a static non-$\Phi$-derivable approximate $U$-flow that was overlooked
in Ref.~\onlinecite{Ren15} and which turns out to be remarkably accurate for the
Anderson impurity model. 

We also consider the combined $C$- and $U$-flow.  We find that the
corresponding $\Phi$-derivable fRG approximations are identical to
those of the pure $U$-flow.  Furthermore we construct a
non-$\Phi$-derivable combined $C$- and $U$-flow approximation for the
flow of the physical values with a parameter that allows to smoothly
interpolate between pure $C$-flow and pure $U$-flow approximations.
At a suitably chosen parameter value the range of applicability of the
combined method to the Anderson impurity model is slightly larger than
that of the pure $U$-fow.

We apply all our conserving and non-conserving approximations
to the Anderson impurity model in equilibrium and study the effective
mass, the spin susceptibility, the charge susceptibility and the
linear conductance. In this
way, we provide a comprehensive application of 2PI fRG
approximations to a condensed matter quantum many-body
problem.

In Ref.~\onlinecite{Ren15} we studied the performance of 2PI fRG approximations
on the toy model of the anharmonic oscillator.  There we identified a
non-$\Phi$-derivable ``modified'' variant of the $U$-flow going back
to Ref.~\onlinecite{Dup14} as more precise and faster than 1PI fRG with flowing
four-point vertex; the modified $U$-flow variant has the self-consistent
Hartree-Fock approximation as starting point which provides already a
fairly good approximation for the case of the anharmonic oscillator.  The question
arises as to whether the high efficiency of this flow scheme pertains as
well to actual many-body problems.  The Anderson impurity model
provides a test of particular interest, since self-consistent
Hartree-Fock predicts for this model an unphysical breaking of spin
symmetry at increased interactions.  Does the modified $U$-flow
restore the symmetry which is violated in its initial conditions?  In
this paper we show that this is not the case and that the modified
$U$-flow performs comparably bad. Furthermore, we prove the plain and
modified $U$-flow approximations of Ref.~\onlinecite{Ren15} to be non-$\Phi$-derivable
by comparing numerical results for the dot occupancy obtained from
different approaches. Consequently, there is no reason to expect these methods to preserve
conservation laws; therefore, we will frequently refer to them as
non-conserving methods.

Concerning our $\Phi$-derivable and thus conserving approximation to
the $U$-flow, the numerical results turn out to be quite similar to
those of the FLEX approximation for all studied observables.  In
particular, the effective mass is quickly overestimated as $U$
increases.  There exists an analytic prediction~\cite{Ham69} that
another approximation similar to FLEX produces a characteristic
temperature scale $\sim \exp(-cU^2)$ [as opposed to the correct Kondo
temperature $\sim \exp(-c'U)$].  The consequent presumption that our approximate
effective mass correspondingly shows an $\exp(cU^2)$-behavior is
however not confirmed by the data.

The paper is organized as follows: In Sec.~\ref{sec:2PI-formalism},
we briefly repeat the notation and the main definitions of
Ref.~\onlinecite{Ren15}. Section~\ref{sec:Phi-symmetry} then summarizes important
aspects of $\Phi$-derivability discussed in the literature. In
Sec.~\ref{sec:Cons_and_non-cons_appr}, we present our main
analytical findings. We show how to obtain a conserving $U$-flow fRG
scheme and specify its relation to FLEX. We then discuss how to view
the $U$-flow schemes of Ref.~\onlinecite{Ren15} as approximations to the
conserving $U$-flow. Moreover we show how to obtain a static
(i.e. frequency-independent) $U$-flow scheme. We then describe an
approximate 2PI fRG scheme that combines $C$- and $U$-flow. In
Sec.~\ref{sec:Anderson_model_in_2PI}, we apply the different
methods to the Anderson impurity model in equilibrium; for a concise
presentation in the main part, we discuss many details in the
appendices. We present numerical results for the Anderson model in
Sec.~\ref{sec:Numerical_results} which is followed by the
concluding Sec.~\ref{sec:conclusion}.  Throughout the paper we set
$\hbar=1$ and $k_\ix{B}=1$.


\section{Fermionic 2PI formalism: notation and definitions}
\label{sec:2PI-formalism}

In this paper we use the same notation and definitions as
in Ref.~\onlinecite{Ren15}.  In this section, we only summarize the most important ones,
restricting ourselves to the case of some fermionic many-body
system.  For details we refer to Ref.~\onlinecite{Ren15}, in particular
sections~2, 3 and 5. At the end of this section, we comment on the existence and uniqueness of the Luttinger-Ward functional. 

We construct suitable generating functionals for equilibrium Green
functions from the grand canonical partition function $Z[J]$ furnished
with a source term. The source is chosen to be quadratic in the
fields,
\begin{equation}
  \label{eq:Z}
  Z[J] = \int\! D[\psi] e^{-S[\psi] + \frac{1}{2} \sum_{\alpha
      \alpha'} 
    \psi_{\alpha} J_{\alpha \alpha'} \psi_{\alpha'}}.
\end{equation}
Here, $\int \! D[\psi]$ is a functional integral over imaginary time
Grassmann variables, and the action is given by
\begin{align}
  \label{eq:action}
  S[\psi]
  =& 
  -\frac{1}{2} \sum_{\alpha \alpha'} \psi_{\alpha}
  \left(C^{-1}\right)_{\alpha \alpha'} \psi_{\alpha'}  
  \\ \notag &+
  \frac{1}{4!}\! \sum_{\alpha_1 \alpha'_1 \alpha_2 \alpha'_2} \!
  U_{\alpha_1 \alpha'_1 \alpha_2 \alpha'_2} \psi_{\alpha_1}
  \psi_{\alpha'_1} \psi_{\alpha_2}\psi_{\alpha'_2}. 
\end{align}
We use multi-indices $\alpha=(c,s,\tau)$.  The charge index $c$
determines whether a field is creating ($+$) or annihilating
($-$). For the Anderson model below, the state index $s$ will be the
spin, $s = \sigma = \;\uparrow$ or $\downarrow$.  As usual in thermal
equilibrium, we can switch from imaginary times $\tau$ to Matsubara
frequencies $\nu_n=\frac{\pi}{\beta}(2n+1)$. One-particle quantities
such as the free propagator $C$ are antisymmetric under exchange of
the indices, $C_{\alpha \alpha^\prime}=-C_{\alpha^\prime \alpha}$. The
two-particle interaction $U$ is fully antisymmetrized,
$U_{\alpha_1 \alpha_2 \alpha_3 \alpha_4} = \sgn(P)
U_{\alpha_{P1} \alpha_{P2} \alpha_{P3} \alpha_{P4}}$ for any
permutation $P$.

The charge index notation is advantageous
for the methodological part of the paper.  It allows for a compact
notation with e.g. a single expression representing different channels
of pair propagation.  Furthermore, it applies in the same form to
models which do or do not conserve particle number.  We use it at the
cost of obtaining at first unwieldy matrices with many zero components (which 
we then reduce to simpler objects) once we apply the formalism to
the Anderson model with conserved particle number.

A Legendre transformation leads from $W[J]=\ln Z[J]$ to the 2PI
effective action
\begin{equation}
  \label{eq:Gamma-def}
 \Gamma[G]=\left.\left\{-W[J]-J \cdot G\right\}\right|_{J[G]},
\end{equation}
cf.~Refs.~\onlinecite{Dom64a, Dom64b} and~\onlinecite{Wet07}.  The new independent variable is
the full propagator $G$ with components $G_{\alpha \alpha'} = - \delta
W /
\delta J_{\alpha \alpha'} = - \langle T \psi_\alpha \psi_{\alpha'}
\rangle_J$. In Eq.~\eqref{eq:Gamma-def} we employed the dot
product notation
\begin{equation}
  J \cdot G 
  = \frac{1}{2} \sum_{\alpha \alpha'} J_{\alpha \alpha'} G_{\alpha
    \alpha'}
  = \frac{1}{2} \sum_\gamma J_\gamma G_\gamma 
  = -\frac{1}{2} \tr JG.
\end{equation}
Later we will use as well a trace based on the combined index
$\gamma=(\alpha,\alpha')$,
\begin{equation}
  \Tr X = \frac{1}{2} \sum_\gamma X_{\gamma \gamma},
\end{equation}
which is to be distinguished from the single-index trace $\tr
Y=\sum_\alpha Y_{\alpha \alpha}$.  Furthermore, we will use the dot
product inverse of a four-point function. It satisfies $(X \cdot
X^\inv)_{\gamma_1 \gamma_2} = I_{\gamma_1 \gamma_2}$.  Here
$I_{\gamma_1 \gamma_2}=\delta_{\alpha_1 \alpha_2}\delta_{\alpha'_1
  \alpha'_2} - \delta_{\alpha_1 \alpha'_2} \delta_{\alpha'_1
  \alpha_2}$ is the neutral element with respect to the dot product,
$X \cdot I = I \cdot X = X$.
 
The Luttinger-Ward functional is the difference between the 2PI
effective action in the interacting and noninteracting case,
\begin{align}
 \label{eq:Def_Phi}
  \Phi[G] 
  &= \Gamma[G]-\Gamma_0[G] 
  \\ \notag
  &= \Gamma[G] - \frac{1}{2} \left[\tr \ln (-G) - \tr(C^{-1}G-1)
  \right].
\end{align}
Diagrammatically, it is given by minus the sum of all
skeleton (2PI) diagrams contributing to the partition function, using
full propagators as lines. Its functional derivative with respect to
$G$ is minus the self-energy
\begin{equation}
  \label{eq:Phi1-Sigma}
  \Phi^{(1)}_{\gamma}[G] = \frac{\delta \Phi[G]}{\delta G_\gamma} =
  G^{-1}_\gamma - C^{-1}_\gamma  -J_{\gamma}[G] = - \Sigma_\gamma[G].
\end{equation}
We use a superscript ``$(n)$'' to indicate the
$n$-th functional derivative, for example $W^{(2)}_{\gamma_1 \gamma_2}
= \delta^2 W / \delta J_{\gamma_1} \delta J_{\gamma_2}$.  The
derivatives of $\Phi$ and of other functionals obey the symmetry
relations
\begin{align}
 \label{eq:symm_interchange_alphas}
 \begin{split}
 \Phi^{(n)}_{\gamma_1 \ldots \gamma_n}&=\Phi^{(n)}_{\gamma_{P1} \ldots
   \gamma_{Pn}},
 \\ \Phi^{(n)}_{\gamma_1 \ldots (\alpha_i,\alpha_i^\prime)\ldots
   \gamma_n}&=-\Phi^{(n)}_{\gamma_1 \ldots
   (\alpha_i^\prime,\alpha_i)\ldots \gamma_n}.
 \end{split}
\end{align}

One obtains the physical quantities (denoted by a bar) by setting the
external source $J$ to zero, for example $\overline{J}=0, \overline G
= G[\overline J], \overline \Sigma = - \overline \Phi {}^{(1)} =
- \Phi^{(1)}[\overline G]$.

An important quantity is the pair propagator
\begin{equation}
  \label{eq:def-pair-prop}
  \Pi_{\gamma_1 \gamma_2}[G]
  = -\frac{\delta G_{\gamma_2}}{\delta G^{-1}_{\gamma_1}} 
  = G_{\alpha'_1 \alpha_2} G_{\alpha_1 \alpha'_2} - G_{\alpha_1
    \alpha_2} G_{\alpha'_1 \alpha'_2}.
\end{equation}
It arises for instance in the flow Eq.~\eqref{eq:flow-Phi} of
$\Phi$ and in the Bethe-Salpeter equation
\begin{equation}
  \label{eq:Bethe-Salpeter}
 W^{(2)}=(\Pi^\inv + \Phi^{(2)})^\inv = \Pi - \Pi \cdot \Phi^{(2)} \cdot W^{(2)}.
\end{equation}

To conclude this section, let us briefly comment on the questions of
existence and uniqueness of the Luttinger-Ward functional. 

For some systems, the physical Green function has zeros, such that
$\det \overline G = 0$. Then, $\tr \ln(-\overline
G)=\ln \det(-\overline G)$ in Eq.~\eqref{eq:Def_Phi} is not
defined and $\Phi[\overline G]$ does not exist; the formalism is not
applicable in this case. This happens for gapped systems~\cite{Dav13}
which we do not investigate here.

The Legendre transformation in Eq.~\eqref{eq:Gamma-def} requires
the functional $J[G]$. A recent numerical study revealed the existence
of a $\tilde J \neq 0$ with $G[\tilde J]=G[0]=\overline G$ for some
models with on-site interaction.\cite{Koz15} This includes the
Anderson model which we study below. This finding means that there
exist two (or more) branches $J_i[G]$, $i=1,2$ of the functional
$J[G]$.  Two branches $\Gamma_i[G]$ of the 2PI effective action arise,
as well as two branches $\Phi_i[G]=\Gamma_i[G]-\Gamma_0[G]$ of the
Luttinger-Ward functional (the noninteracting $\Gamma_0[G]$ is unique)
and two branches $\Sigma_i[G] = - \Phi^{(1)}_i[G] = -G^{-1} + C^{-1}
+J_i[G]$ of the self-energy functional.  The physical state is
correctly described by the branch which satisfies $J_i[\overline G]=0$
or, equivalently, the self-consistency equation $\Sigma_i[\overline G]
= -\overline G{}^{-1} + C^{-1}$.  Which branch does so may depend on the
strength of the interaction, cf.~Ref.~\onlinecite{Koz15} as well
as Refs.~\onlinecite{Sta15} and~\onlinecite{Ros15} for toy model studies.  On any branch,
$\overline G$ is the only possible solution of the self-consistency
equation $\Sigma_i[G] = -G^{-1}+C^{-1}$, since $J_i[G]=0$ entails
$G=G[J_i[G]]=G[0]=\overline G$.  Therefore, the self-consistency
equation has the unique solution $\overline G$ on the physical branch
and no solution on the other branches.

Below we study approximate functionals $\Phi^\ix{app}[G]$ and
$\Sigma^\ix{app}[G]=-G^{-1} + C^{-1} + J^\ix{app}[G]$.  For these it
may occur that the self-consistency equation
$\Sigma^\ix{app}[G]=-G^{-1} + C^{-1}$ has several solutions $\overline
G{}^\ix{app}_j$ with $J^\ix{app}[\overline G{}^\ix{app}_j]=0$,
$j=1,2,\dots$.  They indicate the existence of several branches of the
functional $G^\ix{app}_j[J]$.  A prominent example are magnetic and
non-magnetic solutions of the self-consistent Hartree-Fock
approximation for the Anderson impurity model.\cite{And61}  All
$\overline G{}^\ix{app}_j = G{}^\ix{app}_j[J=0]$ are approximations to
the physical Green function $\overline G$ at vanishing external source
$J$.


\section{$\Phi$-derivable approximations}
\label{sec:Phi-symmetry}

In Refs.~\onlinecite{Bay61} and~\onlinecite{Bay62}, Baym and Kadanoff establish a method to
construct a class of conserving approximations referred to as
``$\Phi$-derivable''.  Here, we summarize those aspects of the method
which are most relevant to devise and apply a conserving 2PI fRG
approximation in the following sections.

A $\Phi$-derivable approximation is established in two steps.  The
first step is to set up an approximation to the Luttinger-Ward
functional $\Phi[G]$ that is invariant under the relevant symmetry
transformations of $G$.  The second step is to determine the physical
self-energy from a self-consistency equation.

Let us first discuss what it means if $\Phi$ is invariant under respective
symmetry transformations of $G$. Let $\theta$ represent the parameters
of the respective transformation. Then the invariance implies
\begin{equation}
  \label{eq:invariance-of-Phi}
  0 = \frac{\delta}{\delta \theta} \Phi[G[\theta]].
\end{equation}
For an illustration, we switch to the notation of Ref.~\onlinecite{Bay62} in
which real time arguments and no charge indices are used. In the case
of particle number conservation, $\theta=\theta(\vec{r},t)$ and
$G[\theta]$ is given by a gauge transformation
\begin{equation}
  G[\theta]_{\vec{r},\vec{r}'}(t,t') =
  e^{i\theta(\vec{r},t)} G_{\vec{r},\vec{r}'}(t,t')e^{-i\theta(\vec{r}',t')}.
\end{equation}
Reference~\onlinecite{Bay62} is concerned with diagrammatic approximations
to $\Phi[G]$ in terms of closed skeleton diagrams. For such
approximations, the invariance of $\Phi[G[\theta]]$ results from a
symmetry of the interaction vertices. For example, a density-density
interaction is invariant under a gauge transformation,
\begin{multline}
  \label{eq:U-inv}
  e^{-i\theta(\vec r'_1,t)} e^{-i\theta(\vec r'_2,t)}
  \langle \vec r'_1 \vec r'_2  | V | \vec r_1 \vec r_2 \rangle
  e^{i\theta(\vec r_1,t)} e^{i\theta(\vec r_2,t)}
  \\=   \langle \vec r'_1 \vec r'_2  | V | \vec r_1 \vec r_2 \rangle .
\end{multline}
Given a diagram to $\Phi$, one can combine each vertex with the
transformations belonging to the ends of the attached propagator lines
and obtain an invariant expression; this argument of Ref.~\onlinecite{Bay62}
proves the invariance of $\Phi$. It can be formulated as well in
charge index notation and with imaginary time arguments instead of
real ones. Thus, a simple way to set up an invariant approximation to
the Luttinger-Ward functional is to construct a diagrammatic
approximation.  Although this was not considered in Ref.~\onlinecite{Bay62}, one
can construct as well non-diagrammatic approximations to $\Phi[G]$
which are invariant.

Now, let us discuss the second step. Given an invariant approximate
$\Phi[G]$, a conserving approximation results when the physical
self-energy is determined from the self-consistency equation
$\overline \Sigma = -\Phi^{(1)}[G[\overline \Sigma]]$, in which
$G[\overline \Sigma]=(C^{-1} - \overline \Sigma)^{-1}$.
The physical two-particle Green function $\overline
W{}^{(2)}=-\left.\delta G/\delta J\right|_{J=0}$ can be obtained from
$\overline \Phi{}^{(2)}$ via the Bethe-Salpeter
equation~\eqref{eq:Bethe-Salpeter}. (Baym and Kadanoff use a
Bethe-Salpeter equation in the particle-hole channel
only.\cite{Bay61}) Physical quantities computed from $\overline
\Sigma$ and $\overline W {}^{(2)}$ respect conservation laws for
particle number, momentum, and energy.

In Sec.~\ref{sec:2PI-formalism} we mentioned that for some models
(including the Anderson impurity model) there exist unphysical
branches of the Luttinger-Ward functional.  By solving the
self-consistency equation one ensures that $\Phi$-derivable
approximations are indeed always on the physical branch.

A problem of $\Phi$-derivable approximations is that their
two-particle functions violate the crossing symmetry.\cite{Sta85,
Bic91a, Bic91b}  The exact solution for $W^{(2)}$ obeys the crossing
symmetry relation
\begin{multline}
  \left(W^{(2)}-\Pi\right)_{\alpha_1 \alpha_2 \alpha_3 \alpha_4}
  \\ =
  \sgn(P)
  \left(W^{(2)}-\Pi\right)_{\alpha_{P1} \alpha_{P2} \alpha_{P3} \alpha_{P4}}.
\end{multline}
for any $P$.  This relation is a consequence of the anti-commutativity
of fermionic field operators; it can hence be considered as a
manifestation of the Pauli principle.  By comparing different channels
of the Bethe-Salpeter equation~\eqref{eq:Bethe-Salpeter} one can show
that typical $\Phi$-derivable approximations violate the crossing
symmetry.  For instance, crossing symmetry is broken in the
self-consistent Hartree Fock and in the FLEX approximation, and also
in the conserving flow scheme which we derive in
Sec.~\ref{sec:conserving-approximations}.  In our application to
the Anderson model below we avoid the problem of violated crossing
symmetry by studying only quantities that can be derived from $\Sigma$
alone, without computing $W^{(2)}$.

In order to compute physical observables which directly benefit from the
conserving nature of $\Phi$-derivable approximations, one usually
needs to determine $W^{(2)}$. For example, suppose to split the lead
of the Anderson model studied below into a right and a left one.  Then
we could compute a left and a right conductance in linear response
from a four-point vertex like $\overline{W}{}^{(2)}$, and both
conductance values would be equal in $\Phi$-derivable
approximations. Although we here do not access $W^{(2)}$ and derived
observables, we call all $\Phi$-derivable approximations discussed
below ``conserving''.  This is appropriate as such observables could
be calculated, the conservation laws being guaranteed to hold (but the
crossing symmetry being broken).  $\Phi$-derivable approximations do
not only preserve conservation laws.  They have as well advantages for
quantities that can be derived from the self-energy alone. As examples
we now describe that they maintain the equivalence of different
approaches to the mean occupancy and that they preserve the Friedel
sum rule.  In the applicaton to the Anderson model below we return to
these issues, see Sec.~\ref{sec:Testing_PUF_MUF_non-cons} and Fig.~\ref{fig:SIAMeq_log_test_non-cons}(b).

\paragraph*{Mean occupancy.} The mean occupancy $\langle n_s \rangle_{J=0} =
\Tr a^\dagger_s a_s e^{-(H-\mu N)/T}/ \overline Z$ of a
single-particle state $s$ in the physical ($J=0$) thermal equilibrium
can on the one hand be computed from the imaginary time Green function with equal time
arguments,
\begin{align}
  \notag
  \overline G_{(-,\tau s)(+,\tau s)} 
  &=
  - \langle T a_s(\tau) a^\dagger_s(\tau) \rangle_{J=0} 
  \\ \label{eq:n-from-G}
   &=
  \langle a^\dagger_s a_s \rangle_{J=0}
  =
  \langle n_s \rangle_{J=0}.
\end{align}
On the other hand, $\langle n_s \rangle_{J=0}$ can be computed from
the grand potential $\overline \Omega = - T \ln \overline Z$.  For
this purpose we use a source term $\epsilon_s a^\dagger_s a_s$ in the
Hamiltonian, which is either present on physical grounds or added as
an auxiliary term.  Given hence a Hamiltonian of the form $H=\epsilon_s
a^\dagger_s a_s + H'$, we find
\begin{align}
  \notag 
  \frac{d \overline \Omega}{d \epsilon_s} 
  &=
  -\frac{T}{\overline Z} \frac{d}{d \epsilon_s} \Tr
  e^{-(\epsilon_s a^\dagger_s a_s + H'-\mu N)/T}
  \\ \label{eq:n-from-Omega}
  &=
  \frac{1}{\overline Z} \Tr a^\dagger_s a_s e^{-(H-\mu N)/T} 
  = 
  \langle n_s \rangle_{J=0}.
\end{align}
This holds even if $a^\dagger_s a_s$ and $H'$ do not commute.

Equations~\eqref{eq:n-from-G} and~\eqref{eq:n-from-Omega} are
equivalent; in an exact calculation, they would yield the same
result. However, for approximate calculations, this is in general not
guaranteed.  Truncated 1PI fRG, for instance, was found to spuriously
break the equivalence of the two equations in an application to the
Anderson impurity model.\cite{Kar08} For $\Phi$-derivable
approximations, in contrast, the thermodynamic consistency proven in Sec.~IV of
Ref.~\onlinecite{Bay62} ensures that both ways to determine $\langle
n_s \rangle_{J=0}$ yield the same result. We sketch the
argument only briefly.  

The parameter $\epsilon_s$ enters the generating functionals via the
free propagator $C$.  Therefore, the derivative $d \overline \Omega /
d \epsilon_s$ is formally given by the fRG flow equation for
$\overline \Omega$ with a flow parameter in $C$.  This is
Eq.~(46) of Ref.~\onlinecite{Ren15} and reads in the present context
\begin{equation}
  \label{eq:dOmega-depsilon}
  \frac{d \overline \Omega}{d \epsilon_s}
  =
  T \overline G \cdot \frac{d C^{-1} }{d \epsilon_s}.
\end{equation}
Its validity depends on the self-consistency $\overline \Sigma = -
\Phi{}^{(1)}[G[\overline \Sigma]]$ which is satisfied by construction
in $\Phi$-derivable approximations.  We insert $d C^{-1}_{++}/d
\epsilon_s = d C^{-1}_{--}/d \epsilon_s = 0$, $d C^{-1}_{-+}/d
\epsilon_s = - d C^{-1}_{+-}/d \epsilon_s$ and
\begin{equation}
  \frac{d}{d \epsilon_s} C^{-1}_{(+,\tau_1 s_1)(-,\tau_2 s_2)}
  =
  -\delta(\tau_1 - \tau_2) \delta_{s_1 s_2} \delta_{s_1 s}
\end{equation}
to find
\begin{equation}
  \frac{d \overline \Omega}{d \epsilon_s}
  =
  \overline G_{(-,\tau  s)(+,\tau s)}.
\end{equation}
Therefore, Eqs.~\eqref{eq:n-from-G} and \eqref{eq:n-from-Omega}
are equivalent for $\Phi$-derivable approximations like the conserving
fRG scheme from Sec.~\ref{sec:conserving-approximations} below.

\paragraph*{Friedel sum rule.} As a second example we consider the
Friedel sum rule.  It holds for an impurity in a host at zero
temperature and relates the scattering off the impurity to the charge
displacement which it induces. In approximate calculations, the Friedel sum rule is not guaranteed to
be preserved.  Truncated 1PI fRG, for instance, was found to
spuriously break the Friedel sum rule in an application to the
Anderson impurity model.\cite{Kar08} In contrast, $\Phi$-derivable
approximations keep the Friedel sum rule valid, as explained
now.

The rule was proven for interacting
systems by Langer and Ambegaokar in Ref.~\onlinecite{Lan61}. Their proof relies on the identity
\begin{equation}
  \label{eq:Langer-Ambegaokar-require}
  \tr \int \! d \nu \, e^{i \nu 0^+} \overline G_{-+}(i\nu)
  \frac{\partial \overline \Sigma_{+-}(i\nu)}{\partial \nu} = 0,
\end{equation}
in which trace and matrix multiplication indicate a summation over
single-particle states.  The argument $\nu$ denotes the Matsubara
frequency obtained by the usual Fourier transform [later we employ a
different convention for the Fourier transform in
Eqs.~\eqref{eq:FT_G} and~\eqref{eq:FT_Sigma}].
Equation~\eqref{eq:Langer-Ambegaokar-require} in turn was proven by
Luttinger and Ward in Ref.~\onlinecite{Lut60a} by exploiting that the vertices in
the diagrams to $\Phi$ conserve frequency.  This is a consequence of
the interaction being local in time and of the conservation of
particle number. Equation~\eqref{eq:Langer-Ambegaokar-require} holds
indeed in any $\Phi$-derivable approximation as long as the global
gauge transformation
\begin{equation}
  G_{(c_1 s_1 \tau_1)(c_2 s_2 \tau_2)}(\theta) = e^{i c_1 \tau_1
    \theta} G_{(c_1 s_1 \tau_1)(c_2 s_2 \tau_2)} e^{i c_2 \tau_2
    \theta}
\end{equation}
leaves the approximate $\Phi[\overline G(\theta)]$ invariant,
\begin{equation}
  0 
  =
  \left. \frac{d \Phi[\overline G(\theta)]}{d\theta}
  \right|_{\theta=0} 
  =
  \overline \Phi{}^{(1)} \cdot \left. \frac{d \overline G(\theta)}{d
      \theta} \right|_{\theta=0}.
\end{equation}
In fact, as particle number conservation entails $\overline G_{++} =
\overline G_{--} = 0$, this invariance equation can be written as
\begin{align}
  \notag 
  0 &= 
  -\frac{i}{2} \sum_{c_1 c_2} \sum_{s_1 s_2} \int \! d\tau_1 d\tau_2
  \, \overline \Sigma_{(c_1 s_1 \tau_1)(c_2 s_2 \tau_2)} 
  \\ & \qquad \qquad \qquad \quad  \times (c_1 \tau_1 + c_2 \tau_2) \overline G_{(c_1 s_1 \tau_1)(c_2 s_2 \tau_2)} 
  \\
  &= i \tr \int \! d\tau_1 d\tau_2
  \, \overline \Sigma_{+-}(\tau_1,\tau_2)(\tau_1 -
  \tau_2) \overline G_{-+}(\tau_2,\tau_1).
\end{align}
This leads to
\begin{align}
 \notag
  &0 =
  -\frac{1}{\beta^2} \tr \sum_{\nu_n,\nu_m} \overline
  \Sigma_{+-}(i\nu_n) \overline G_{-+}(i\nu_m) e^{i(\nu_n + \nu_m)0^+}
  \\ & \qquad \qquad \qquad \quad \times \frac{\partial}{\partial \nu_n} \int_{-\beta/2}^{\beta/2}d\tau \,
  e^{-i(\nu_n-\nu_m)\tau} 
  \\ 
  &\rightarrow - \tr \!\int \! \frac{d\nu_1 d\nu_2}{2\pi} \, \overline
  \Sigma_{+-}(i\nu_1) \overline G_{-+}(i\nu_2) e^{i(\nu_1 \!+
    \nu_2)0^+} \delta'\!(\nu_1\!-\!\nu_2)
\end{align}
in the limit $\beta = 1/T \rightarrow \infty$, which entails
Eq.~\eqref{eq:Langer-Ambegaokar-require}.  We thus indirectly confirmed the
validity of the Friedel sum rule in $\Phi$-derivable approximations
like the conserving fRG scheme from
Sec.~\ref{sec:conserving-approximations}.

The same reasoning holds for Luttinger's theorem. This
theorem applies to bulk systems at zero temperature and states that
the volume in momentum space in which the real part of the physical
Green function at zero frequency is positive is given by the average
particle number. Its derivation in Refs.~\onlinecite{Lut60a} and~\onlinecite{Lut60b} is based on
the same identity~\eqref{eq:Langer-Ambegaokar-require} as used for the
Friedel sum rule.  Therefore, Luttinger's theorem is preserved in
$\Phi$-derivable approximations.

The proof of Eq.~\eqref{eq:Langer-Ambegaokar-require} described
above obviously requires the existence of the Luttinger-Ward
functional.  As mentioned in Sec.~\ref{sec:2PI-formalism}, there are
systems for which the Luttinger-Ward functional does not exist.  The
Friedel sum rule and Luttinger's theorem may then by violated.
Explicit examples for the breakdown of Luttinger's theorem are
known.\cite{Dav13,Ros07}


\section{Conserving and non-conserving approximations to the $U$-flow}
\label{sec:Cons_and_non-cons_appr}

If a flow parameter $\lambda$ is introduced into the action, the
$\lambda$-derivative of the Luttinger-Ward functional is given by an
fRG flow equation.  In Ref.~\onlinecite{Ren15}, we focused on the hierarchy of
flow equations that emerges for the physical vertex functions
$\overline \Phi{}^{(n)}$.  For the case that the flow parameter is
introduced into the free propagator $C$ (``$C$-flow''), we proved the
equivalence of the truncated hierarchy to the well-known conserving
self-consistent perturbation theory.  For the ``$U$-flow'', where the
flow parameter enters instead the two-particle interaction $U$, we did
not discuss how the truncation schemes relate to conserving
approximations; we do so in this paper.  We show how
$\Phi$-derivable, conserving approximations can be constructed and how
they are connected to the $U$-flow approximations used
in Ref.~\onlinecite{Ren15}.

In this Sec.~\ref{sec:Cons_and_non-cons_appr} we impose only few
restrictions on the form in which the two-particle interaction
$U_\lambda$ depends on the parameter $\lambda$ flowing from
$\lambda_\ix{i}$ to $\lambda_\ix{f}$.  First, we require that the
interaction vanishes at the beginning of the flow,
$U_{\lambda_\ix{i}}=0$.  Consequently, the Luttinger-Ward functional
and its vertex functions vanish there.  Second, the original
interacting problem is fully restored at the end of the flow,
$U_{\lambda_\ix{f}}=U$.  Finally, $U_\lambda$ has the same full index
permutation antisymmetry as the bare interaction.  In the discussion
of the Anderson model below, we choose a multiplicative flow parameter
$U_\lambda=\lambda U$ with $\lambda$ flowing from $0$ to $1$.  This
simple choice of the flow parameter is sufficient; there is no need to
regularize any divergence since perturbation theory in powers of $U$
is well behaved for the Anderson model.\cite{Yam75}  When a flow of
the propagator is considered in addition to the flow of the
interaction, we introduce the flow parameter differently [cf.
Eq.~\eqref{eq:CU-cut-off}].


\subsection{Conserving approximations to the $U$-flow of $\Phi[G]$}
\label{sec:conserving-approximations}

Let us construct conserving approximations to the $U$-flow.  According
to Eq.~(77) of Ref.~\onlinecite{Ren15}, the $U$-flow of the Luttinger-Ward
functional is given by
\begin{equation}
  \label{eq:flow-Phi}
  \dot{\Phi}_\lambda =
  \frac{1}{3!} \Tr \dot{U}_\lambda \cdot \left[\left(\Pi^\inv +
      \Phi_\lambda^{(2)}\right)^\inv +
    \frac{\Pi}{2} \right], 
\end{equation}
in which the dot denotes the derivative with respect to the flow
parameter $\lambda$. This exact flow equation is the first of an
infinite hierarchy: the flow of $\Phi_\lambda$ depends on
$\Phi^{(2)}_\lambda$, that of $\Phi^{(2)}_\lambda$ involves
$\Phi^{(3)}_\lambda$ and $\Phi^{(4)}_\lambda$, and so on.

In order to compute one-particle properties, we require an
approximation to the physical value $\overline \Sigma = - \overline
\Phi{}^{(1)}$ of the self-energy.  According to
Sec.~\ref{sec:Phi-symmetry}, a conserving approximation follows
from the self-consistency equation $\overline \Sigma = -
\Phi{}^{(1)}[G[\overline \Sigma]]$ if $\Phi$ satisfies the invariance
equation $\delta \Phi[G[\theta]] / \delta \theta = 0$.  Let us hence
study how one can obtain such an invariant $\Phi$ from a truncated
flow equation.

The simplest truncation is to set $\Phi^{(2)}_\lambda =
\Phi^{(2)}_{\lambda_\ix{i}}=0$ on the right-hand side of
Eq.~\eqref{eq:flow-Phi}. The resulting flow equation reads as
\begin{eqnarray}
  \label{eq:lowest_flow_eq_Phi}
  \dot{\Phi}_\lambda &=&
  \frac{1}{3!} \Tr \dot{U}_\lambda \cdot \frac{3}{2} \Pi = \frac{1}{2}
  G \cdot \dot U_\lambda \cdot G.
\end{eqnarray}
As the flow starts at $U=0$ and $\Phi=0$, the solution is
\begin{equation}
  \Phi = \frac{1}{2} G \cdot U \cdot G
\end{equation}
which is the first order perturbation theory result for the Luttinger
Ward functional. The self-consistency equation $\overline\Sigma =
-\overline \Phi{}^{(1)} = -U\cdot \overline G$ yields precisely the
well known conserving Hartree-Fock approximation.

Now we consider the next higher order of truncation. We replace
$\Phi^{(2)}_\lambda$ on the right-hand side of Eq.~\eqref{eq:flow-Phi}
by its leading perturbative value $U_\lambda$ (cf.~also Sec.~6.2
of Ref.~\onlinecite{Ren15}). The resulting flow equation is
\begin{align}
 \label{eq:flow_eq_Phi}
 \dot{\Phi}_\lambda &= \frac{1}{3!} \Tr \dot{U}_\lambda \cdot 
  \left[\frac{3}{2} \Pi + \Pi \cdot \sum_{k=1}^\infty (-U_\lambda\cdot \Pi)^k
 \right] 
 \\ 
 & =\frac{d}{d\lambda} \left[ \frac{1}{4} \Tr U_\lambda \cdot \Pi -
   \frac{1}{3!} \sum_{k=1}^\infty \frac{1}{k+1} \Tr
   \left(-U_\lambda\cdot \Pi\right)^{k+1} \right] .
\end{align}
Here, $(U_\lambda \cdot \Pi)^k$ denotes the $k$-fold dot product
$(U_\lambda \cdot \Pi) \cdot (U_\lambda \cdot \Pi) \cdot \ldots \cdot
(U_\lambda \cdot \Pi)$.  If $U_\lambda$ has the same invariance under
symmetry transformations as $U$, then $\delta \dot
\Phi_\lambda[G[\theta]] / \delta \theta = 0$.  To see this one can
apply the same argument as used after Eq.~\eqref{eq:U-inv}: combine
the vertices $\dot U_\lambda$ or $U_\lambda$ with the transformations
belonging to the ends of the attached propagators $G[\theta]$ (hidden
in $\Pi[G[\theta]]$) to invariant expressions.  In this case, the
invariance of $\Phi_\lambda$ is respected during all of the flow.  If,
however, $\dot U_\lambda$ does not have the symmetry, the invariance
equation for $\Phi_\lambda$ is violated during the flow. Nevertheless,
it is reestablished at the end of the flow by the solution
\begin{equation}
  \label{eq:2PI_fRG_anal_result_Phi}
  \Phi^\ix{cfRG} =\frac{1}{4} \Tr U \cdot \Pi - \frac{1}{3!}
  \sum_{k=2}^\infty \frac{1}{k} \Tr \left(-U\cdot
    \Pi\right)^k. 
\end{equation}
We label this conserving approximation scheme by ``cfRG''.\footnote{This conserving 2PI functional RG scheme should not be confused with the ``constrained functional RG'' proposed by Kinza and Honerkamp\cite{Kin15} which they also abbreviate as ``cfRG''.}
$\Phi^\ix{cfRG}$ deviates from the exact Luttinger-Ward functional in
order $U^3 G^6$ and higher.  The corresponding approximate self-energy
functional $\Sigma^\ix{cfRG}[G]=-\Phi^{\ix{cfRG}\,(1)}[G]$ can be
determined from the rule $\Tr A \cdot \delta \Pi / \delta G = 4 A^\R
\cdot G$. Here, $A$ denotes any four-point function with the usual
symmetries $A_{\alpha_1 \alpha'_1 \alpha_2 \alpha'_2} = A_{\alpha_2
  \alpha'_2 \alpha_1 \alpha'_1} = -A_{\alpha'_1 \alpha_1 \alpha_2
  \alpha'_2}$, and we defined $A^\R$ via
\begin{equation}
  A^\R_{\alpha_1 \alpha'_1 \alpha_2 \alpha'_2} =
  A_{\alpha_1 \alpha_2 \alpha'_2 \alpha'_1}.
\end{equation}       
In $A^\R \cdot G$, $G$ connects one index from the left index pair of
$A$ to one index from the right pair.  Applying the differentiation
rule yields
\begin{align}
  \notag
  \Sigma^\ix{cfRG} &= -U^\R \cdot G  - \frac{2}{3} \sum_{k=1}^\infty
  \left[(-U \cdot \Pi)^k \cdot U\right]^\R \cdot G
  \\ &= -U \cdot G + \frac{2}{3} (\Upsilon \cdot U)^\R \cdot G,
  \label{eq:A_find_Sigma}
\end{align}
in which we introduced $\Upsilon=- \sum_{k=1}^\infty (-U\cdot \Pi)^k$.
When we insert $G=(C^{-1} - \Sigma^\ix{cfRG})^{-1}$ and solve
the resulting self-consistency equation we obtain the physical value
$\overline{\Sigma}{}^\ix{cfRG}$.  From $\Phi^\ix{cfRG}=\Phi_\ix{exact}+\mathcal{O}(U^3
G^6)$ follows that $\overline \Sigma{}^\ix{cfRG}$ comprises all diagrams
from second order self-consistent perturbation theory: $\overline
\Sigma{}^\ix{cfRG} = \overline \Sigma_\ix{exact} + \mathcal{O}(U^3 \overline{G}
{}_\ix{2SCPT}^5)$, where $\overline G_\ix{2SCPT}$ denotes the full
propagator of the physical state in second order self-consistent
perturbation theory.

We do not discuss higher order conserving truncation schemes since
their analytic structure becomes increasingly complicated.  Also their
numerical solutions are difficult to realize; as the flowing objects
are functionals, their numerical sampling would require a grid in the
infinite dimensional space of functions.

Apart from the $U$-flow scheme given by Eq.~\eqref{eq:flow-Phi},
we studied in Ref.~\onlinecite{Ren15} as well a modification originally developed
in Ref.~\onlinecite{Dup14}. In the definition of the modified variant, the first
order contribution to $\Phi$ is excluded from the replacement $U \to
U_\lambda$. As a consequence, the RG flow does not start at
$\Phi_{\lambda_\ix{i}}=0$ but at the Hartree-Fock solution,
$\Phi_{\lambda_\ix{i}}=\frac{1}{2}G\cdot U\cdot G$. If one truncates
the corresponding flow equation for $\Phi_\lambda$ by setting
$\Phi^{(2)}_\lambda=\Phi^{(2)}_{\lambda_\ix{i}} = U$, the final
solution is again the conserving approximation $\Phi^\ix{cfRG}$ given
in Eq.~\eqref{eq:2PI_fRG_anal_result_Phi}. Hence, both approaches
in their respective truncations are equivalent.


\subsection{Similarity between the cfRG and the FLEX
  approximation}
\label{sec:Similarity_2PI_fRG_FLEX}

\begin{figure*}
 \includegraphics[width=0.8\linewidth,clip]{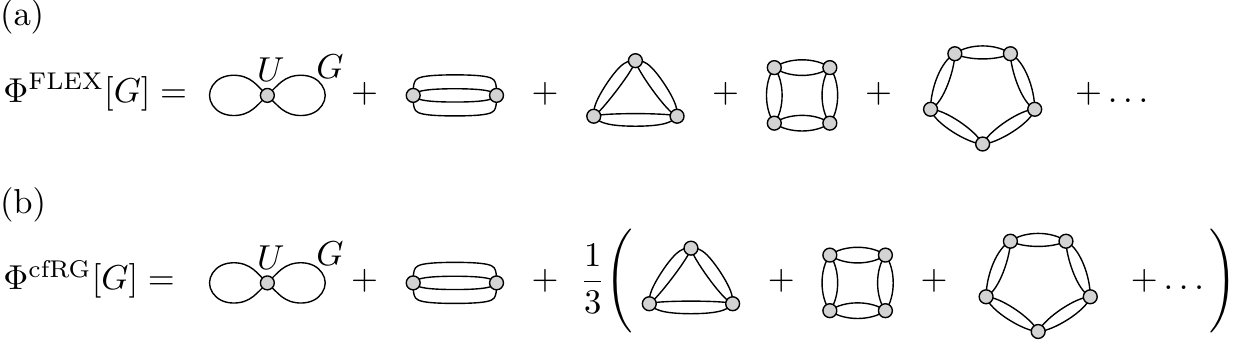}
 \caption{\label{fig:FLEX-diagrams} (a) Diagrammatic representation of
   $\Phi^\ix{FLEX}$.  Our antisymmetrized charge index notation leads
   to diagrams with undirected lines and dot-like Hugenholtz vertices.
   When $\Phi^\ix{FLEX}$ is expressed in terms of diagrams composed of
   directed propagator lines and Feynman interaction lines, one can
   see that it comprises three different channels: one with
   longitudinal spin fluctuations and density fluctuations, one with
   transverse spin fluctuations, and one with particle-particle
   fluctuations, cf.~Refs.~\onlinecite{Bic89a} and~\onlinecite{Bic89b}. (b) Diagrammatic
   representation of $\Phi^\ix{cfRG}$.}
\end{figure*}

The cfRG approximation of Sec.~\ref{sec:conserving-approximations}
is closely related to the FLEX approximation of Refs.~\onlinecite{Bic89a} and~\onlinecite{Bic89b}
which was heavily used to study high
temperature superconductivity.\cite{Man04} The
FLEX approximation is as well $\Phi$-derivable. The approximate
Luttinger-Ward functional $\Phi^\ix{FLEX}$ is computed from a series
of diagrams that describe ringlike pair-propagation, see
Fig.~\ref{fig:FLEX-diagrams}(a). The motivation for this
approximation is to incorporate effects resulting from the exchange of
spin, density and particle-particle fluctuations.  Compared to the
expansion of the exact Luttinger-Ward functional, the first missing
diagram is of order $U^4 G^8$.

Each of the diagrams in Fig.~\ref{fig:FLEX-diagrams} represents
several total index pairings according to the Wick theorem. A pairing
$P$ that contributes to a diagram of order $U^n G^{2n}$ has the value
\begin{equation}
  \frac{\sgn(P)(-1)^{n+1}}{n!4!^n} U\dots U G\dots G,
\end{equation}
in which $U\dots U G\dots G$ is a short hand notation for the
appropriate index contractions.  Summing up all diagrammatic
contributions leads to
\begin{equation}
  \Phi^\ix{FLEX}\!=\!\frac{1}{4} \!\Tr U \cdot \Pi -\frac{1}{12} \!\Tr U \cdot
  \Pi \cdot U \cdot \Pi -\frac{1}{2} \sum_{k=3}^\infty \frac{1}{k} \Tr
  \left(-U\cdot \Pi\right)^k
\end{equation}
[see Eqs.~(148) --- (151) of Ref.~\onlinecite{Mai05} for an expression in
charge-index free notation]. The functional $\Phi^\ix{cfRG}$ from the
cfRG approximation given in
Eq.~\eqref{eq:2PI_fRG_anal_result_Phi} is identical to
$\Phi^\ix{FLEX}$ except for a factor $\frac{1}{3}$ in front of all
diagrams of third order in $U$ and higher.  We conclude that
$\Phi^\ix{cfRG}$ accounts precisely for the FLEX diagrams, including
however only a part of the weight of the higher order diagrams, see
Fig.~\ref{fig:FLEX-diagrams}(b).  We
verified explicitly that the missing $\frac{2}{3}$ of those
diagrams are generated in the conserving fRG by terms which are
neglected in our truncation scheme.

In FLEX, the self-consistency equation for the self-energy reads
\begin{equation}
  \label{eq:FLEX_eq_for_Sigma}
  \overline \Sigma{}^\ix{FLEX} = - U \cdot \overline G - \frac{4}{3}
  (U \cdot \overline \Pi \cdot U)^\R \cdot \overline G + 2 (\overline
  \Upsilon \cdot U)^\R \cdot \overline G. 
\end{equation}


\subsection{Non-conserving approximations to the $U$-flow of $\overline \Sigma$}
\label{sec:non-conserving-approximations}

Let us study how the cfRG approximation of
Sec.~\ref{sec:conserving-approximations} is related to the
truncations of $U$-flow described in Ref.~\onlinecite{Ren15}.  Instead of
computing the flow of the whole functional $\Phi_\lambda[G]$ we now
consider only the flow of our quantity of interest, namely the
physical value of the self-energy $\overline \Sigma_\lambda =
-\overline \Phi {}^{(1)}_\lambda$.  The corresponding flow equation is
\begin{equation}
  \label{eq:flow-Phi-bar-1}
  \dot{\overline \Sigma}_\lambda = -\overline{\dot \Phi}{}^{(1)}_\lambda
  - \overline \Phi{}^{(2)}_\lambda \cdot \dot{\overline G}_\lambda,
\end{equation}
in which according to Ref.~\onlinecite{Ren15}
\begin{align}
  \label{eq:flow-Phi1}
  \dot \Phi {}_\lambda^{(1)} &=
  \frac{1}{3!} \Tr \dot{U}_\lambda \cdot \left[\frac{1}{2}\frac{\delta \Pi}{\delta G}\right.
  \\ \notag
   &\qquad \quad \left.+ W_\lambda^{(2)} \cdot
    \left(\Pi^\inv \cdot \frac{\delta \Pi}{\delta G} \cdot \Pi^\inv -
      \Phi_\lambda^{(3)} \right) \cdot W_\lambda^{(2)} \right], 
  \\
  \label{eq:flow-G-bar}
  \dot{\overline G}_\lambda &= \overline \Pi_\lambda \cdot
  \dot{\overline \Sigma}_\lambda, 
\end{align}
with $W^{(2)} = \left(\Pi^\inv +
  \Phi^{(2)}\right)^\inv$. The right-hand sides depend on $\overline
\Phi {}^{(2)}_\lambda$ and $\overline \Phi{}^{(3)}_\lambda$.
Nontrivial truncations of the flow Eq.~\eqref{eq:flow-Phi} for
$\Phi_\lambda[G]$ produce approximate functionals $\Phi_\lambda[G]$
with non-vanishing $\overline \Phi {}^{(n)}_\lambda \neq 0$, $n\ge 1$,
compare for example Eq.~\eqref{eq:2PI_fRG_anal_result_Phi}.
[Only the most basic truncation Eq.~\eqref{eq:lowest_flow_eq_Phi}
produces the Hartree-Fock solution with $\overline \Phi
{}^{(n)}_\lambda=0$ for $n\ge 3$.]  Consequently, we need
$\overline \Phi {}^{(2)}_\lambda$ and $\overline
\Phi {}^{(3)}_\lambda$ in order to determine the flow of $\overline
\Sigma_\lambda$.  However, the flow of $\overline \Phi
{}^{(2)}_\lambda$ and $\overline \Phi {}^{(3)}_\lambda$ depends on
higher $\overline \Phi {}^{(n)}_\lambda$, and so on. We face a new
infinite hierarchy of coupled flow equations describing the flow of
the physical values that correspond to the approximate
$\Phi_\lambda[G]$.

Let us examine the idea to truncate as well the new hierarchy. In this
way one obtains an approximation to the conserving approximation,
which we expect to be in general non-conserving.

Consider for example the approximate flow of the functional $\Phi[G]$
described by Eq.~\eqref{eq:flow_eq_Phi}. It results from setting
$\Phi^{(2)}_\lambda = U_\lambda$ on the right-hand side of
Eq.~\eqref{eq:flow-Phi}.  Let us hence truncate as well the new
hierarchy for the physical values by setting $\overline \Phi
{}^{(2)}_\lambda = U_\lambda$, $\overline \Phi {}^{(3)}_\lambda = 0$
on the right-hand side of Eqs.~\eqref{eq:flow-Phi-bar-1}
and~\eqref{eq:flow-Phi1}. This leads precisely to the approximation
scheme for the plain $U$-flow derived and used in Ref.~\onlinecite{Ren15}. The
resulting flow equation is 
\begin{multline}
  \dot{\overline \Sigma}{}^\ix{PUF}_{\lambda}
  = 
  -\frac{2}{3} \left(\overline{\Upsilon}_\lambda
    \cdot \dot U_\lambda 
    \cdot \overline{\Upsilon}_\lambda^\ix{T} 
    - \overline{\Upsilon}_\lambda \cdot \dot{U}_\lambda -
    \dot{U}_\lambda \cdot \overline{\Upsilon}_\lambda^\ix{T}  
  \right)^\R \cdot  \overline G^\lambda
  \\ - \dot U_\lambda \cdot \overline G_\lambda - U_\lambda \cdot
  \dot{\overline G}_\lambda, 
  \label{eq:PUF_flow_eq}
\end{multline}
with $\Upsilon^\ix{T}_{\gamma_1 \gamma_2}=\Upsilon_{\gamma_2
  \gamma_1}$.  We label this approximation scheme by ``PUF''.  The
corresponding initial conditions are $U_{\lambda_\ix{i}}=0$ and
$\overline{\Sigma}{}^\ix{PUF}_{\lambda_\ix{i}}=0$.  We have identified
this scheme as a probably non-conserving approximation to the cfRG
approximation of Sec.~\ref{sec:conserving-approximations}.
Numerical results for the impurity occupancy of the Anderson model
show that the PUF approximation is non-$\Phi$-derivable, cf.
Sec.~\ref{sec:PUF_StUF_MUF_num_non-cons}. We therefore expect possible 
extensions of this approximation scheme which access two-particle functions to be non-conserving.

The same strategy can be applied in the framework of the modified
$U$-flow.  Then it leads to the approximation for the modified
$U$-flow derived and used in Ref.~\onlinecite{Ren15}.  It obviously constitutes
another non-$\Phi$-derivable approximation to the cfRG approximation.  The
corresponding flow equation is
\begin{multline}
  \label{eq:MUF_flow_eq}
  \dot{\overline \Sigma}{}^\ix{MUF}_{\lambda}
  =
  -\frac{2}{3} \left(\overline{\Upsilon}_\lambda
    \cdot \dot U_\lambda \cdot
    \overline{\Upsilon}_\lambda^\ix{T} -
    \overline{\Upsilon}_\lambda \cdot \dot{U}_\lambda  
    - \dot{U}_\lambda \cdot \overline{\Upsilon}_\lambda^\ix{T} 
  \right)^\R \cdot \overline G^\lambda 
  \\ - U \cdot  \dot{\overline{G}}_\lambda, 
\end{multline}
with the self-consistent Hartree-Fock self-energy as starting point,
$\overline{\Sigma} {}^\ix{MUF}_{\lambda_\ix{i}}=\overline{\Sigma}
{}^{\ix{HF}}=-U \cdot G[\overline \Sigma {}^\ix{HF}]$.  We use the
label ``MUF'' for this specific approximation.

Let us apply the idea of a second truncation as well to the most basic
truncation scheme from Eq.~\eqref{eq:lowest_flow_eq_Phi}. This has
been constructed by setting $\Phi^{(2)}_\lambda = 0$ on the right-hand
side of the flow equation~\eqref{eq:flow-Phi} for the
functional. Accordingly, we truncate the new hierarchy for the
physical values by setting $\overline \Phi {}^{(2)}_\lambda = 0$,
$\overline \Phi {}^{(3)}_\lambda = 0$ on the right-hand side of
Eqs.~\eqref{eq:flow-Phi-bar-1} and~\eqref{eq:flow-Phi1}. This leads to the
flow equation
\begin{equation}
  \label{eq:StUF_flow_eq}
  \dot{\overline \Sigma}{}^\ix{StUF}_\lambda = -\dot U_\lambda \cdot \overline G
\end{equation}
with initial condition
$\overline{\Sigma}{}^\ix{StUF}_{\lambda_\ix{i}}=0$. It provides a
simple, static approximation to the physical self-energy, which we
refer to as ``StUF''.  The existence of this approximation was
overlooked in Ref.~\onlinecite{Ren15}.


\subsection{Combined $C$- and $U$-flow}

It was shown in Ref.~\onlinecite{Ren15} that straightforward truncations of
$C$-flow lead to standard self-consistent perturbation theory.  For
the Anderson model which we study below, Ref.~\onlinecite{Whi92} provides data from second order self-consistent
perturbation theory, equivalent to $C$-flow truncated at level $4$.
We observe that these data typically deviate from the exact result in
the opposite direction than those obtained by the $U$-flow methods
from the previous sections.  Therefore, we suspect that mixing both
schemes could improve the approximation.  The idea of introducing a
flow parameter into both, $C$ and $U$, was already formulated
in Ref.~\onlinecite{Dup14}.  Here, we refer to this approach as $CU$-flow.

The Luttinger-Ward functional $\Phi[G]$ does not depend on $C$. As a
consequence, the flow equation for $\Phi_\lambda [G]$ in the $CU$-flow
is identical to the $U$-flow case and given by
Eq.~\eqref{eq:flow-Phi}.  Accordingly, the conserving approximations
to the $U$-flow of $\Phi[G]$ from
Sec.~\ref{sec:conserving-approximations} pertain as well to the
$CU$-flow.

In contrast, new (non-$\Phi$-derivable) approximations arise for the
$CU$-flow of the physical value of the self-energy, $\dot{\overline
  \Sigma}_\lambda = -\overline{\dot \Phi}{}^{(1)}_\lambda - \overline
\Phi{}^{(2)}_\lambda \cdot \dot{\overline G}_\lambda$.  While $\dot
\Phi {}_\lambda^{(1)}$ is still given by Eq.~\eqref{eq:flow-Phi1},
$\dot{\overline G}_\lambda$ now satisfies
\begin{equation}
   \label{eq:CUflow-G-bar}
  \dot{\overline G}_\lambda = \overline \Pi_\lambda \cdot
  \left(\dot{\overline \Sigma}_\lambda - \dot C^{-1}_\lambda  \right)
\end{equation}
instead of Eq.~\eqref{eq:flow-G-bar}. Here, $\dot{C} {}_\lambda^{-1}$ denotes $d(C_\lambda^{-1})/d\lambda$.
Let us apply the same truncation as in
Sec.~\ref{sec:non-conserving-approximations} and set $\overline
\Phi{}^{(2)}_\lambda=U_\lambda$ and $\overline \Phi{}^{(3)}_\lambda =
0$ on the right-hand sides. This results in an equation that is
formally identical to Eq.~\eqref{eq:PUF_flow_eq} from the
$U$-flow, however with $\dot{\overline G}_\lambda$ now given by
Eq.~\eqref{eq:CUflow-G-bar}.

For the Anderson impurity model studied below it is known that second
order perturbation theory provides good approximations for $U\ll \pi
\Gamma$,\cite{Yam75} with $\Gamma$ being a measure for the coupling between impurity level and lead.  On that account we demand that the approximate
$\overline \Sigma$ obtained from the truncated $CU$-flow is exact up
to second order in $U$ (as are $\overline{\Sigma}{}^\ix{cfRG}$,
$\overline{\Sigma}{}^\ix{FLEX}$, $\overline{\Sigma}{}^\ix{PUF}$ and
$\overline{\Sigma}{}^\ix{MUF}$). The truncation described above does
not satisfy this condition. The perturbative expansion of the exact
physical value of the self-energy in the presence of a flow parameter
is given by
\begin{equation}
  \overline \Sigma {}_\lambda^\ix{exact}
  =
  D^\ix{1st}_\lambda + D^\ix{2ndHF}_\lambda + D^\ix{2ndS}_\lambda + \mathcal{O}(U^3),
\end{equation}
in which $D^\ix{1st}_\lambda = - U_\lambda \cdot C_\lambda$ denotes
the value of the first order diagram, $D^\ix{2ndHF}_\lambda =
U_\lambda \cdot \Pi^0_\lambda \cdot U_\lambda \cdot C_\lambda$ that of
the (non-skeleton) second order diagram contained in self-consistent
Hartree-Fock, and $D^\ix{2ndS}_{\lambda} = \frac{2}{3} (U_\lambda
\cdot \Pi^0_\lambda \cdot U_\lambda)^\R \cdot C_{_\lambda}$ that of
the skeleton second order diagram. [$\Pi^0$ is defined by
Eq.~\eqref{eq:def-pair-prop} with $C$ replacing $G$.]  One can
show that the above truncation satisfies
\begin{align}
  \label{eq:Sigma_CUF_before_redefinition}
  &\dot{\overline \Sigma}_{\lambda,\gamma_1} 
  =
  \dot D{}^\ix{1st}_{\lambda,\gamma_1} 
  + \dot D{}^\ix{2ndHF}_{\lambda,\gamma_1}
  \\ \notag
  &+  \frac{2}{3} \left(U_\lambda \cdot \Pi^0_\lambda
    \cdot \dot U_\lambda + \dot U_\lambda \cdot \Pi^0_\lambda \cdot
    U_\lambda \right)^{\!\R} \!\!\cdot C_\lambda
  + \mathcal{O}(U_\lambda^2 \dot U_\lambda, U^3_\lambda).
\end{align}
Obviously, the last addend of
\begin{multline}
  \dot D{}^\ix{2ndS}_{\lambda}
  =
  \frac{2}{3}  \left(U_\lambda \cdot \Pi^0_\lambda
    \cdot \dot U_\lambda + \dot U_\lambda \cdot \Pi^0_\lambda \cdot
    U_\lambda \right)^\R \cdot C_{\lambda}
  \\+
  2(U_\lambda \cdot \Pi^0_\lambda \cdot U_\lambda)^\R \cdot \dot C_{\lambda}
  \label{eq:D2ndS-dot-parts}
\end{multline}
is missing.

Let us formulate a minimal extension of the above truncation scheme
which makes $\overline \Sigma$ exact up to second order in $U$.  As it
is insufficient to truncate the flow equation by the first order
approximation $\overline \Phi{}^{(2)}_\lambda=U_\lambda$, we consider
the second order approximation,
\begin{align}
  &\left.\Phi^{(2)}_{\lambda}[G]\right|_\ix{2nd}
  = U_{\lambda} 
  +
  V_{\lambda}[G]
  \\
  &V_{\lambda}[G]
  =
  - (U_\lambda \cdot \Pi[G] \cdot U_\lambda)^\R
  - {(U_\lambda \cdot \Pi[G] \cdot U_\lambda)^\R}^\R.
\end{align}
For $\overline
\Sigma$ to be exact in second order, it is indeed sufficient to use
$\overline \Phi{}^{(2)}_\lambda = \overline
\Phi{}^{(2)}_\lambda\!\!\left.\vphantom{\Phi^\lambda}\right|_\ix{2nd}$
for one particular $\overline \Phi{}^{(2)}_\lambda$ in the flow
equation only.  The other vertex functions can be truncated as before
by $\overline \Phi{}^{(2)}_\lambda=U_\lambda$ and $\overline
\Phi{}^{(3)}_\lambda=0$.  In this way, the numerical effort for
solving the flow equations does not increase
significantly. Specifically, we truncate the flow equation
\begin{equation}
  \dot{\overline \Sigma}_\lambda 
  =
  -\overline{\dot \Phi}{}_\lambda^{(1)} 
  - 
  \overline \Phi{}^{(2)}_\lambda \cdot \overline \Pi_\lambda \cdot
  \dot{\overline \Sigma}_\lambda
  + \overline
  \Phi{}^{(2)}_\lambda \cdot \overline \Pi_\lambda \cdot \dot
  C_\lambda^{-1}
\end{equation}
to
\begin{multline}
  \label{eq:CU_flow_eq}
  \dot{\overline \Sigma}{}_\lambda^\ix{CUF}
  =
  -\left.\overline{\dot \Phi}{}_\lambda^{(1)}\right|_{\overline
    \Phi{}^{(2)}_\lambda \rightarrow U_\lambda, \overline \Phi{}^{(3)}
    \rightarrow 0}
  \\- 
  U_\lambda \cdot \overline \Pi_\lambda \cdot
  \dot{\overline \Sigma}{}_\lambda^\ix{CUF}
  + \left. \overline
    \Phi{}^{(2)}_{\lambda}\right|_\ix{2nd}\cdot \overline \Pi_\lambda \cdot
  \dot C_\lambda^{-1}.
\end{multline}
We label this approximation by ``CUF''.  Compared to
Eq.~\eqref{eq:Sigma_CUF_before_redefinition}, it includes the
additional addend
\begin{equation}
 \label{eq:Def_dot_Delta_Sigma}
  \overline{V}_{\lambda} \cdot \overline \Pi_\lambda \cdot
  \dot C_\lambda^{-1}
  =
  2(U_\lambda \cdot \Pi^0_\lambda \cdot U_\lambda)^\R \cdot \dot
  C_{\lambda} + \mathcal{O}(U^3). 
\end{equation}
which contains indeed the missing part of $\dot D{}^\ix{2ndS}_\lambda$
from Eq.~\eqref{eq:D2ndS-dot-parts}.

So far we have discussed a combination of the $C$-flow with the plain
$U$-flow.  Likewise it is possible to combine the $C$-flow with the
modified $U$-flow.  As described in Ref.~\onlinecite{Dup14}, the starting point
of the flow is then $\overline \Sigma_{\lambda_\ix{i}}=0$, whereas the
original modified $U$-flow intentionally starts at $\overline
\Sigma_{\lambda_\ix{i}}=\overline \Sigma{}^\ix{HF}$.  For the Anderson
impurity model, we have implemented both, the combination of $C$-flow
with plain and with modified $U$-flow.  As the results are
qualitatively similar, we do not present further details on the
combination with modified $U$-flow.

Concerning the flow parameter, we combine a sharp infrared cut-off of
the imaginary frequency in the free propagator with an exponential
rescaling of the interaction amplitude,
\begin{equation}
  \label{eq:CU-cut-off}
 C_\lambda(\nu_n)=C(\nu_n) \theta(|\nu_n|-\lambda), \qquad 
 U_\lambda=e^{-\lambda/\Lambda} U,
\end{equation}
in which $\lambda$ flows from infinity to zero.  The resulting initial
conditions are $C_{\lambda_\ix{i}}=0$, $U_{\lambda_\ix{i}}=0$,
$\overline \Sigma{}_{\lambda_\ix{i}}^{\ix{CUF}_\Lambda}=0$.  The
superscript ``CUF$_\Lambda$'' now comprises a reference to the
constant $\Lambda > 0$ which appears in
Eq.~\eqref{eq:CU-cut-off}.  This constant determines how fast $U$
is turned on in comparison to $C$.  Indeed, it allows to interpolate
between the pure $U$-flow and $C$-flow methods.  If $\Lambda$ is
small, the largest part of the flow of the free propagator happens
while the interaction is still negligibly small.  Only then, given an
almost completely restored propagator, $U$ flows to considerable
values. Hence, we expect the method to produce data close to the
$U$-flow result, $\overline
\Sigma{}^{\ix{CUF}_\Lambda} \rightarrow \overline \Sigma{}^\ix{PUF}$
for $\Lambda \rightarrow 0$.  If $\Lambda$ is large, we expect in turn
results close to that of the pure $C$-flow.  The scale $\Lambda_0$
which separates the two regimes depends on the model and is difficult
to determine \emph{a priori}.  In the limiting case of infinite
$\Lambda$, that is $U_\lambda = U$, we reproduce a pure $C$-flow,
$\overline \Sigma{}^{\ix{CUF}_\infty}=\overline
\Sigma{}^\ix{CF}$, in which
\begin{equation}
  \label{eq:CF}
  \dot{\overline \Sigma}{}^{\ix{CF}}_\lambda = -U\cdot \dot{\overline G}_\lambda +
  \left.\overline V_{\lambda} \right|_{U_\lambda \to U} \cdot \overline \Pi_\lambda
  \cdot \dot C^{-1}_\lambda.
\end{equation}
We refer to this specific approximation as ``CF''. The underlying
truncation, which is partly based on the approximation $\overline
\Phi^{(2)}_\lambda=U$ and partly on $\overline
\Phi^{(2)}_\lambda=\overline
\Phi^{(2)}_{\lambda}\!\!\left.\vphantom{\Phi^\lambda}\right|_\ix{2nd}$,
is not among the truncations described in Ref.~\onlinecite{Ren15} and is not
equivalent to self-consistent perturbation theory.

Numerical computations cannot start at $\lambda_\ix{i}=\infty$ but
only at some finite $\lambda^\ix{num}_\ix{i}$.  If
$\lambda^\ix{num}_\ix{i}$ is chosen sufficiently large, the flow from
$\lambda=\infty$ to $\lambda^\ix{num}_\ix{i}$ does not contribute
significantly to $\overline \Sigma{}^{\ix{CUF}_\Lambda}$.
However, there is an important contribution to $\overline{\Sigma}
{}^{\ix{CF}}$ given by
\begin{align}
  \notag 
  \overline{\Sigma} {}^{\ix{CF}}_{\lambda_\ix{i}^\ix{num}} 
  &= 
  \lim_{\eta \to 0^+} \int_\infty^{\lambda_\ix{i}^\ix{num}} d\lambda
  \, (- U \cdot \overline G_\lambda)
  \\ \label{eq:num_init_cond_CMUF}
  &\approx
  \lim_{\lambda_0 \to \infty} \lim_{\eta \to 0^+}
  \int_\infty^{\lambda_0} d\lambda \, (- U \cdot \overline G_\lambda).
\end{align}
Here, $\eta$ is the infinitesimal shift of imaginary time which
ensures that creators are ordered to the left of annihilators with
equal time arguments.  A similar contribution due to the flow from
$\lambda=\infty$ to $\lambda_\ix{i}^\ix{num}$ is known from
1PI fRG with imaginary frequency cut-off, cf.
e.g. Ref.~\onlinecite{And04}.


\subsection{Summary of methods}

In the following sections we apply the different approximation schemes
to the single impurity Anderson model. For a better overview, we list
the methods that we have introduced: the conserving fRG approximation
$\overline
\Sigma{}^\ix{cfRG}$ which is the self-consistent solution of
Eq.~\eqref{eq:A_find_Sigma}, and the FLEX approximation $\overline
\Sigma{}^\ix{FLEX}$ which follows from
Eq.~\eqref{eq:FLEX_eq_for_Sigma}; furthermore, as non-conserving
approximations to $\overline \Sigma{}^\ix{cfRG}$, the plain $U$-flow and
modified $U$-flow approximations $\overline \Sigma{}^\ix{PUF},
\overline \Sigma{}^\ix{MUF}$ given by
Eqs.~\eqref{eq:PUF_flow_eq} and~\eqref{eq:MUF_flow_eq}; additionally, the
static variant $\overline \Sigma{}^\ix{StUF}$ from
Eq.~\eqref{eq:StUF_flow_eq}; finally, the $CU$-flow approximation
$\overline \Sigma{}^\ix{CUF_\Lambda}$ from Eq.~\eqref{eq:CU_flow_eq}
[with $\Lambda$ referring to the definition of the cut-off in
Eq.~\eqref{eq:CU-cut-off}] and the $C$-flow
approximation $\overline \Sigma{}^{\ix{CF}}$ according to
Eq.~\eqref{eq:CF} (which is not equivalent to self-consistent perturbation theory).


\section{Application to the Anderson impurity model}
\label{sec:Anderson_model_in_2PI}


\subsection{Hamiltonian and action}
\label{sec:Ham_and_action_SIAM}

The dot Hamiltonian of the single impurity Anderson model is
\begin{align}
\notag
H_\ix{dot} 
&=
\sum_\sigma \left(V_\ix{g} + \sigma B \right)
d_\sigma^\dagger d_\sigma + U \left( d_\uparrow^\dagger d_\uparrow -
  \frac{1}{2} \right)\left( d_\downarrow^\dagger d_\downarrow -
  \frac{1}{2} \right) 
\\ 
&=
\sum_\sigma \epsilon_\sigma d_\sigma^\dagger d_\sigma -
\sum_\sigma \frac{U}{2}  d_\sigma^\dagger
d_\sigma+ U d^\dagger_{\uparrow} d_{\uparrow}
d^\dagger_{\downarrow} d_{\downarrow} +\textrm{const.}
\label{eq:SIAM_dot_Hamiltonian}
\end{align}
We combine the gate voltage $V_\ix{g}$ and the magnetic field $B$ in a
single-particle energy $\epsilon_\sigma = V_{\ix{g}} + \sigma B$, with
$\sigma=\pm 1 = \,\uparrow, \downarrow$.  The interaction $U$ is
introduced in a particle-hole symmetric way which entails an
additional single-particle term as illustrated in the second line.

The dot is coupled to a semi-infinite lead of non-interacting fermions
by a momentum- and spin-independent coupling $t$.  We perform the
wide-band limit, which means to assume a constant lead density of
states on the whole energy axis.  As a consequence, we can account for
the lead by a constant hybridization $\Gamma=\pi \rho_\ix{lead}(0)
|t|^2$ in the free dot propagator, in which $\rho_\ix{lead}(0)$
denotes the density of states at the end of the lead.  The whole
system is prepared in grand canonical equilibrium with temperature
$T=1/\beta$ and chemical potential $\mu=0$.  For the numerical
evaluation we choose $T=0$.  The action entering the
formula~\eqref{eq:Z} for the partition function has the form
\begin{align}
  \notag
  S[\psi] 
  =&   
  \frac{1}{2} \sum_{\alpha \alpha'} \psi_{\alpha}
  \left(-C^{-1}_{\alpha \alpha'} + U^{(1)}_{\alpha \alpha'}\right)
  \psi_{\alpha'}   
  \\ &+
  \frac{1}{4!}\! \sum_{\alpha_1 \alpha_2 \alpha_3 \alpha_4} \!
  U^{(2)}_{\alpha_1 \alpha_2 \alpha_3 \alpha_4} \psi_{\alpha_1}
  \psi_{\alpha_2} \psi_{\alpha_3}\psi_{\alpha_4}. 
\end{align}
Compared to Eq.~\eqref{eq:action} we have an additional quadratic
contribution due to the interaction.  This could as well be absorbed
into the inverse free propagator.  However, we prefer a quadratic
contribution to the interaction, as it allows for a more transparent
treatment of particle-hole symmetry and for a clear distinction of
$U$-flow and $C$-flow.  As the leads are integrated out, the
multi-indices $\alpha=(c,\sigma,\nu_n)$ contain no lead states but
only dot states $\sigma$.  Furthermore, we switch to Matsubara
frequencies $\nu_n=\frac{\pi}{\beta}(2n+1)$ instead of imaginary times
$\tau$.  In the following we discuss the constituents of the action.

\paragraph*{Free propagator and Fourier transform.} In the usual
Fourier transform without charge indices ($\int_\tau=\int_0^\beta
d\tau$)
\begin{align}
 \notag
 (C_\ix{reg}^{-1})^{\sigma_1 \sigma_1^\prime}_{n_1
   n_1^\prime}
 &=
 \int_{\tau_1} \int_{\tau_1^\prime} e^{i \nu_{n_1}
   \tau_1} (C^{-1})^{\sigma_1
   \sigma_1^\prime}\left(\tau_1,\tau_1^\prime\right) e^{-i
   \nu_{n_1^\prime} \tau_1^\prime}
 \\ \label{eq:reg_FT}
 &= \beta \delta_{n_1 ,n_1^\prime}
 \delta_{\sigma_1 ,\sigma_1^\prime} C_{\ix{reg},\sigma}^{-1}(\nu_n), 
\end{align}
the inverse lead-dressed free propagator on the dot reads as
\begin{equation}
  C_{\ix{reg},\sigma}^{-1}(\nu_n)
  =
  i\nu_n - \epsilon_\sigma +i \sgn(\nu_n)\Gamma.
\end{equation}
In this work based on the charge index notation, we use a different
convention for the Fourier transform [$y=(\sigma,c)$]:
\begin{align}
 \label{eq:FT_G}
 G^{y_1 y_1^\prime}_{n_1 n_1^\prime}&=\int_{\tau_1}
 \int_{\tau_1^\prime} e^{- i \nu_{n_1} \tau_1} G^{y_1
   y_1^\prime}\left(\tau_1,\tau_1^\prime\right) e^{-i \nu_{n_1^\prime}
   \tau_1^\prime}, 
 \\ \label{eq:FT_Sigma}
 \Sigma^{y_1 y_1^\prime}_{n_1 n_1^\prime}&=\int_{\tau_1}
 \int_{\tau_1^\prime} e^{i \nu_{n_1} \tau_1}  \Sigma^{y_1
   y_1^\prime}\left(\tau_1,\tau_1^\prime\right) e^{i \nu_{n_1^\prime}
   \tau_1^\prime}. 
\end{align}
For the vertex-like $C^{-1}$, the two conventions are connected in the
following way ($\bar{c}=-c$):
\begin{equation}
  \left(C^{-1}\right)^{c c^\prime}_{\sigma \sigma^\prime,n n^\prime}
  =
  c \delta_{c,\bar{c}^\prime} \delta_{\sigma,\sigma^\prime}
  \delta_{n+n^\prime,0}C_{\ix{reg},\sigma}^{-1}(\nu_n). 
\end{equation}
Here, $\delta_{n+n^\prime,0}$ is a sloppy short-hand notation for
$\delta_{\nu_n+\nu_{n^\prime},0}$, that is for the requirement
$\frac{\pi}{\beta}(2 n + 1 + 2 n^\prime +1)=0$ or $n+n^\prime+1=0$.
From
\begin{equation}
  \left(C^{-1}\right)^{c c^\prime}_{\sigma \sigma^\prime,n n^\prime}=
  \beta \delta_{c,\bar{c}^\prime} \delta_{\sigma,\sigma^\prime} 
  \delta_{n+n^\prime,0}\left(C^{-1}\right)^{c}_{\sigma,n}, 
\end{equation}
we deduce
\begin{equation}
  \left(C^{-1}\right)^{c}_{\sigma,n}=c \left[i\nu_n -
    \epsilon_\sigma +i \sgn(\nu_n)\Gamma\right].
\end{equation}
In order to derive a rule for inversion, let $A$ denote a self-energy
or propagator. From
\begin{multline}
   \sum_{\sigma^\prime,c^\prime} \int_0^\beta d\tau^\prime
  \left(A\right)^{c c^\prime}_{\sigma \sigma^\prime}(\tau,\tau^\prime)
  \left(A^{-1}\right)^{c^\prime c^{\prime \prime}}_{\sigma^\prime
    \sigma^{\prime\prime}}(\tau^\prime,\tau^{\prime\prime})
    \\=
  \delta(\tau-\tau^{\prime\prime})
  \delta_{\sigma,\sigma^{\prime\prime}} \delta_{c,c^{\prime\prime}}   
\end{multline}
in time-space follows
\begin{equation}
 \frac{1}{\beta} \sum_{n^\prime,\sigma^\prime,c^\prime} \left(A\right)^{c c^\prime}_{\sigma \sigma^\prime, n n^\prime} \left(A^{-1}\right)^{c^\prime c^{\prime \prime}}_{\sigma^\prime \sigma^{\prime\prime},n^\prime n^{\prime \prime}}= \beta \delta_{n,n^{\prime \prime}}\delta_{\sigma,\sigma^{\prime\prime}}\delta_{c,c^{\prime\prime}}.
\end{equation}
We thus find
\begin{equation}
 A^c_{\sigma,n}=\frac{1}{\left(A^{-1}\right)^{\bar{c}}_{\sigma,-n}}.
\end{equation}
For the Anderson model, the antisymmetry of two-point functions means $A^{cc^\prime}_{\sigma \sigma^\prime,n n^\prime}=-A^{c^\prime c}_{\sigma^\prime \sigma,n^\prime n}$. Due to $A^{cc^\prime} \sim \delta_{c,\bar{c}^\prime}$, it is thus sufficient to use either the $c=+$ or the $c=-$ component. We choose to use $c=+$ for the self-energy and $c=-$ for propagators, i.e.
\begin{align}
  C^{-}_{\sigma,n} &= \frac{1}{\left(C^{-1}\right)^{+}_{\sigma,-n}}=-\frac{1}{i\nu_n
    + \epsilon_\sigma +i \sgn(\nu_n)\Gamma} , 
  \\ 
  G^{-}_{\sigma,n} &= \frac{1}{\left(G^{-1}\right)^{+}_{\sigma,-n}}=\frac{1}{\left(C^{-1}\right)^{+}_{\sigma,-n}-\Sigma^+_{\sigma,-n}}.
\end{align}

\paragraph*{Interaction part of the action.}   The quadratic interaction
contribution to the action is
\begin{equation}
  -\frac{U}{2} \int_0^\beta d\tau \sum_\sigma \overline{\psi}_\sigma
  (\tau) \psi_\sigma(\tau) 
  =
  \frac{1}{2} \sum_{\alpha  \alpha^\prime} \psi_\alpha U^{(1)}_{\alpha
    \alpha^\prime} \psi_{\alpha^\prime} 
\end{equation}
with
\begin{equation}
 U^{(1),c c^\prime}_{\sigma \sigma^\prime, n n^\prime}=c \beta
 \delta_{n+n^\prime,0} \delta_{\sigma \sigma^\prime}
 \delta_{c\bar{c}^\prime} \frac{U}{2}.
\end{equation}
The quartic interaction contribution to the action is
\begin{multline}
 U \int_0^\beta d\tau \overline{\psi}_\uparrow (\tau)
\overline{\psi}_\downarrow (\tau) \psi_\downarrow (\tau) \psi_\uparrow
(\tau) 
\\ = \frac{1}{4!} \sum_{\alpha_1 \ldots \alpha_4} U^{(2)}_{\alpha_1
   \alpha_2 \alpha_3 \alpha_4} \psi_{\alpha_1} \psi_{\alpha_2}
 \psi_{\alpha_3} \psi_{\alpha_4}.
\end{multline}
In order to determine $U^{(2)}$ we use that
\begin{equation}
  U d^\dagger_{\uparrow} d_{\uparrow} d^\dagger_{\downarrow}
  d_{\downarrow}=\frac{1}{2!^2} \sum_{\sigma_1 \sigma_2 \sigma_3
    \sigma_4} U^-_{\sigma_1 \sigma_2 \sigma_3 \sigma_4}
  d_{\sigma_1}^\dagger d_{\sigma_2}^\dagger d_{\sigma_4} d_{\sigma_3} 
\end{equation}
with
\begin{equation}
 U^-_{\sigma_1 \sigma_2 \sigma_3 \sigma_4}=\delta_{\sigma_1 ,
   \overline{\sigma}_2} \delta_{\sigma_3 , \overline{\sigma}_4} \left[
   \delta_{\sigma_1 , \sigma_3} - \delta_{\sigma_1 ,
     \overline{\sigma}_3}\right] U .
\end{equation}
Hence,
\begin{equation}
 U^{(2) c_1 c_2 c_3 c_4}_{\sigma_1 \sigma_2 \sigma_3 \sigma_4,n_1 n_2
   n_3 n_4}=\beta \delta_{n_1+n_2+n_3+n_4,0} U^{(2) c_1 c_2 c_3
   c_4}_{\sigma_1 \sigma_2 \sigma_3 \sigma_4} 
\end{equation}
with
\begin{eqnarray}
 \label{eq:U(2)_charge_index_notation}
 U^{(2) c_1 c_2 c_3 c_4}_{\sigma_1 \sigma_2 \sigma_3 \sigma_4}=
 \left\{ \begin{array}{cl} 
 -U^-_{\sigma_1 \sigma_2 \sigma_3 \sigma_4}  & \textrm{if } c_1=c_2=\overline{c}_3=\overline{c}_4 \\
 U^-_{\sigma_1 \sigma_3 \sigma_2 \sigma_4} & \textrm{if } c_1=c_3=\overline{c}_2=\overline{c}_4 \\
 -U^-_{\sigma_2 \sigma_3 \sigma_1 \sigma_4} & \textrm{if } c_2=c_3=\overline{c}_1=\overline{c}_4 \\
 0 & \textrm{else}
 \end{array} \right. .
\end{eqnarray}
We note that the majority of the $2^8=256$ components in
Eq.~\eqref{eq:U(2)_charge_index_notation} are zero.


\subsection{The quadratic interaction part in the self-energy
  equations and (un-)restricted MUF}
\label{sec:quadr_int_part_in_self_energy_eqs}

The quadratic interaction contribution to the action causes a few
minor changes to the equations for the self-energy, which we summarize now.
First of all, we replace Eq.~\eqref{eq:Def_Phi} for the
definition of the Luttinger-Ward functional by
\begin{equation}
 \label{eq:formula_free_eff_action}
 \Phi[G] = \Gamma[G]  -\frac{1}{2} \tr \ln(-G) +\frac{1}{2}
 \tr\left[\left(C^{-1}+U^{(1)}\right)G-1 \right]. 
\end{equation}
Then $\Phi$ is again minus the sum of all closed skeleton
diagrams made of two-particle vertices $U^{(2)}$ and full propagator
lines $G$.  In particular, $\Phi$ does not depend on the one-particle
vertex $U^{(1)}$.  Equation~\eqref{eq:Phi1-Sigma} for the self-energy
functional now reads as
\begin{equation}
  \Sigma[G] = -\Phi^{(1)}[G] - U^{(1)}.
\end{equation}

We induce the $U$-flow by a flow parameter in the two-particle
interaction, $U^{(2)}\rightarrow U^{(2)}_\lambda$.  For all our
$U$-flow schemes except MUF, particle-hole symmetry during all of the
flow is ensured by dressing $U^{(1)}=U^{(1)}_\lambda$ with the
corresponding $\lambda$-dependence.  As $\Phi$ does not depend on
$U^{(1)}$, the flow equations for $\Phi$ and $\Phi^{(1)}$ maintain the
form derived in Sec.~\ref{sec:Cons_and_non-cons_appr} (now with the
notation $U^{(2)}$ instead of $U$ for the two-particle vertex).
However, the single-particle vertex $U^{(1)}$ enters the self-energy
$\Sigma = -\Phi^{(1)} - U^{(1)}$.  Consequently, an addend $-U^{(1)}$
must be added to the self-consistency
equations~\eqref{eq:A_find_Sigma} and~\eqref{eq:FLEX_eq_for_Sigma} of
cfRG and FLEX.  For instance, Eq.~\eqref{eq:FLEX_eq_for_Sigma} is
replaced by
\begin{multline}
 \label{eq:FLEX_after_redefinition}
 \Sigma^\ix{FLEX}=
 - U^{(2)} \cdot G - \frac{4}{3} \left(U^{(2)} \cdot \Pi \cdot U^{(2)}\right)^\R \cdot G 
 \\+ 2 \left(\Upsilon \cdot
   U^{(2)}\right)^\R \cdot G  - U^{(1)}.
\end{multline}
Similarly, an addend $-\dot U{}^{(1)}_\lambda$ enters the flow
equations~\eqref{eq:PUF_flow_eq}, \eqref{eq:StUF_flow_eq}
and~\eqref{eq:CU_flow_eq} for the PUF, StUF and CUF
approximations.  For instance, Eq.~\eqref{eq:PUF_flow_eq} is
replaced by
\begin{align}
  \notag
  \dot{\overline \Sigma}{}^\ix{PUF}_{\lambda}
  =& -\!\frac{2}{3}\!\left(\overline{\Upsilon}_\lambda
    \!\cdot\! \dot U_\lambda^{(2)} 
    \!\cdot\! \overline{\Upsilon}_\lambda^\ix{T} 
    - \overline{\Upsilon}_\lambda \!\cdot\! \dot{U}_\lambda^{(2)} -
    \dot{U}_\lambda^{(2)} \!\cdot\! \overline{\Upsilon}_\lambda^\ix{T}  
  \right)^\R \!\cdot \overline G^\lambda  
  \\ \label{eq:PUF_flow_eq_SIAM}
  &- \dot U_\lambda^{(2)} \cdot \overline G_\lambda
    - U_\lambda^{(2)} \cdot \dot{\overline G}_\lambda
    -\dot{U}^{(1)}_\lambda.
\end{align}
The initial conditions for these three flow schemes remain unchanged
because $U_{\lambda_\ix{i}}^{(1)}=0$.

For the MUF approximation, we leave $U^{(1)}$ independent of $\lambda$
to ensure particle-hole symmetry. The flow
equation~\eqref{eq:MUF_flow_eq} remains unchanged. However, the
self-consistent Hartree-Fock initial condition now reads
$\overline{\Sigma}_{\lambda_\ix{i}}=-U^{(2)} \cdot
\overline{G}_{\lambda_\ix{i}} - U^{(1)}$.  For the Anderson model, the
self-consistent Hartree-Fock method predicts an unphysical
spin-symmetry breaking for $U>U_\ix{crit} =\pi \Gamma$ (at
$V_\ix{g}=0=B$): there are two ``unrestricted'' magnetic solutions
which can be mapped onto one another by flipping the spins. There is
yet another, ``restricted'', solution which is non-magnetic but
responds unphysically to infinitesimal magnetic fields, having a negative
spin susceptibility.  We can choose any of these solutions as starting
point of the modified $U$-flow. Accordingly, we obtain two different
MUF schemes for $U>U_\ix{crit}$ which we call ``restricted MUF'' and
``unrestricted MUF''.  The question arises as to whether the flow is able to
eliminate the artifacts introduced by the initial conditions.  The
numerical results described in Sec.~\ref{sec:results_PM_UF} show
that this is not the case.

For the CF approximation, neither $U^{(2)}$ nor $U^{(1)}$ is made
$\lambda$-dependent and the flow equation~\eqref{eq:CF} stays the
same. However, the numerical initial
condition~\eqref{eq:num_init_cond_CMUF} is changed to
\begin{equation}
  \overline{\Sigma} {}^{\ix{CF}}_{\lambda_\ix{i}^\ix{num}} 
  = 
  \lim_{\lambda_0 \to \infty} \lim_{\eta \to 0^+}
  \int_\infty^{\lambda_0} d\lambda \, (- U^{(2)} \cdot \overline
  G_\lambda) 
  - U^{(1)}.
\end{equation}


\subsection{Steps towards implementable equations for the Anderson
  impurity model}

In Appendix~\ref{sec:Der_impl_eqs}, we derive specifically for the
Anderson model the relevant equations for the numerical computation of
the self-energy. Here, we summarize the important steps.

In Appendix~\ref{sec:Reducing_number_indices}, a suitable reduced index
notation is defined.  It exploits that the number of indices on four-point functions can be reduced significantly by making use of symmetry
relations. Furthermore, many components can be shown to be zero due to
particle-number and spin conservation.

In Appendix~\ref{sec:calc_Ups_for_SIAM}, it is shown how to calculate
$\overline{\Upsilon}$. While the four-point function
$\overline{\Upsilon}$ depends on four frequencies or rather on three
independent frequencies, we find that one frequency is always summed
over independently. We thus define a $\widetilde{\Upsilon}$ which
depends only on the two remaining frequencies. $\widetilde{\Upsilon}$
then turns out to depend only on the sum of the two frequencies which
is only one composite (bosonic) frequency. The non-zero components of
$\widetilde{\Upsilon}$ are identified with a particular channel
(particle-particle, direct or exchange particle-hole) and are labeled
accordingly.

In Appendix~\ref{sec:impl_flow_eq_for_c_schemes}, the self-consistency
equations for the self-energy of cfRG and FLEX are cast into a form
suitable for numerical implementation.  When we evaluate the dot
products in the self-consistency equations~\eqref{eq:A_find_Sigma}
and~\eqref{eq:FLEX_eq_for_Sigma} (adapted according to
Sec.~\ref{sec:quadr_int_part_in_self_energy_eqs}), we exploit the
sparseness of the components mentioned in the preceding paragraphs.
We then perform the $T=0$ limit. The final resulting equation is
\begin{align}
\label{eq:cons_Sigma_flow_eq}
\overline{\Sigma}_{\sigma}(\nu) \!
=&U \int_{0}^\infty \frac{d\nu^\prime}{\pi} \ix{Re}\left[\overline{G}_{\bar{\sigma}}(\nu^\prime)\right]
\\  \nonumber &+ U \!\int_{-\infty}^\infty \!\frac{d\omega}{2\pi}\!
\left\{ \kappa_\ix{p}\left[\kappa_0
    \widetilde{\Psi}^{\ix{p}}(\omega)\!-\!\widetilde{\Upsilon}^{\ix{p}}(\omega)
  \right]\overline{G}_{\bar{\sigma}}(\nu\!-\!\omega)\right. 
\\ \notag  & \qquad \quad \left.+\kappa_\ix{d}
  \left[\widetilde{\Upsilon}^{\ix{d}\sigma}\!(\omega) \!-\! \kappa_0
    \widetilde{\Psi}^{\ix{d}\sigma}\!	(\omega)
  \right]\overline{G}_{\sigma}(\omega\!-\!\nu) \right.
 \\ \notag
 &\qquad \quad \left.+\kappa_\ix{x}\left[\kappa_0
    \widetilde{\Psi}^{\ix{x}\sigma}\!(\omega)\!
    -\!\widetilde{\Upsilon}^{\ix{x}\sigma}\!(\omega)
  \right]\overline{G}_{\bar{\sigma}}(\omega\!-\!\nu)\right\} . 
\end{align}
It describes either FLEX or cfRG, depending on the choice of the newly
introduced coefficients $\kappa_i,
i=0,\ix{p},\ix{d},\ix{x}$. We solve the equation numerically by iteration. The details
of the numerical implementation, e.g. the use of frequency grids, are
discussed in Appendix~\ref{sec:details_num_impl}. We take zero as the
initial guess for the iteration of the self-energy. If we plainly
iterated over Eq.~\eqref{eq:cons_Sigma_flow_eq}, the value of $U$
would be limited by the critical value $U_\ix{crit}=\pi \Gamma$ known
from the self-consistent Hartree-Fock solution.\cite{Whi92} In order
to circumvent this problem, we gradually increase $U$ in each step of
the iteration up to the desired value; this idea was already applied in Ref.~\onlinecite{Whi92}.  In addition, we calculate
the next guess of an iteration step by combining the last guess and
the outcome of the self-consistency equation in a weighted manner. We
found empirically that this makes the iteration faster and more
stable.

In App.~\ref{sec:impl_flow_eq_for_nc_schemes}, we turn to the flow
equations for the self-energy for the various non-conserving
methods. One proceeds as for the conserving case and obtains
\begin{align}
 \notag
 \dot{\overline{\Sigma}}{}^{\ix{PUF}}_{\sigma,\lambda}(\nu)=& \frac{\dot{U}_\lambda}{3} \int_{-\infty}^\infty \frac{d\omega}{2\pi}\left\{ \left[\widetilde{\Upsilon}^\ix{p}_{\lambda}(\omega)^2 -2 \widetilde{\Upsilon}^\ix{p}_{\lambda}(\omega)\right] \overline{G}{}^{\lambda}_{\bar{\sigma}}(\nu\!-\!\omega) \right.
\\  \notag  
& \qquad \qquad \left. +2\widetilde{\Upsilon}^{\ix{d}\sigma}_{\lambda}\!(-\omega)\left[1- \widetilde{\Upsilon}^{\overline{\ix{d}}}_{\lambda}\!(\omega)\right] \overline{G}{}^{\lambda}_{\sigma}(\omega\!-\!\nu) \right.
\\ \notag
& \qquad \qquad \left. +\left[\widetilde{\Upsilon}^{\ix{x}\sigma}_{\lambda}\!(\omega)^2-2\widetilde{\Upsilon}^{\ix{x}\sigma}_{\lambda}\!(\omega)\right] \overline{G}{}^{\lambda}_{\bar{\sigma}}(\omega\!-\!\nu)  \right\}
 \\ \notag
 &+\dot{U}_\lambda \int_{-\infty}^\infty \frac{d\nu^\prime}{\pi} \ix{Re}\left[\overline{G}{}^{\lambda}_{\bar{\sigma}}(\nu^\prime)\right]
 \\ &+ U_\lambda \int_{-\infty}^\infty \frac{d\nu^\prime}{2\pi} \dot{\overline{G}}{}^{\lambda}_{\bar{\sigma}} (\nu^\prime)
\label{eq:PUF_fl_eq_Sigma_T=0}
\end{align}
for the PUF,
\begin{align}
 \notag
 \dot{\overline{\Sigma}}{}^{\ix{MUF}}_{\sigma,\lambda}(\nu)=& \frac{\dot{U}_\lambda}{3} \int_{-\infty}^\infty \frac{d\omega}{2\pi}\left\{ \left[\widetilde{\Upsilon}^\ix{p}_{\lambda}(\omega)^2 -2 \widetilde{\Upsilon}^\ix{p}_{\lambda}(\omega)\right] \overline{G}{}^{\lambda}_{\bar{\sigma}}(\nu\!-\!\omega) \right.
\\  \notag
& \qquad \qquad \left. +2\widetilde{\Upsilon}^{\ix{d}\sigma}_{\lambda}\!(-\omega)\left[1- \widetilde{\Upsilon}^{\overline{\ix{d}}}_{\lambda}\!(\omega)\right] \overline{G}{}^{\lambda}_{\sigma}(\omega\!-\!\nu) \right.
\\ \notag
& \qquad \qquad \left. +\left[\widetilde{\Upsilon}^{\ix{x}\sigma}_{\lambda}\!(\omega)^2-2\widetilde{\Upsilon}^{\ix{x}\sigma}_{\lambda}\!(\omega)\right] \overline{G}{}^{\lambda}_{\bar{\sigma}}(\omega\!-\!\nu)  \right\}
 \\  &+ U \int_{-\infty}^\infty \frac{d\nu^\prime}{2\pi} \dot{\overline{G}}{}^{\lambda}_{\bar{\sigma}} (\nu^\prime)
\label{eq:MUF_fl_eq_Sigma_T=0}
\end{align}
for the MUF,
\begin{equation}
 \label{eq:CUF_fl_eq_Sigma_T=0}
 \dot{\overline{\Sigma}}{}^{\ix{CUF}_\Lambda}_{\sigma} (\nu)=\dot{\overline{\Sigma}}{}^\ix{PUF}_{\sigma} (\nu)+\Delta^{\lambda}_\sigma (\nu)
\end{equation}
for the CUF and
\begin{equation}
\label{eq:CF_fl_eq_Sigma_T=0}
 \dot{\overline{\Sigma}}{}^\ix{CF}_{\sigma} (\nu)=U\int_{-\infty}^\infty \frac{d\nu^\prime}{2\pi} \dot{\overline{G}}{}^{\lambda}_{\bar{\sigma}} (\nu^\prime)+\left.\Delta^{\lambda}_\sigma (\nu)\right|_{U_\lambda \to U}
\end{equation}
for the CF approximation. In
case of the StUF approximation, the frequency integral on the
right-hand side can be evaluated analytically and one finds the
compact equation
\begin{equation}
  \label{eq:flow_eq_StUF_RHS_integrated}
  \dot{\overline{\Sigma}}{}^{\ix{StUF}}_{\sigma,\lambda}=-
  \frac{\dot{U}_\lambda}{\pi}
  \ix{atan}\left(\frac{\epsilon_{\bar\sigma} +
      \overline{\Sigma}{}^\ix{StUF}_{\bar{\sigma},\lambda}}{\Gamma}\right). 
\end{equation}

In our numerics, we evolve the self-energy according to the respective
flow equation by a standard differential equation solver.  For the
frequency dependent schemes we use the same frequency grid as in the
conserving case. For more details on the implementation, see also
Appendix~\ref{sec:details_num_impl}.

In all methods (except for StUF), the right-hand side contains an
integral over a bosonic frequency which must be carried out
numerically. This must be done for each fermionic frequency of the
self-energy in each step of the iteration or flow. As the fermionic
grid is given by $n_\ix{len}$ frequencies and the bosonic grid by
$m_\ix{len}=2 n_\ix{len}$ frequencies
(cf.~Appendix~\ref{sec:details_num_impl}), the effort of the methods
scales as $\mathcal{O} (n_\ix{len}^2)$ in each step. The
same scaling behavior is known from a 1PI vertex expansion Matsubara
fRG applied to the Anderson model which uses a flowing
frequency dependent two-particle vertex in channel
decomposition.\cite{Kar08}


\subsection{The computation of observables}

For both, conserving and non-conserving methods, we use the numerical
solution $\overline{\Sigma}{}_\sigma(\nu)$ to compute observables.
The occupancy can be obtained from the propagator according to
Eq.~\eqref{eq:n-from-G} via
\begin{equation}
  \label{eq:occupancy_intergal}
  \left<n_\sigma\right>_\ix{prop}\!=\!\int_{-\infty}^{\infty}
  \!\frac{d\nu}{2\pi}\overline{G}_{\sigma}(\nu) e^{-i\nu 0^+}
  \!=\frac{1}{2} + \int_{0}^{\infty} \!\frac{d\nu}{\pi}
  \ix{Re}\!\left[\overline{G}_{\sigma}(\nu)\right]\!. 
\end{equation}
Even though we investigate an equilibrium setup in which we couple the dot to one lead by $\Gamma$, we can calculate the (linear-response) conductance which a system coupled to two leads by $\Gamma/2$ would have at zero bias voltage.\cite{Hewson} The conductance is given by
\begin{equation}
  G^\ix{cond} = \frac{e^2}{h} \Gamma \sum_\sigma
  \ix{Im}\left[\overline{G}_\sigma(0^+)\right] 
\end{equation}
with $h=2\pi \hbar=2\pi$. At $T = B = V_\ix{g} = 0$, the so-called effective mass is defined via
\begin{equation}
  m^\ast =1-\lim_{\nu \searrow 0} \frac{d \ix{Im} \overline{\Sigma}_\sigma(\nu)}{d \nu}
\end{equation}
which is independent of $\sigma$ due to $B=0$.  We are also interested
in the static spin and charge susceptibility given by the derivatives
\begin{align}
 \begin{split}
  \chi_\ix{s}&=-\left. \frac{d\left< n_\uparrow -n_\downarrow\right>_\ix{prop}}{dB}\right|_{B=0}, 
  \\ \chi_\ix{c}&=-\left. \frac{d\left<n_\uparrow +n_\downarrow\right>_\ix{prop}}{d V_\ix{g}}\right|_{V_\ix{g}=0}.
  \end{split}
\end{align}
Numerically, we probe by a very small magnetic field ($B/
\Gamma=10^{-5}$) or a shift of the gate voltage
($V_\ix{g}/\Gamma=10^{-4}$) and compute the finite difference
approximations
\begin{align}
  \chi_\ix{s} &\approx \frac{\left<n_\downarrow
      -n_\uparrow\right>_\ix{prop}-\left<n_\downarrow
      -n_\uparrow\right>_\ix{prop}|_{B=0}}{B},
  \\
  \chi_\ix{c} &\approx -\frac{\left<n_\downarrow
      +n_\uparrow\right>_\ix{prop}-\left<n_\downarrow
      +n_\uparrow\right>_\ix{prop}|_{V_\ix{g}=0}}{V_\ix{g}}.   
\end{align}


\subsection{Alternative approaches to the occupancy as test for
  conserving approximations}
\label{sec:Testing_PUF_MUF_non-cons}

In Sec.~\ref{sec:Phi-symmetry} we explained that $\Phi$-derivable
approximations (such as FLEX and cfRG) are
thermodynamically consistent and preserve the Friedel sum rule.  This
assures coinciding results when the impurity occupancy is computed
either from the propagator [Eq.~\eqref{eq:occupancy_intergal}] or
from the grand potential [Eq.~\eqref{eq:n-from-Omega}] or from
the Friedel sum rule.  The latter reads as
\begin{equation}
  \left< n_\sigma \right>_\ix{FSR}= \frac{1}{2} - \frac{1}{\pi}
  \ix{atan}\left(
    \frac{\epsilon_\sigma+\ix{Re}\left[\overline{\Sigma}{}_\sigma(0^+)\right]}{\Gamma}
  \right) 
\end{equation}
for the Anderson model at zero temperature and in the wide band
limit.\cite{Hewson}  Based on
Sec.~\ref{sec:Cons_and_non-cons_appr}, we expect the PUF, StUF and
MUF schemes to be non-$\Phi$-derivable methods.  In the results
Sec.~\ref{sec:PUF_StUF_MUF_num_non-cons}, we will indeed see that
for these methods the three ways to the occupancy lead to disagreeing
results.

Let us describe in more detail how we evaluate the occupancy from the
grand potential.  From Eq.~\eqref{eq:n-from-Omega} follows that
\begin{equation}
 \left<n_\uparrow + n_\downarrow\right>_\ix{gp}=\frac{d}{dV_\ix{g}} \overline{\Omega}.
\end{equation}
For the $\Phi$-derivable schemes, we obtain $\overline{\Omega}$ as the sum
of $\overline{\Omega}|_{U=0}$ and $\Delta
\overline{\Omega}=\overline{\Omega}-\overline{\Omega}|_{U=0}$ which we
can calculate directly from $\overline{\Sigma}$.  This yields
\begin{align}
 \notag \left<n_\uparrow + n_\downarrow\right>_\ix{gp}&=\frac{d}{dV_\ix{g}} \left(\Delta \overline{\Omega} +\overline{\Omega}|_{U=0}\right)
 \\ &=\frac{d}{dV_\ix{g}} \Delta \overline{\Omega}+ \left<n_\uparrow + n_\downarrow\right>_{U=0} .
\end{align}
The non-interacting occupancy is given by
\begin{equation}
 \left<n_\uparrow + n_\downarrow\right>_{U=0} =\sum_\sigma \left[ \frac{1}{2} - \frac{1}{\pi} \ix{atan}\left( \frac{\epsilon_\sigma}{\Gamma} \right) \right].
\end{equation}
For the flow schemes, $\overline{\Omega}=\overline{\Omega}{}_{\lambda_\ix{f}}$ leads to
\begin{align}
  \notag \left<n_\uparrow + n_\downarrow\right>_\ix{gp}&=\frac{d}{dV_\ix{g}} \overline{\Omega}_{\lambda_\ix{f}} =\frac{d}{dV_\ix{g}} \int_{\lambda_\ix{i}}^{\lambda_\ix{f}}\!\!\! d\lambda \dot{\overline{\Omega}}_\lambda + \frac{d}{dV_\ix{g}} \overline{\Omega}_{\lambda_\ix{i}}
  \\ &=\frac{d}{dV_\ix{g}} \int_{\lambda_\ix{i}}^{\lambda_\ix{f}} \!\!\! d\lambda \dot{\overline{\Omega}}_\lambda+ \left<n_\uparrow + n_\downarrow\right>_{\lambda_\ix{i}} .
\end{align}
Here, the second addend refers to
\begin{equation}
 \left<n_\uparrow + n_\downarrow\right>_{\lambda_\ix{i}} =\sum_\sigma \left[ \frac{1}{2} - \frac{1}{\pi} \ix{atan}\left( \frac{\epsilon_\sigma+\overline{\Sigma}_{\sigma,\lambda_\ix{i}}}{\Gamma} \right) \right]
\end{equation}
in which we exploit that for all schemes
$\overline{\Sigma}{}_{\sigma,\lambda_\ix{i}} \in \mathbb{R}$ is
frequency independent. The expressions for $\Delta\overline{\Omega}$
and $\dot{\overline{\Omega}}$ are provided in
Appendix~\ref{sec:occupancy_from_Omega}.  Numerically, the derivative
with respect to the gate voltage is carried out by an interpolation
routine.


\section{Numerical results}
\label{sec:Numerical_results}

For the numerical investigations, we resort, as mentioned above, to the $T=0$ limit.  The parameters for the
frequency grids (see Appendix~\ref{sec:details_num_impl}) are
$n_\ix{len}=120$, $d\nu=10^{-6}\Gamma$, $\nu_\ix{max}=10^8 \Gamma$. 
At selected values of the model parameters, we checked that this choice is sufficient to reach numerical convergence on the scale of the plots.


\subsection{Results for the conserving schemes}

\begin{figure*}
 \includegraphics[width=0.395\textwidth,clip]{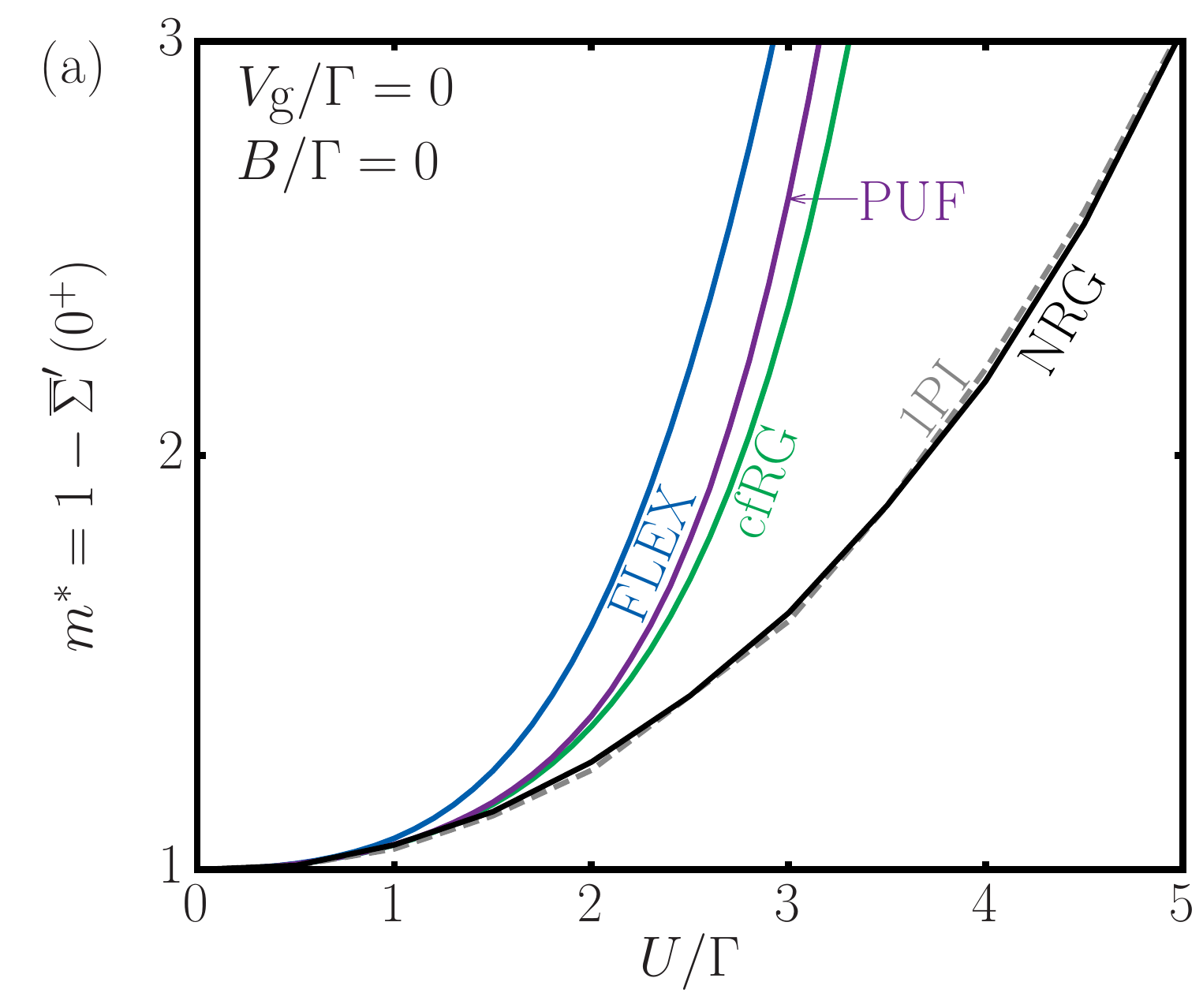}
 \includegraphics[width=0.395\textwidth,clip]{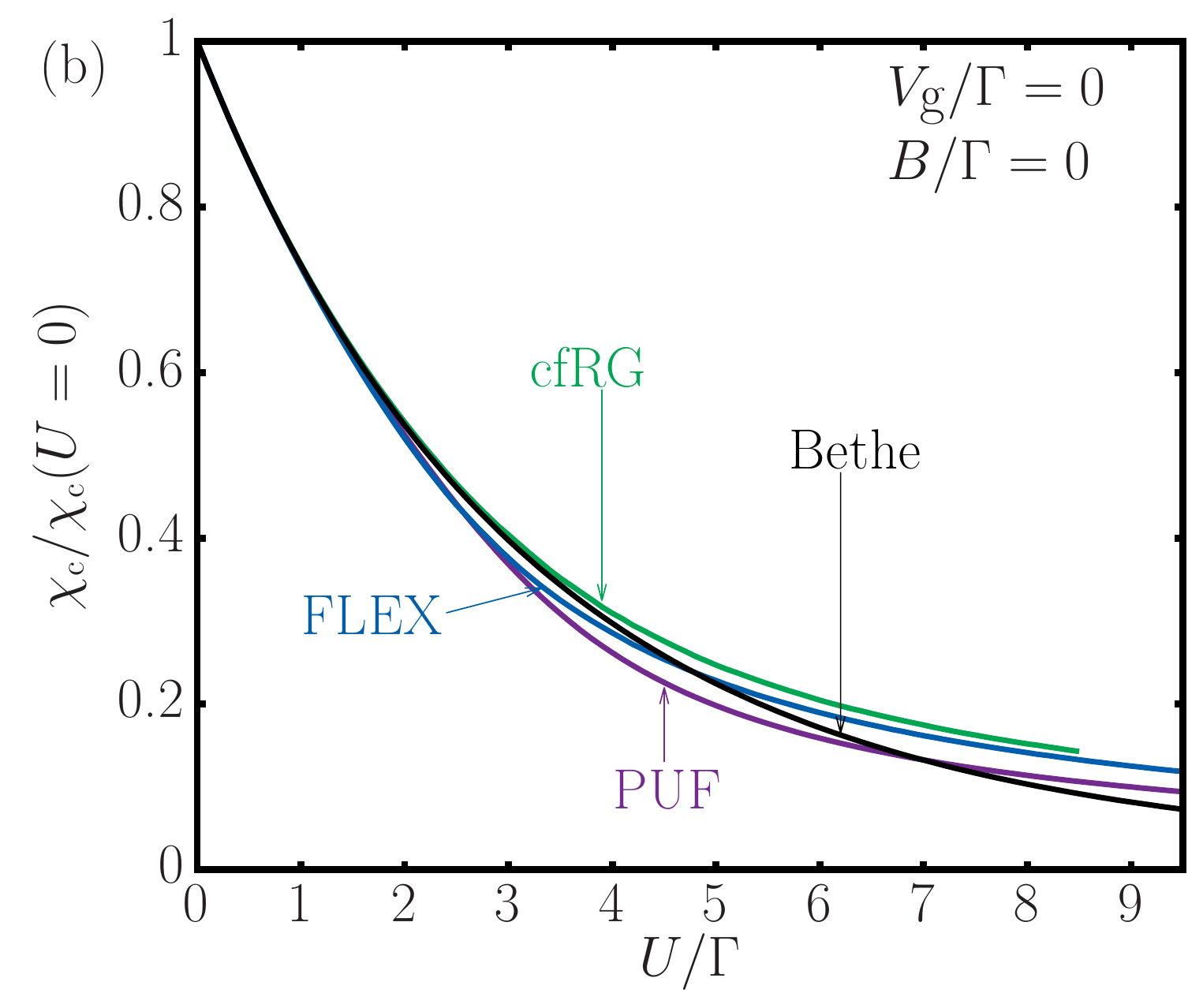}
 \includegraphics[width=0.395\textwidth,clip]{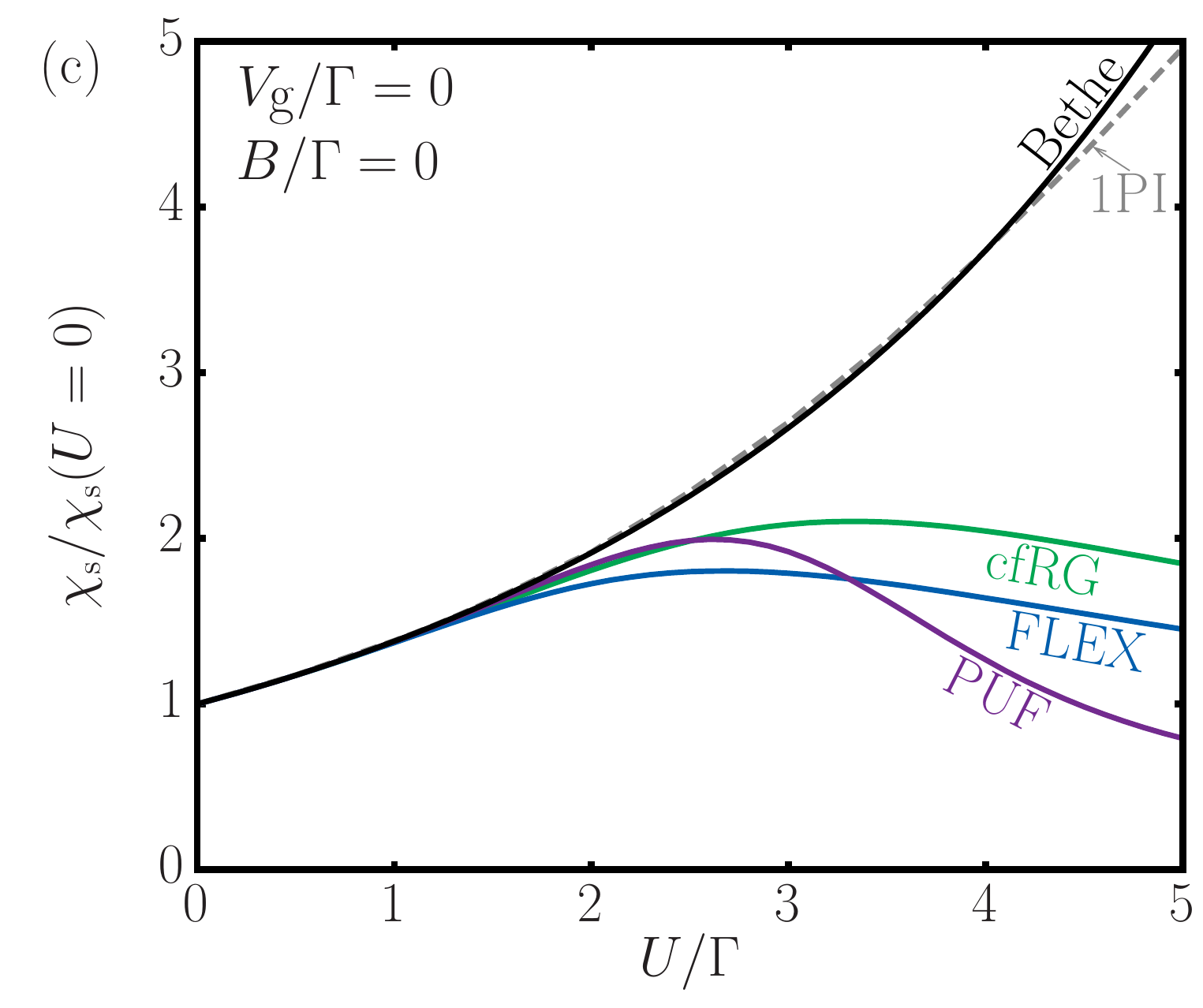}
 \includegraphics[width=0.395\textwidth,clip]{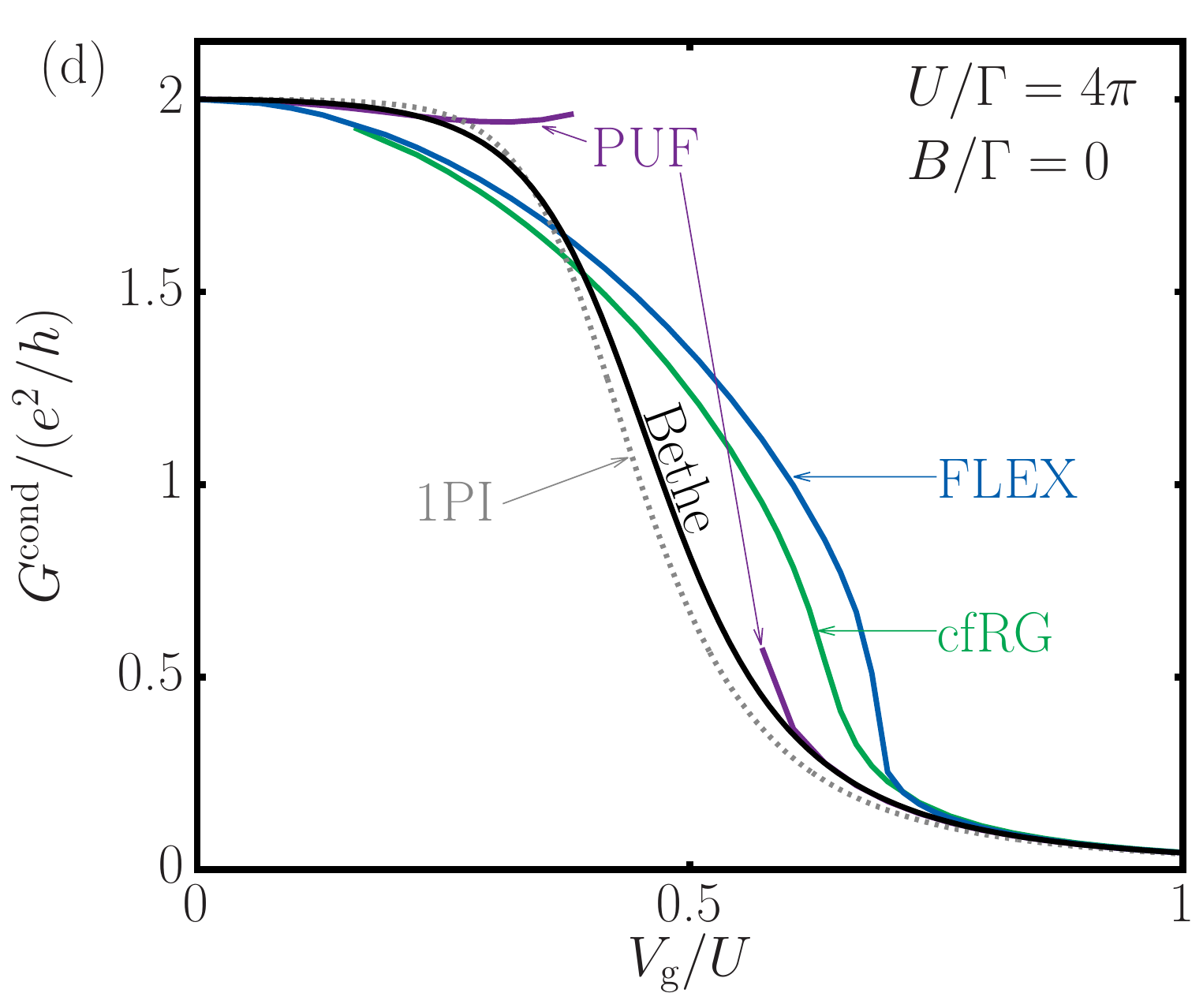}
 \caption{\label{fig:SIAMeq_c_1} (Color online) Numerical data for the cfRG, PUF and
   FLEX approximations.  (a) Results for the effective mass are
   compared to 1PI fRG and NRG results from Fig.~6(a)
   in Ref.~\onlinecite{Kar08}. 
   (b) Results for the charge
   susceptibility are compared to Bethe ansatz results.  We found
   well-converged solutions of the cfRG self-consistency equation only
   for $U < 8.5 \Gamma$.
   (c) Results for the spin susceptibility are
   compared to 1PI fRG data from Fig.~6(b) in Ref.~\onlinecite{Kar08}
   and to Bethe ansatz results. 
   (d) Results for the conductance as function of the gate voltage are
   compared to 1PI fRG data from Fig.~3
   in Ref.~\onlinecite{Kar06} and Bethe ansatz data from Ref.~\onlinecite{Ger00}. 
   We found well-converged solutions of the cfRG
   self-consistency equation only for $V_\ix{g} \ge 2 \Gamma$.}
\end{figure*}

In this section, we discuss
the numerical results for each observable obtained with the cfRG
approximation and with FLEX.  The plots also show the PUF curves for
comparison.  These will be compared to the MUF results in
Sec.~\ref{sec:results_PM_UF}.  Figure~\ref{fig:SIAMeq_c_1} shows
the effective mass, the charge and spin susceptibility as function of
$U$, as well as the conductance as function of the gate voltage.  We
have chosen the parameters such that we can compare with published
data.\cite{Kar08,Whi92,Kar06}  For this purpose, some plots include
data obtained with 1PI vertex expansion Matsubara fRG which takes into account at least a static flow of the 1PI two-particle vertex. For the effective mass and the spin susceptibility, we compare to more elaborate schemes which take into account the frequency dependence of the vertex (in its full or in a channel-decomposed form). Furthermore,
we compare to numerically exact data from the numerical RG (NRG) or to
exact Bethe ansatz results.  The Bethe ansatz data were calculated via
the formulas indicated in Ref.~\onlinecite{Zla83} or in case of the conductance taken from Ref.~\onlinecite{Ger00} and in case of the occupancy taken from Ref.~\onlinecite{Kar08}.

Our FLEX data coincide with the FLEX data presented by
White in Ref.~\onlinecite{Whi92}.\footnote{There is a minor deviation for the charge
  susceptibility at large values of $U/\Gamma$.  In Fig.~5
  of Ref.~\onlinecite{Whi92}, the three points plotted for the largest $U/\Gamma$
  seem to indicate the presence of a very slight shoulder; our data
  shown in Fig.~\ref{fig:SIAMeq_c_1}(b) does not do so.  We find
  agreement to the middle one of these three points (the one at $U/\pi
  \Gamma\approx 2.3$) but slight deviations from the other two points.
  We suspect that those two points of White are not fully converged.}
This confirms that the FLEX data are correctly determined, in
particular as our implementation differs from that of White.  White
transformed the frequency integrations to the real axis while we work
entirely on the imaginary axis.

cfRG, PUF and FLEX correctly describe the observables at very small
$U$.  The reason is that they are exact up to order $U^2$.  However,
when $U$ is increased, they deviate much earlier from the NRG
or Bethe ansatz results than 1PI fRG.  The cfRG approximation performs
slightly better than FLEX and PUF.  Generally, the results of all
three approximations are similar.  This is plausible since we found in
Sec.~\ref{sec:Similarity_2PI_fRG_FLEX} that cfRG and FLEX are
closely related, and since we identified PUF in
Sec.~\ref{sec:non-conserving-approximations} as an approximation to
cfRG.

Let us now discuss each plot in more detail.  Concerning the effective
mass shown in Fig.~\ref{fig:SIAMeq_c_1}(a), cfRG, PUF and FLEX
quickly overestimate the correct value.  The FLEX data are reasonably
precise up to $U\approx \Gamma$, those of cfRG and PUF up to $U
\approx 1.5 \Gamma$.  (For comparison, the shown 1PI fRG which employs channel
decomposition provides good results up to $U\approx 5.5\Gamma$.\cite{Kar08})  In Sec.~\ref{sec:Hamann}, we study a possible
exponential behavior of the approximate effective mass.

We now turn to the charge susceptibility in
Fig.~\ref{fig:SIAMeq_c_1}(b).  The FLEX and PUF data turn out to be
trustworthy up to $U\approx 2 \Gamma$, those of cfRG up to $U \approx
3 \Gamma$.  The FLEX and PUF curves lie below the Bethe ansatz curve at low
$U$ and cross it as $U$ increases, while the cfRG curve always lies
above.  Nevertheless, all three approximations are roughly similar.

Let us proceed to the spin susceptibility in
Fig.~\ref{fig:SIAMeq_c_1}(c).  Here, visible deviations of the FLEX
data from the Bethe ansatz result start at $U\approx 1.5 \Gamma$; cfRG
and PUF deviate only slightly later.  All three approximations have in
common that they produce values that are far too low for larger
interaction strengths.  They even show decreasing values instead of an
exponential growth.

In Fig.~\ref{fig:SIAMeq_c_1}(d), the conductance is shown.  The cfRG
and the FLEX data are again quite similar.  Around zero gate voltage,
they do not yield a conductance plateau, but instead a wide curved
region.  The PUF curve is remarkably distinct, with an overpronounced
plateau and convergence problems around the plateau edge. This
exceptional behavior is to be attributed to the large value of the
interaction $U/\Gamma=4\pi$ and it is lifted when turning to smaller values of $U/\Gamma$.
We chose this large value in order to compare to existing data. Only at such large values, the conductance plateau is clearly visible.


\subsection{Hamann's prediction not confirmed}
\label{sec:Hamann}

\begin{figure*}
 \includegraphics[width=0.395\textwidth,clip]{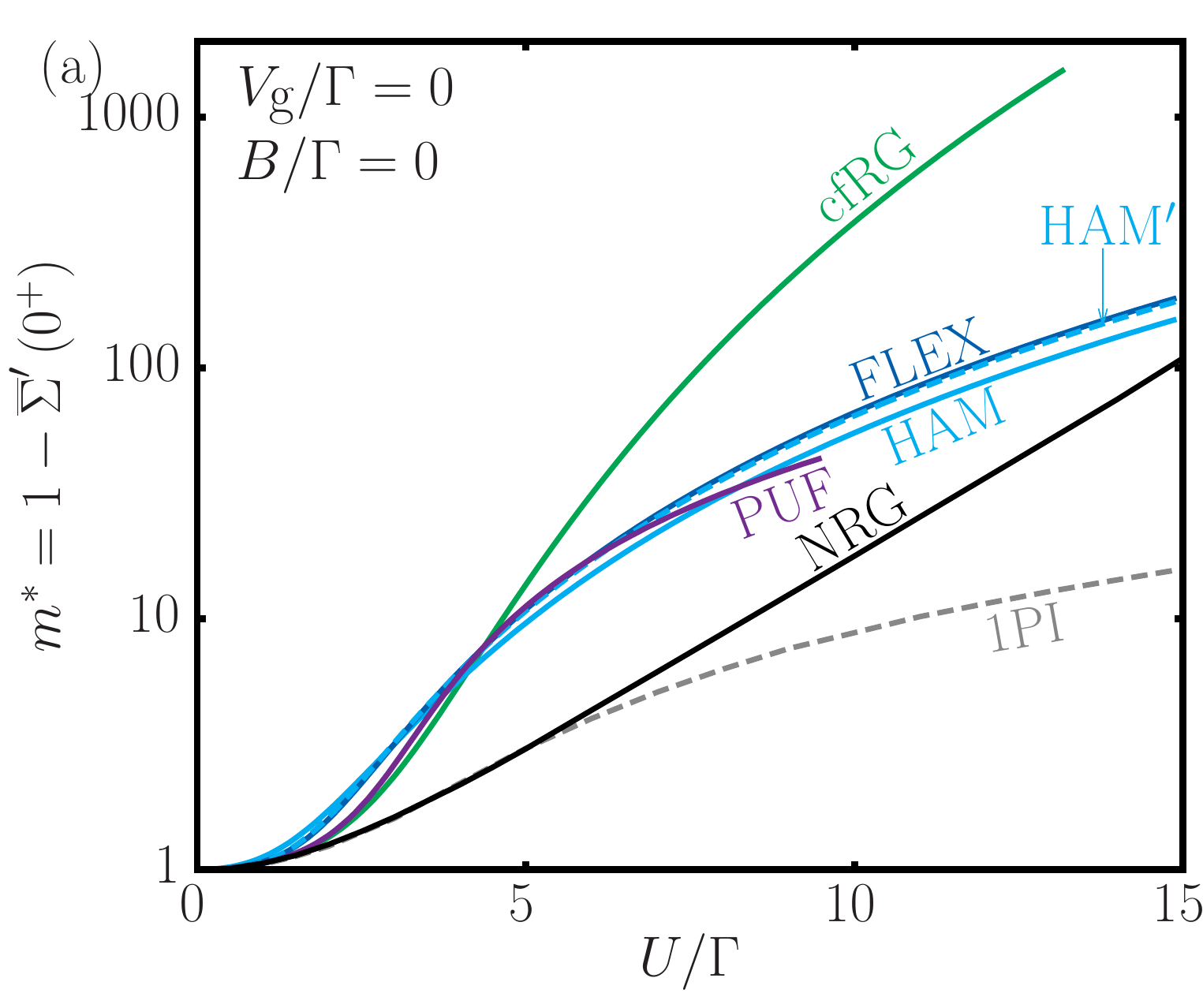}
 \includegraphics[width=0.395\textwidth,clip]{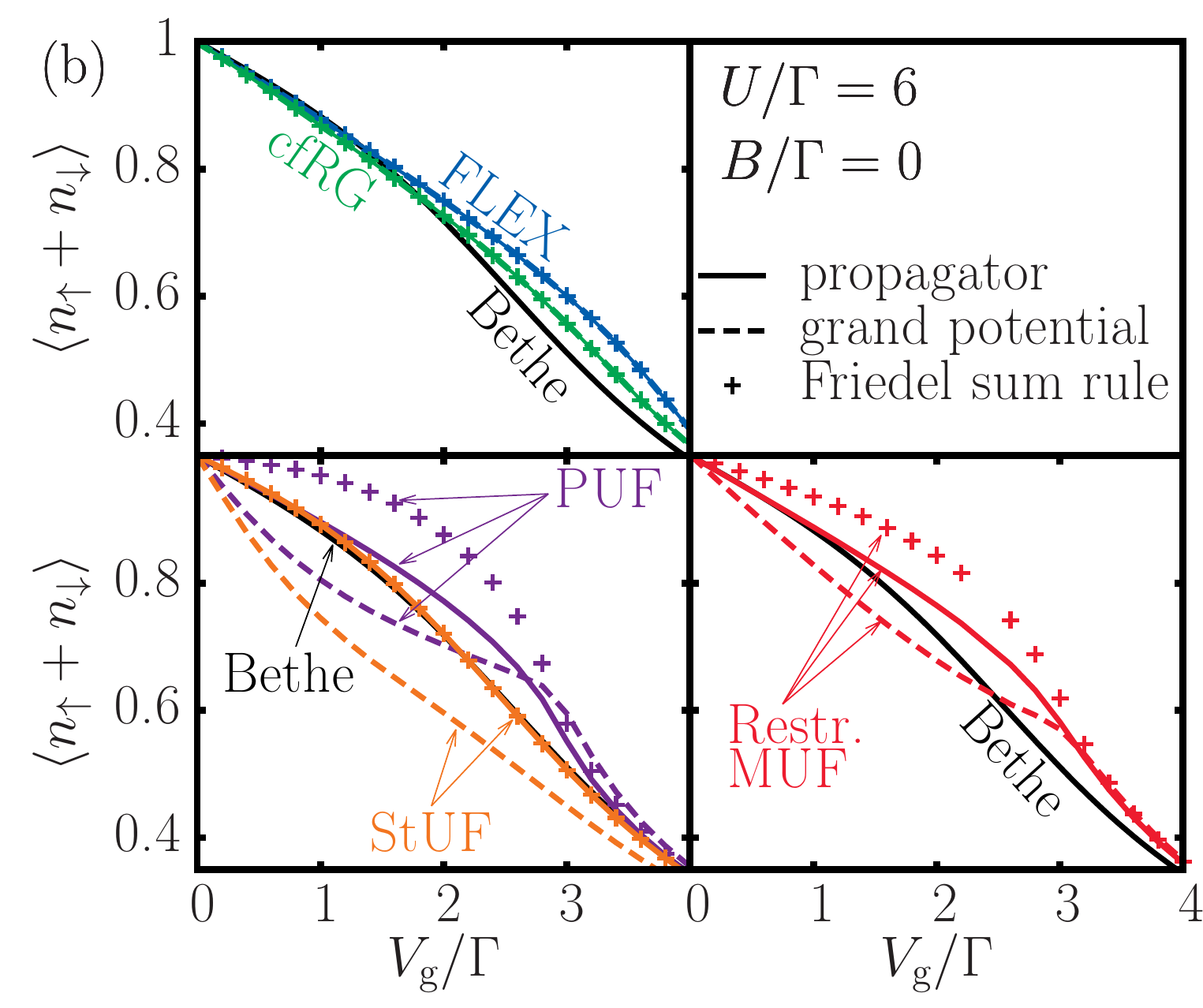}
 \caption{\label{fig:SIAMeq_log_test_non-cons}(Color online) (a) The same numerical
   data for the effective mass as in Fig.~\ref{fig:SIAMeq_c_1}(a),
   but on a logarithmic scale and including higher interaction
   strengths.  The plot shows additionally data obtained with Hamann's
   approximation.  (b) Numerical data for the occupancy of the dot,
   calculated for certain $\Phi$-derivable and non-$\Phi$-derivable schemes in
   various ways. The Bethe ansatz result (from Fig.~5
   in Ref.~\onlinecite{Kar08}) is plotted for comparison.}
\end{figure*}

In Ref.~\onlinecite{Ham69}, Hamann investigates analytically an approximation to
the self-energy of the Anderson model which can be considered an
ancestor of the FLEX method.  For this approximation, which he
attributes to Suhl~\cite{Su67}, he predicts the occurrence of a
characteristic temperature $\sim \exp \left[-\frac{1}{3} (\frac{U}{\pi
    \Gamma})^2\right]$, as opposed to the exact Kondo temperature
$\sim \exp \left(-\frac{\pi}{8} \frac{U}{\Gamma} \right)$.  As the
approximation is similar to FLEX, the characteristic temperature with
quadratic exponent might as well appear for FLEX and the related cfRG
and PUF.  In this case the approximate effective mass should be
proportional to $\exp \left[\frac{1}{3} (\frac{U}{\pi
    \Gamma})^2\right]$.  Here we show that the numerical data do not
confirm this expectation, neither for FLEX and the fRG schemes, nor
for Hamann's approximation itself.

Hamann's approximation~\cite{Ham69} to the self-energy can be derived
from an approximate Luttinger-Ward functional with a diagrammatic
representation almost identical to that of FLEX in
Fig.~\ref{fig:FLEX-diagrams}.  The difference is that the diagrams
with particle-particle ladders are neglected and that the sum of
diagrams with bubble chains (direct particle-hole channel) is
approximated by $\frac{1}{2}$ the sum of diagrams with particle-hole
ladders (exchange particle-hole channel).  Effectively, only the
particle-hole ladder contribution is used, multiplied by a factor of
$\frac{3}{2}$ for all diagrams from second order on.  This yields a
conserving approximation for the self-energy which does not capture
second order perturbation theory with bare lines as the skeleton
second order diagram is multiplied by $\frac{3}{2}$.  By setting
$\kappa_0=0$, $\kappa_\ix{x}=\frac{3}{2}$ and
$\kappa_\ix{p}=\kappa_\ix{d}=0$ in
Eq.~\eqref{eq:cons_Sigma_flow_eq}, we can calculate data
according to this approach.  We refer to this scheme by the index
``HAM''.  A variant of Hamann's idea that takes into account the
natural structure of Eq.~\eqref{eq:FLEX_Sigma_flow_eq_SIAM} for
FLEX is to set $\kappa_0=\frac{2}{3}$, $\kappa_\ix{x}=\frac{3}{2}$ and
$\kappa_\ix{p}=\kappa_\ix{d}=0$. We thus define an alternative scheme
``HAM$^\prime$~'' according to this choice (which also does not
capture second order perturbation theory correctly).

Figure~\ref{fig:SIAMeq_log_test_non-cons}(a) presents again the
effective mass data from Fig.~\ref{fig:SIAMeq_c_1}(a), but on a
logarithmic scale and up to larger values of $U/\Gamma$, now including
HAM and HAM$^\prime$ data.  We observe that the curves for FLEX and
for HAM behave similarly. This confirms that
Hamann's approach to replace all three FLEX channels by $3/2$ the
particle-hole ladder is reasonable.  We observe even better agreement
(almost coincidence on the scale of the plot) of the alternative
proposal HAM$^\prime$ with FLEX. For large $U/\Gamma$, the NRG
effective mass follows the exact result $\sim \exp \left(\frac{\pi}{8}
  \frac{U}{\Gamma} \right)$ which occurs as a straight line in the
log-linear plot.  According to Hamann's prediction, the curve
corresponding to his approximation should increase quadratically in
the log-linear plot at high $U/\Gamma$.  This is obviously not the
case; also the FLEX and the fRG (and HAM$^\prime$) curves do not show
this behavior.  On the contrary, based on the data we expect that the
NRG effective mass even surpasses the Hamann and the FLEX one from
about $U\approx 18 \Gamma$ on.  The reason for this discrepancy to
Hamann's prediction remains to be clarified.


\subsection{Establishing that PUF, StUF and MUF are non-$\Phi$-derivable}
\label{sec:PUF_StUF_MUF_num_non-cons}

In this section we present numerical results which
illustrate that the PUF, StUF and MUF schemes are non-$\Phi$-derivable
approximations.  Fig.~\ref{fig:SIAMeq_log_test_non-cons}(b) shows
the occupancy of the dot calculated for each scheme by the three ways
suggested in Sec.~\ref{sec:Testing_PUF_MUF_non-cons}: from the
propagator, from the grand potential and from the Friedel sum rule.
For the cfRG and the FLEX method, the three ways correctly produce
coinciding results, as expected for $\Phi$-derivable schemes. In contrast,
each way produces a distinctly different result for the PUF, StUF and
MUF schemes.  As a single exception, the
Friedel sum rule and integration of the propagator lead to coinciding
results for the StUF approximation. This, however, is true for all static methods; for these, the
propagator can be integrated analytically to yield the Friedel sum
rule. We have thus provided strong numerical evidence that PUF, StUF
and MUF are indeed non-$\Phi$-derivable approximations. We remark that the same quantities were used to
illustrate that truncated 1PI fRG is not thermodynamically consistent, cf.~Fig.~5 of Ref.~\onlinecite{Kar08}.


\subsection{Results for PUF and MUF}
\label{sec:results_PM_UF}

\begin{figure*}
 \includegraphics[width=0.395\textwidth,clip]{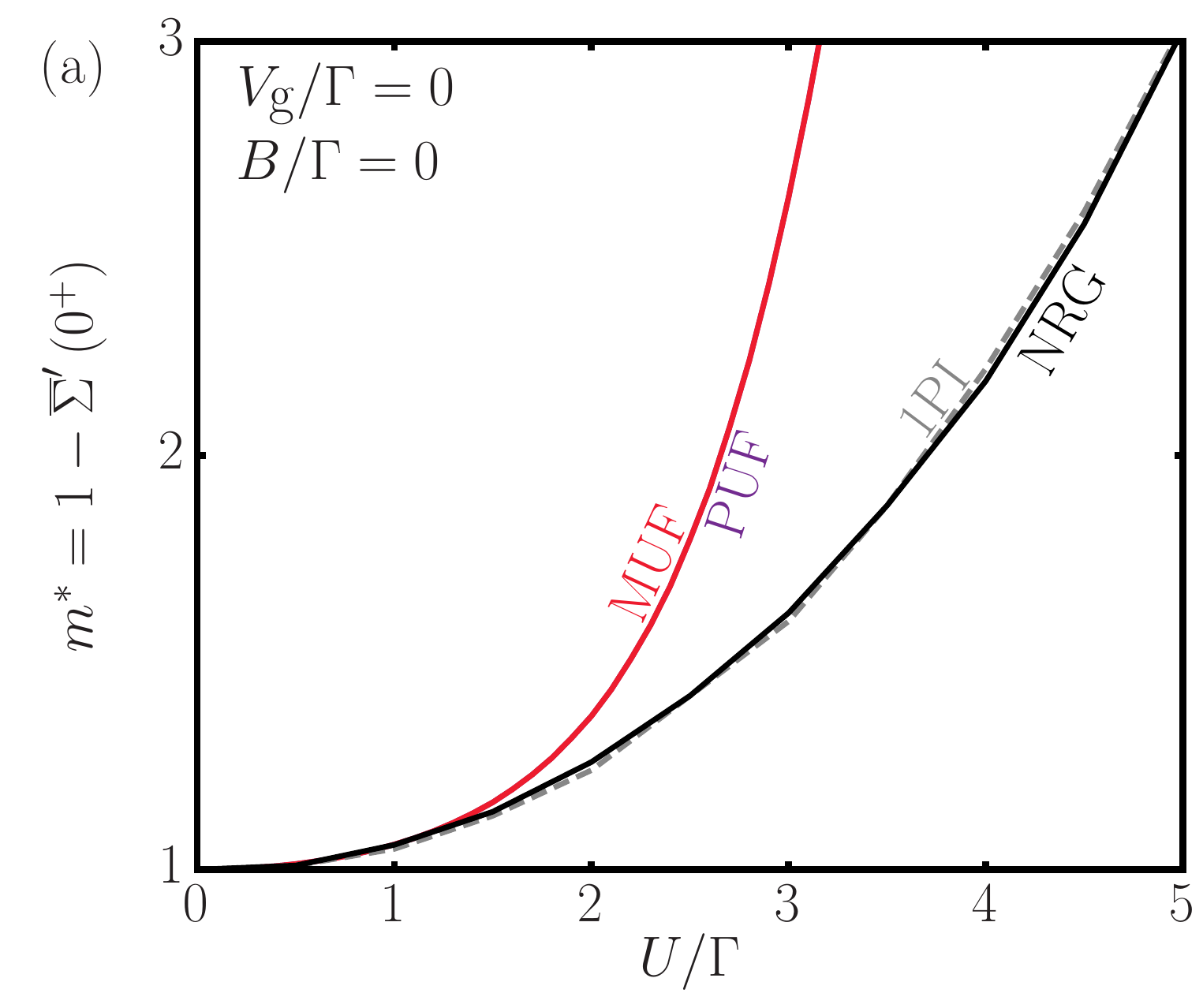}
 \includegraphics[width=0.395\textwidth,clip]{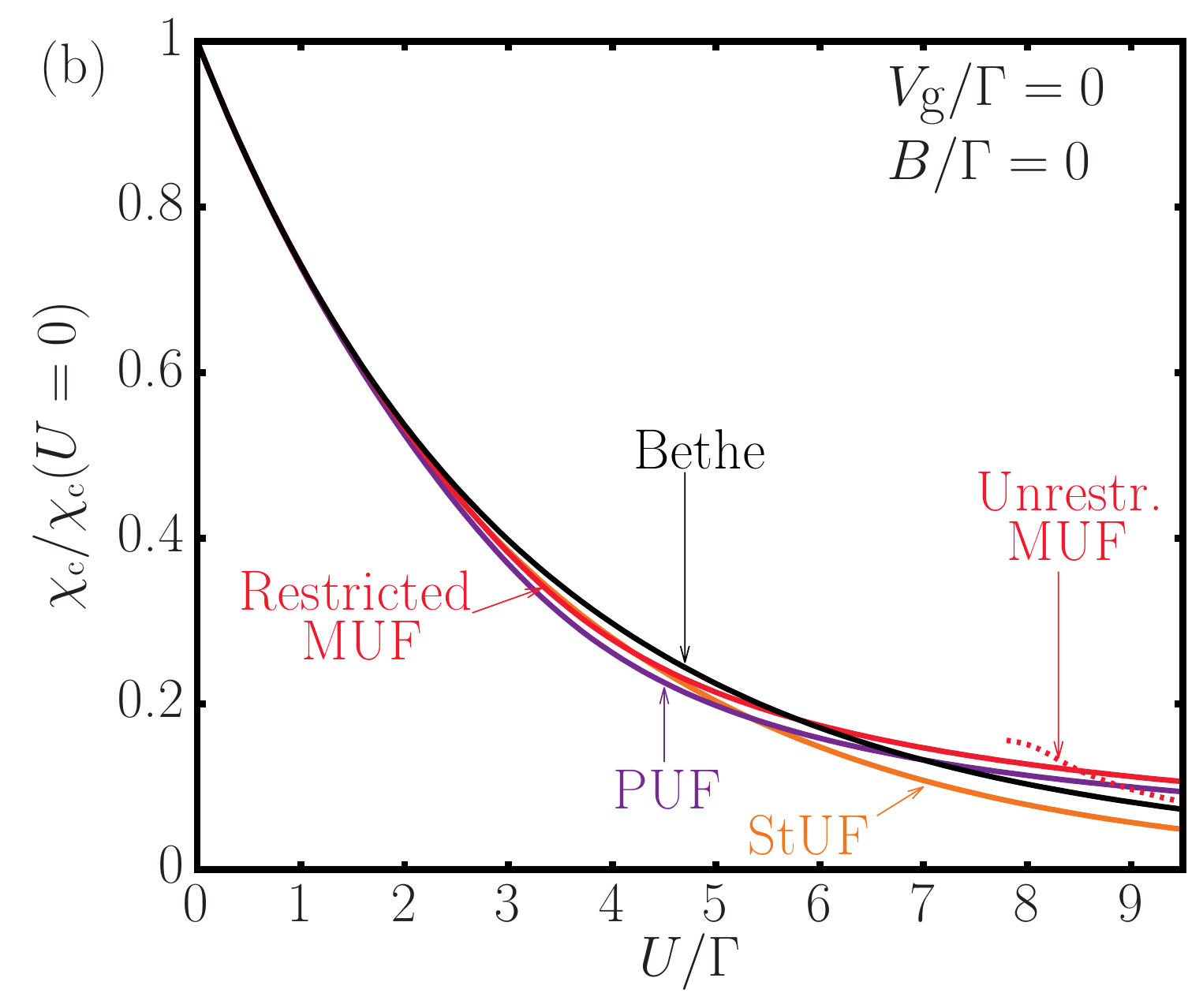}
 \includegraphics[width=0.395\textwidth,clip]{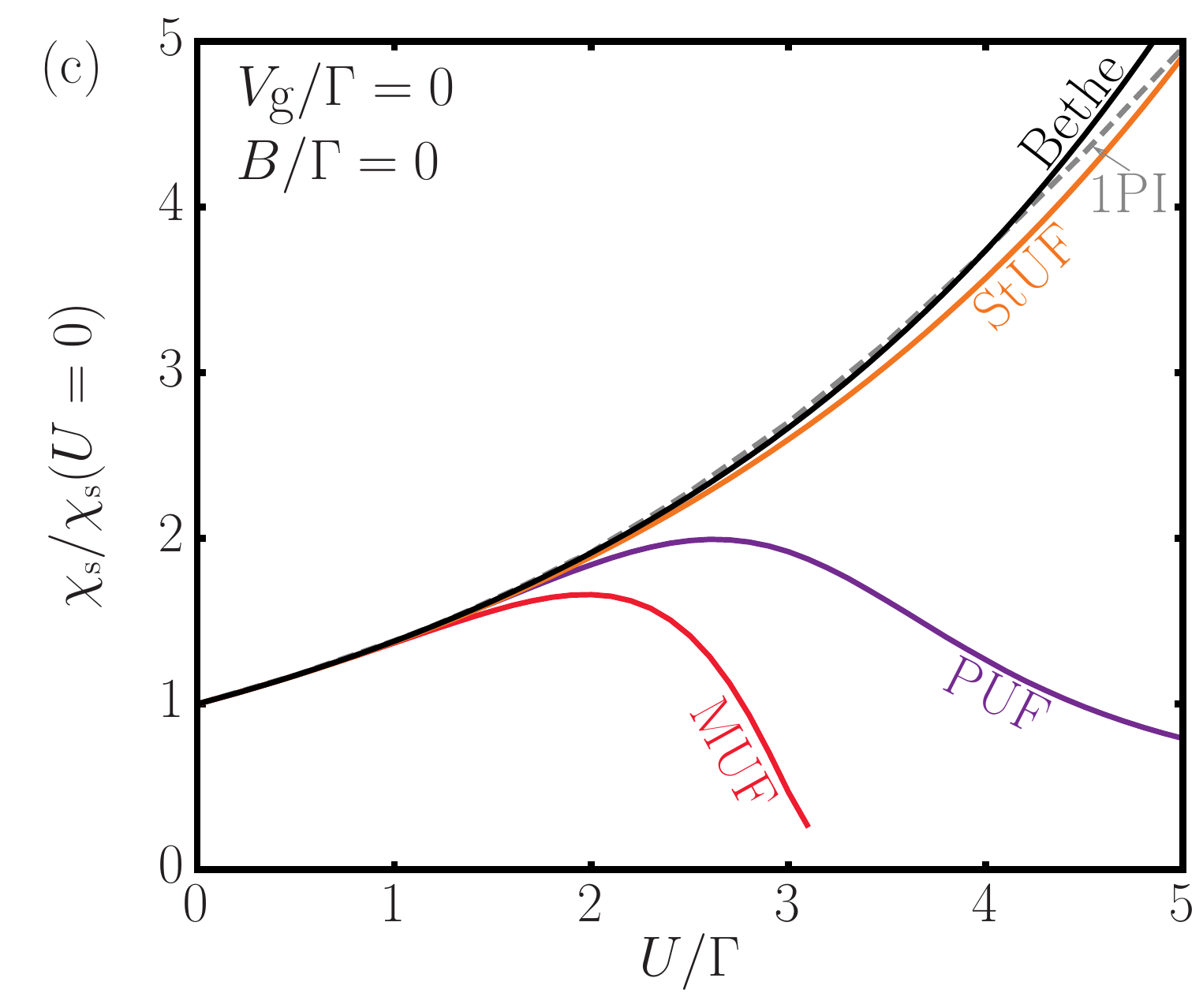}
 \includegraphics[width=0.395\textwidth,clip]{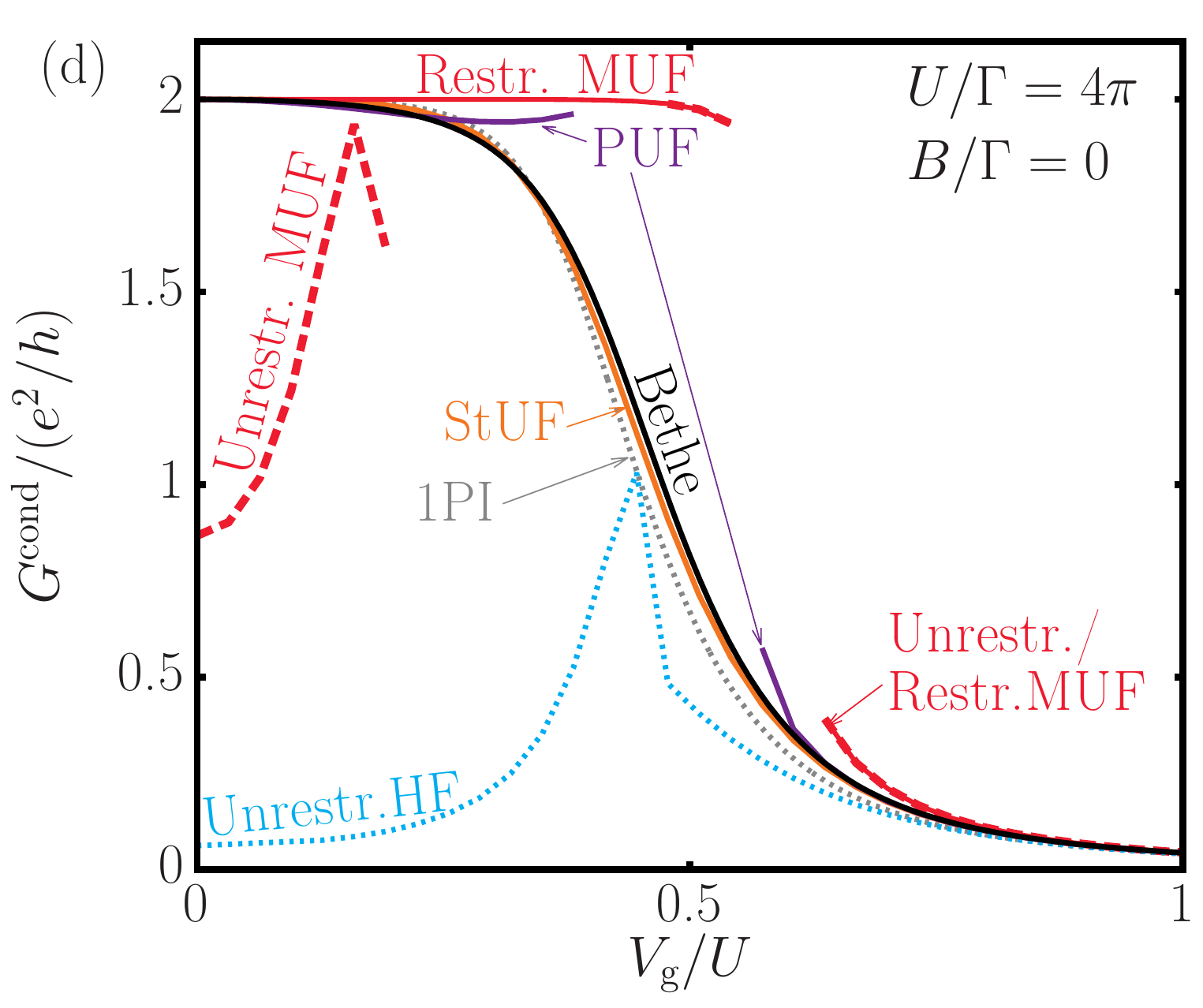}
 \caption{\label{fig:SIAMeq_nc_1} (Color online) Numerical data for the PUF, StUF and
   MUF approximations in comparison to the same NRG, Bethe and 1PI fRG
   curves as in Fig.~\ref{fig:SIAMeq_c_1}. (a) For the effective
   mass, the PUF and MUF curves are nearly indistinguishable in this
   plot. (b) For the charge susceptibility, the flow of unrestricted
   MUF does not come to an end for the intermediate regime
   $U/\Gamma\approx \pi \ldots 8$. (c) For the spin susceptibility,
   neither the flow of restricted MUF nor that of unrestricted MUF
   comes to an end beyond the critical interaction $U>\pi \Gamma$. (d)
   For the conductance, the flow does not come to an end (except for
   StUF) for gate voltages around the plateau edge at such a high
   interaction strength.}
\end{figure*}

In this section, we
discuss the numerical results for the PUF and the MUF approximation.
Figure~\ref{fig:SIAMeq_nc_1} shows the same observables as above for
these schemes.

The PUF and MUF results agree in acceptable limits with the exact ones
only up to rather small $U/ \Gamma \approx 1 \ldots 2$. This becomes
particularly evident in the effective mass and spin
susceptibility. Both, PUF and MUF, quickly overestimate the effective
mass. Similarly, both quickly underestimate the spin
susceptibility. These deviations are reminiscent of the FLEX results
[compare to Fig.~\ref{fig:SIAMeq_c_1}(a) and \ref{fig:SIAMeq_c_1}(c)
or to Ref.~\onlinecite{Whi92}]. In fact, this similarity to FLEX extends to the
charge susceptibility [compare to Figs.~\ref{fig:SIAMeq_c_1}(b) or
to Ref.~\onlinecite{Whi92}].

A poor performance of the PUF scheme at larger interactions is already
known from the quantum anharmonic oscillator which was studied
in Ref.~\onlinecite{Ren15} as a toy model for quantum many-body systems. In
contrast, the MUF approximation performs very well for the anharmonic
oscillator. We attribute the poor performance for the Anderson model
to the following reason: The success or failure of the MUF
approximation is closely related to the success or failure of the
self-consistent Hartree-Fock solution which is used as the starting
point of the flow. This was already anticipated in Ref.~\onlinecite{Ren15}. The
Hartree-Fock method performs well for the anharmonic oscillator
(within 3\% relative error compared to the exact result for a large
range of interaction strengths). For the Anderson model in contrast,
the Hartree-Fock solution is significantly less accurate. This
explains the setback.

For $U>U_\ix{crit}=\pi\Gamma$ the unrestricted Hartree-Fock solution
as starting point of unrestricted MUF unphysically breaks the spin
symmetry.  The numerics indicate that the flow does not restore the
symmetry; on the contrary, it even suffers from convergence
problems. For the effective mass, the charge susceptibility and the
spin susceptibility, the flow of unrestricted MUF does not come to end
for $U\approx U_\ix{crit}\ldots 8\Gamma$. Furthermore, the values calculated
for $U/\Gamma >8$ are not trustworthy. For the effective mass, they
are unconvincingly high (not plotted); for the spin susceptibility,
they are unstable and vary over a large range including negative
values (not plotted); for the charge susceptibility, they are in an
acceptable range but the method predicts a curvature around $U/\Gamma
=8$ contradictory to all other schemes [see
Fig.~\ref{fig:SIAMeq_nc_1}(b)]. Furthermore, the unrestricted MUF
approximation does not reproduce the correct unitary conductance
$G^\ix{cond}=2\frac{e^2}{h}$ at $V_\ix{g}=0$. This deficiency is
shared by the unrestricted Hartree-Fock method [see
Fig.~\ref{fig:SIAMeq_nc_1}(d)] and is obviously not settled by the
flow.

These numerical findings of our
unrestricted MUF scheme do not comply with a prediction made
in Sec.~IV.B of Ref.~\onlinecite{Dup14}. There, it is argued that in
2PI fRG flow schemes a spurious symmetry breaking should decrease and
eventually vanish during the flow due to the influence of Goldstone
modes.  This prediction derives from an analysis of the contribution
$-\overline{\dot \Phi}{}_\lambda^{(1)}$ to the flow of the self-energy
$\dot{\overline \Sigma}_\lambda = -\overline{\dot
\Phi}{}^{(1)}_\lambda - \overline \Phi{}^{(2)}_\lambda \cdot
\dot{\overline G}_\lambda$.  This contribution is argued to reduce the
symmetry breaking with increasing efficiency during the course of the
RG flow.

It can be understood in more detail why the symmetry is not restored in our scheme. For $U$ moderately greater than $\pi\Gamma$, we can observe numerically that the
symmetry breaking indeed starts to decrease during the flow.  However,
a divergence occurs in the flow equation before the symmetry is
restored.  One can understand analytically that this divergence
necessarily occurs in our truncation scheme.  The factor which
diverges becomes apparent when the flow equation for the self-energy
is formulated as non-self-consistent equation, $\dot{\overline
\Sigma}_\lambda = -\left(I+\overline{\Phi}{}^{(2)}_\lambda \cdot
\overline \Pi_\lambda \right)^\inv \cdot \overline{\dot
\Phi}{}^{(1)}_\lambda$, compare Eq.~(98) of Ref.~\onlinecite{Ren15}.
In the MUF truncation, the factor
$\left(I+\overline{\Phi}{}^{(2)}_\lambda \cdot \overline \Pi_\lambda
\right)^\inv$ takes the form $\left(I+U \cdot \overline \Pi_\lambda
\right)^\inv$ with bare interaction $U>\pi\Gamma$.  This RPA-like
series reaches a pole when the symmetry breaking becomes smaller; crucially, this
happens before the symmetry is restored, since $U$ is greater
than the critical value $U_\ix{crit}=\pi\Gamma$ of the
non-symmetry-broken state.  A more detailed analysis of this
divergence is given in Appendix~\ref{sec:details_num_impl}.

For $U$ distinctly greater than $\pi\Gamma$, the flow of the
self-energy in the MUF truncation becomes more complicated. The
emerging frequency dependence and imaginary parts of the self-energy
then play a dominating role and the dressed RPA-like series no longer
has a pole. Therefore, the MUF converges again from $U\approx 8\Gamma$
on. However, only at the beginning of the flow the self-energy is
essentially static and we observe numerically a tendency to suppress
the symmetry breaking. In contrast, at the end of the flow the strongly
frequency dependent self-energies for spin up and down differ largely.

We thus find the prediction of Ref.~\onlinecite{Dup14} that a spurious
symmetry breaking vanishes automatically during the flow not fulfilled
in our MUF scheme.  The
prediction of Ref.~\onlinecite{Dup14} might still be applicable to
more advanced truncation schemes than our MUF.

Let us now turn to the restricted MUF. For the effective mass and the charge susceptibility, it
is able to produce reasonable results above $U_\ix{crit}$ that are
comparable to those of PUF. For the spin susceptibility, in contrast,
the flow does not come to an end for $U>U_\ix{crit}$. This indicates
that the unphysical response of the restricted Hartree Fock starting
point to magnetic fields is not overcome by the RG flow. This is studied in more detail in Appendix~\ref{sec:details_num_impl}. For the
conductance, the restricted MUF approximation predicts the correct
value at $V_\ix{g}=0$.

For the conductance, we observe a problem that is shared by the PUF
and the restricted and unrestricted MUF schemes: The flow does not
come to an end for values of $U/\Gamma$ around the edge of the
conductance plateau. Note that the plateau is calculated at a large
$U/\Gamma=4\pi$. Tuning $U$ to smaller values lifts this problem. The
points that are calculated show a tendency of all three methods to
enlarge the plateau and to make the fall-off at the edge sharper than
in the exact Bethe ansatz solution.

In summary, we find that the unphysical properties of the unrestricted
and restricted Hartree-Fock starting points constitute a major problem
for the MUF approximation.  The unrestricted MUF scheme has proven to
be not trustworthy for $U>U_\ix{crit}$. The restricted MUF scheme
performs better and makes it possible to pass $U_\ix{crit}$ for
$B=0$. The results are comparable to those of the PUF approximation.


\subsection{Results for StUF}
\label{sec:results_StUF}

In this section, we discuss
the numerical results for the StUF approximation.  The corresponding
data are as well shown in Fig.~\ref{fig:SIAMeq_nc_1}.

No reasonable effective mass can be calculated for this scheme, as the
derivative of $\overline{\Sigma}_\sigma(\nu)$ with respect to $\nu$ is
zero. The other observables, however, agree remarkably well with the
exact results. This holds in particular for the conductance data which is even more remarkable at this large $U/\Gamma=4\pi$.
With 1PI fRG employing a static flow of the 1PI two-particle vertex [cf.~Fig.~\ref{fig:SIAMeq_nc_1}(d) and Ref.~\onlinecite{Kar06}], one is already able to obtain agreement with the exact curve at surprisingly large $U/\Gamma$ but StUF even outperforms this scheme. 

The good performance of the StUF scheme is surprising for three
reasons. First, it constitutes a lower order truncation to the 2PI fRG
than the PUF or MUF scheme. Second, the computational effort needed
for solving the scheme is marginal. Third, we find that it does not
produce good results for the quantum anharmonic oscillator.\cite{Ren16}

For $B=0$, we can gain analytical insight in the scaling behavior of
the renormalized single-particle energy.  Let us introduce the
dimensionless renormalized level position
\begin{equation}
  f_\lambda=\frac{\epsilon_{\bar\sigma}+\overline{\Sigma}{}^{\ix{StUF}}_{\bar{\sigma},\lambda}}{\Gamma} .   
\end{equation}
We use $U_\lambda=\lambda U$ and set $u=\frac{U}{\pi \Gamma}$, to obtain [cf.~Eq.~\eqref{eq:flow_eq_StUF_RHS_integrated}]
\begin{equation}
 \dot{f}_\lambda=- u \;\textrm{atan}\left(f_\lambda\right)
\end{equation}
with initial condition $f_{\lambda_\ix{i}}=\frac{V_\ix{g}}{\Gamma} =\!\!\mathop{:} v_\ix{g}$. If $v_\ix{g}=0$, the solution is $f_\lambda=0$. For $v_\ix{g} \neq 0$, separation of variables yields
\begin{equation}
 \int_{v_\ix{g}}^{f_\lambda} \frac{dx}{\ix{atan}(x)} = -u \lambda .
\end{equation}
The integral over $\frac{1}{\ix{atan}(x)}$ yields a scaling of
$f_{\lambda_\ix{f}}(u) \sim v_\ix{g} e^{-u} \sim e^{-\frac{1}{\pi}
  \frac{U}{\Gamma}}$. The (actual) Kondo temperature $T_\ix{K}$ scales
$\sim e^{-\frac{\pi}{8} \frac{U}{\Gamma}}$. This means that the StUF
approximation correctly predicts an exponential scaling with the
interaction strength but yields the wrong prefactor $\frac{1}{\pi}$
instead of $\frac{\pi}{8}$. Lowest order 1PI vertex expansion
Matsubara fRG (without flow of the two-particle vertex) also predicts $\sim
e^{-\frac{1}{\pi} \frac{U}{\Gamma}}$ but reproduces the conductance
plateau much worse than the StUF approximation [cf.
Fig.~\ref{fig:SIAMeq_nc_1}(d) to Fig.~3 in Ref.~\onlinecite{Kar06}].


\subsection{Results for CUF and CF}
\label{sec:results_CUF_CF}

\begin{figure*}
 \includegraphics[width=0.395\textwidth,clip]{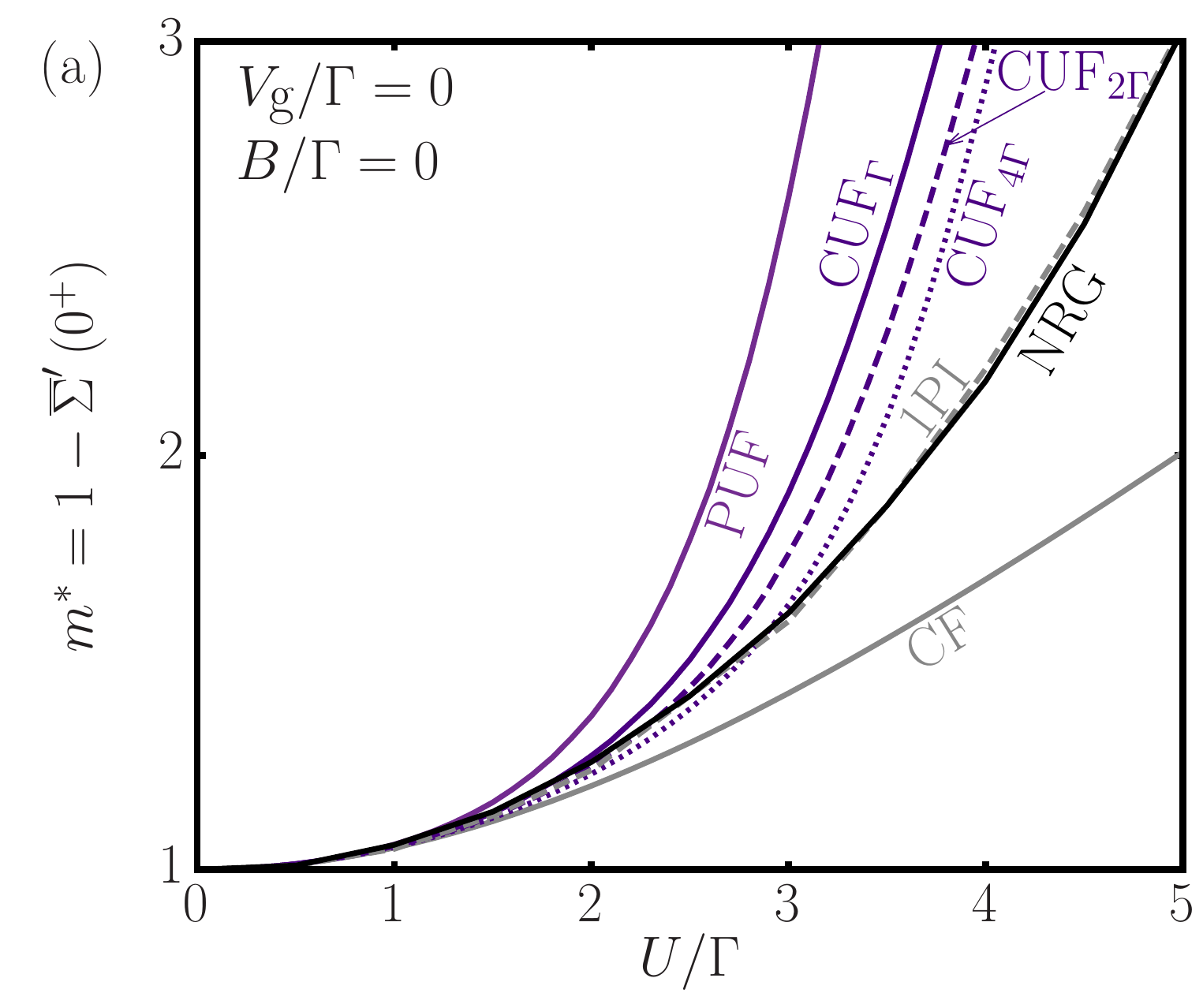}
 \includegraphics[width=0.395\textwidth,clip]{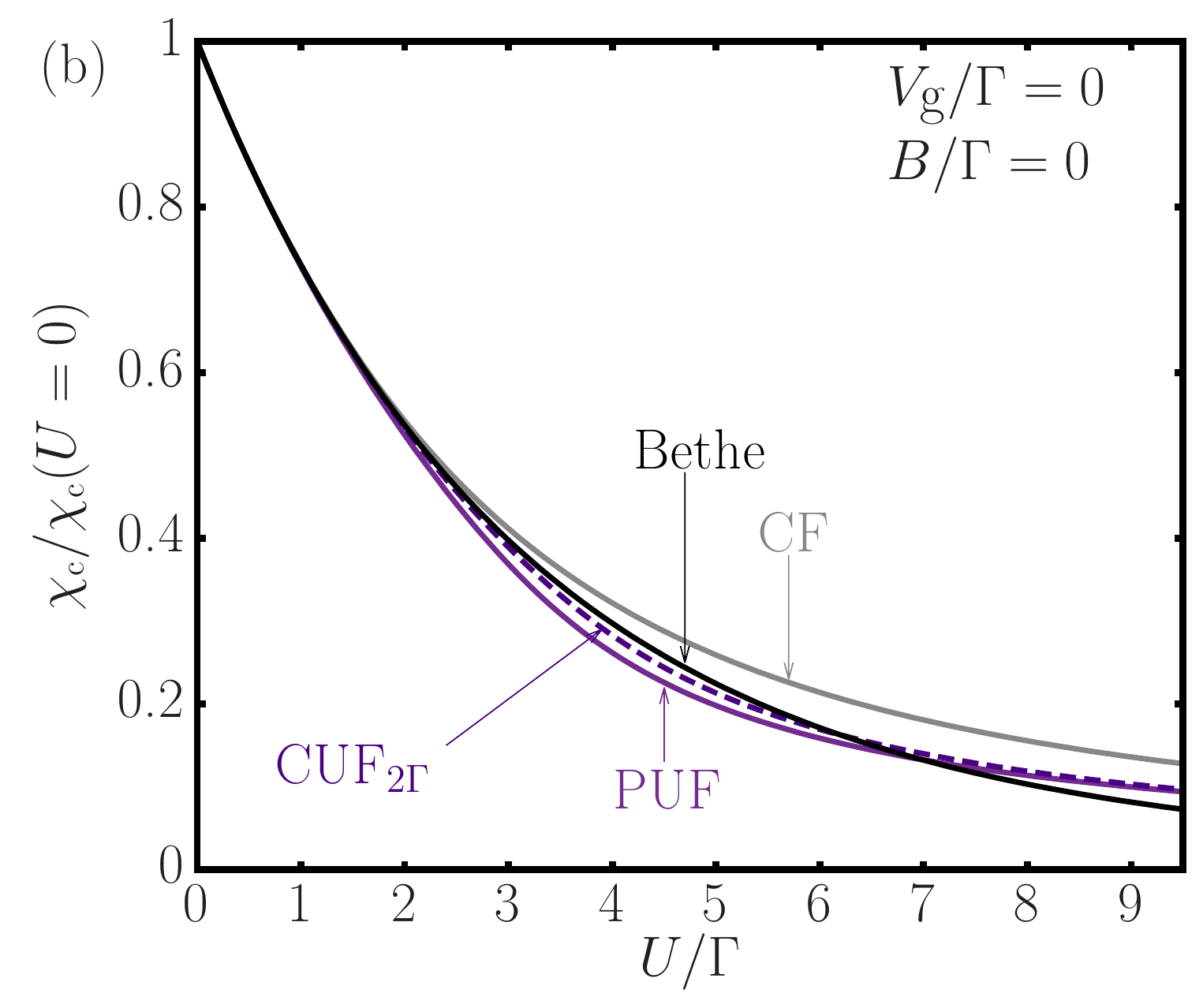}
 \includegraphics[width=0.395\textwidth,clip]{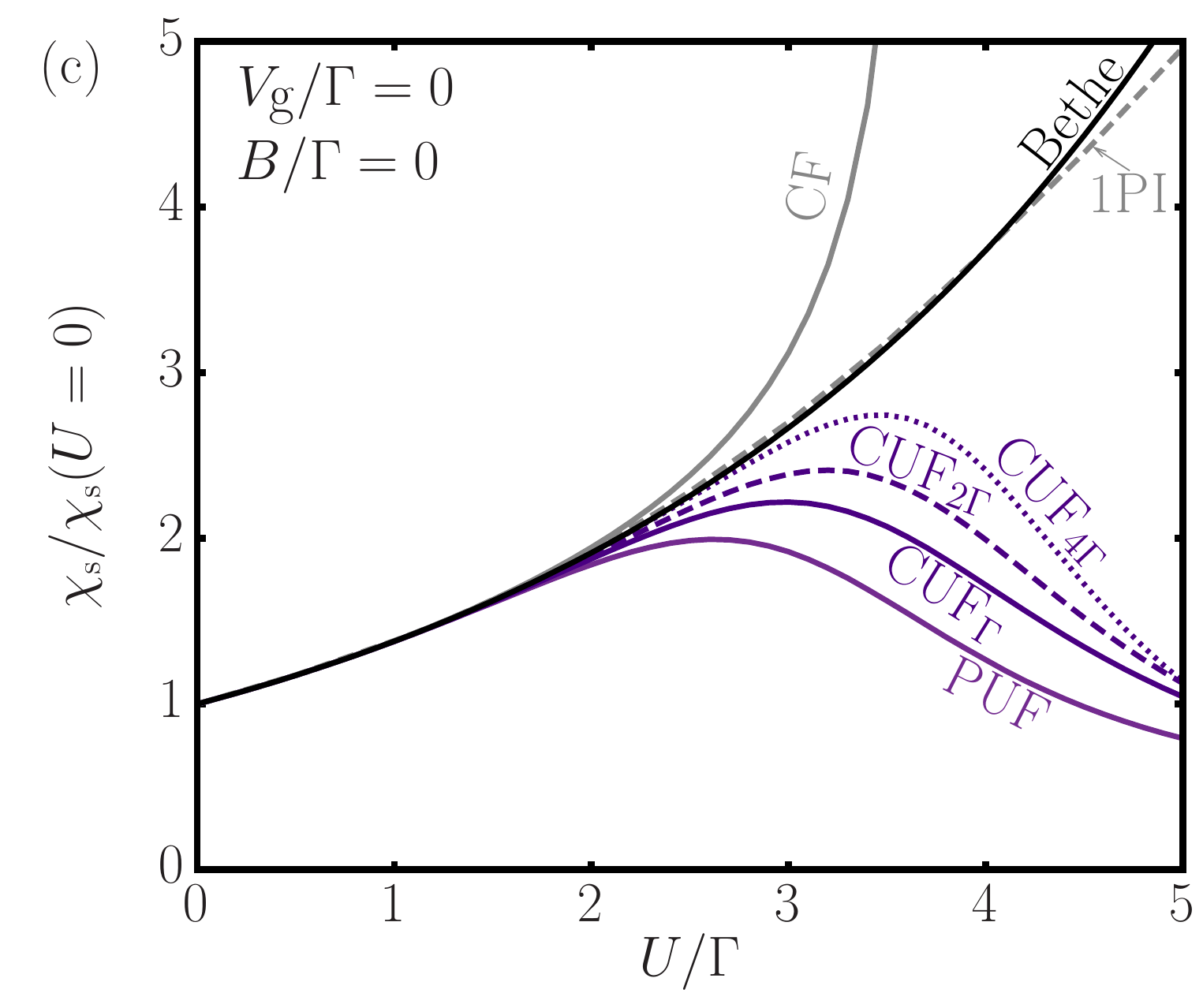}
 \includegraphics[width=0.395\textwidth,clip]{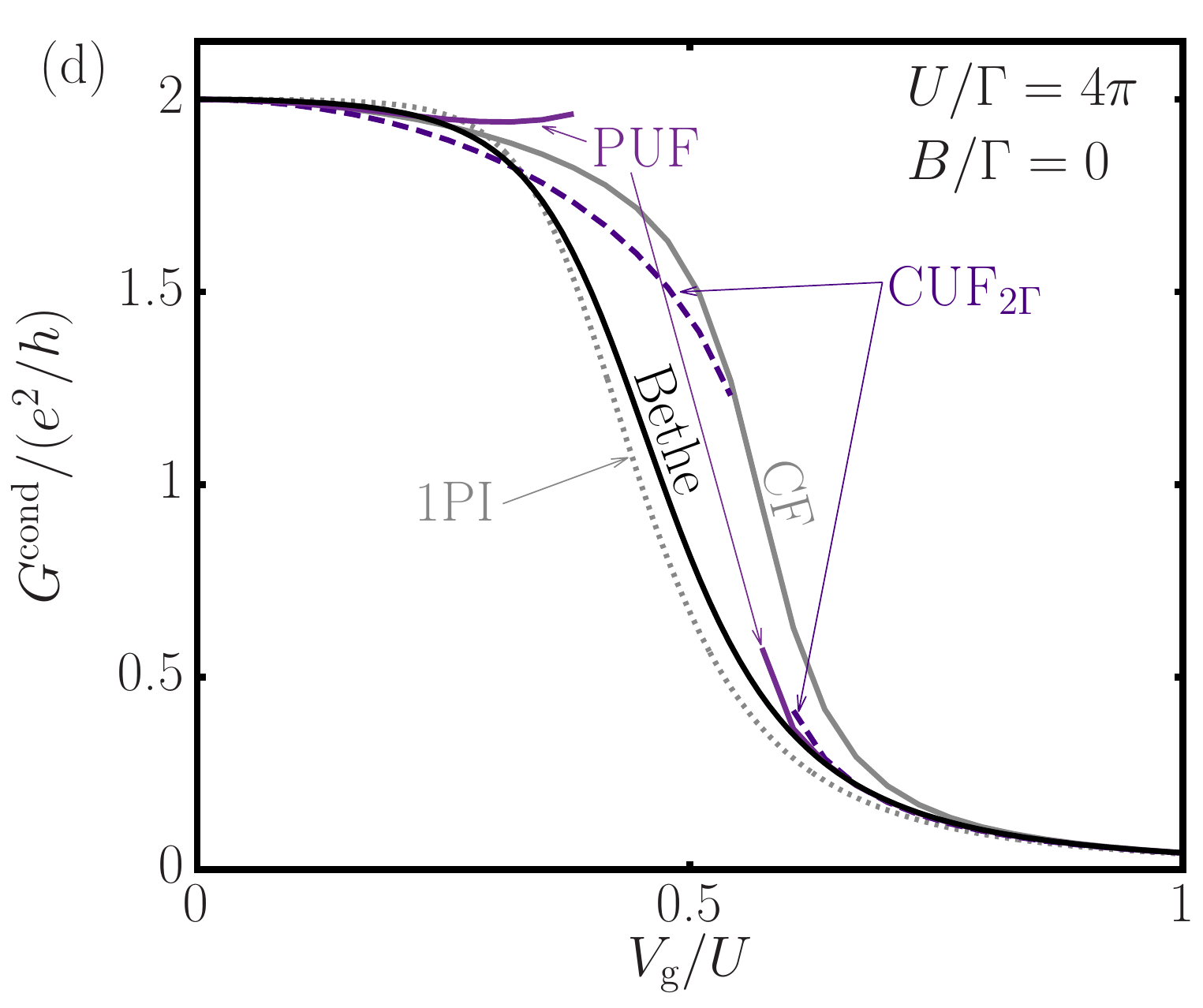}
 \caption{\label{fig:SIAMeq_nc_2} (Color online) Numerical data for the CUF and the CF
   approximations in comparison to the same PUF, NRG, Bethe and 1PI fRG
   curves as in Fig.~\ref{fig:SIAMeq_c_1}. For (a) the effective
   mass and (c) the spin susceptibility, we show CUF data for
   $\Lambda/\Gamma=1,2,4$. For (b) the charge susceptibility and (d)
   the conductance, we only show $\Lambda/\Gamma=2$ data.}
\end{figure*}

In this section, we discuss the results for the CUF and the CF
approximation. Figure~\ref{fig:SIAMeq_nc_2} shows the same observables
as above for these two schemes.

Let us start by discussing the CF approximation. Like PUF and MUF, it
reproduces the exact curves well only up to $U/\Gamma\approx 1\ldots
2$.  For the effective mass and the charge susceptibility, the curves
are close to plain and self-consistent second order perturbation
theory results.\cite{Yam75,Whi92}  This is not surprising.  By
construction the flow equation contains the required terms to generate
plain second order perturbation theory but not enough to generate
self-consistent second order perturbation theory.  For the spin
susceptibility, plain and self-consistent second order perturbation
theory curves lie below the exact curve. The CF curve, in contrast,
lies above. We conjecture that the self-consistent nature of the
lowest order term in the CF scheme induces a strong influence of the
self-consistent Hartree-Fock solution (which also lies above the exact
curve). For the conductance, the CF approximation is able to produce
reasonable data for all gate voltages.

Turning to the CUF method, we find that indeed the results of PUF can
be improved for the effective mass, spin and charge susceptibility by
fine-tuning the value of $\Lambda$. With $\Lambda=2\Gamma$ we are able
to push the boundary for which the CUF data agree acceptably well with
the exact results for all observables up to $U/\Gamma \approx 2 \ldots
3$.  For the spin susceptibility, we see that further increasing
$\Lambda$ to $4\Gamma$ improves the agreement even more. For the
effective mass however, the CUF curve for $\Lambda=4\Gamma$ intersects
the exact curve at $U/\Gamma \approx 3$ but deviates conceivably from
the exact curve for larger and smaller (!)  interaction strength. We
conclude that we can optimize $\Lambda$ in an observable- and
$U$-dependent manner such that we generate agreement to the exact
curve. This is only partially satisfactory.  We would have preferred
the existence of an optimal choice of $\Lambda$ that yields agreement
to the exact curves over a large range of $U$ for all observables.


\section{Conclusion}
\label{sec:conclusion}

In this paper we investigated how the $U$-flow fRG can be used to
construct $\Phi$-derivable approximations, and how different $U$-flow
approximations perform in computing typical observables of the
Anderson impurity model in equilibrium.

Concerning the first question we found it helpful to address the flow
of the Luttinger Ward functional $\Phi[G]$ and the flow of the
physical values $\Phi^{(n)}[\overline G]$ of its vertex functions
separately.  We have seen that elementary truncations of the flow of
the functional lead indeed to approximate $\Phi$'s that are invariant
under symmetry transformations and thus define $\Phi$-derivable
(conserving) approximations.  In the lowest order truncation we
rederived the self-consistent Hartree-Fock approximation while the
next higher truncation (cfRG) led to a $\Phi$ that closely resembles
that of the FLEX approximation. In this sense, the fRG did not provide an approximation of a
fundamentally new structure. In particular, it can again be
understood as a diagrammatic approximation to $\Phi$, except for
prefactors. This might
change in higher order truncations, whose solution is however
analytically quite involved and numerically inaccessible.  It is
remarkable that the analytic integration of the flow in the studied
truncation yielded a result completely independent of the chosen flow
parameter.  This resembles the observation of Ref.~\onlinecite{Ren15} that the
result of truncated $C$-flow for the physical vertex functions is
independent of the flow parameter, namely given by self-consistent
perturbation theory.  As for the $C$-flow, we conclude that for models
with infrared divergencies in perturbation theory the cfRG is only
applicable if the resulting diagrammatic resummation or the
self-consistency has a regularizing effect.

Next we studied the flow of the physical values $\Phi^{(n)}[\overline
G]$ of the vertex functions. It is described by a coupled hierarchy of
flow equations that is infinite even when the hierarchy for the
functional $\Phi[G]$ has been truncated.  Truncating in turn the
hierarchy for the physical values will in general lead to a
non-$\Phi$-derivable approximation.  Therefore we could not obtain new
$\Phi$-derivable fRG approximations from the flow of the physical
values alone; the flow of the whole functional seems to be required.
By truncating the new hierarchy we recovered indeed the plain and modified $U$-flow
approximations of Ref.~\onlinecite{Ren15} (PUF and MUF) as non-$\Phi$-derivable
approximations to the $\Phi$-derivable cfRG-approximation.  We
demonstrated explicitly that they are not thermodynamically consistent by
comparing numerical results for the impurity occupancy obtained from
different approaches.  We truncated the hierarchy for the physical
values also on the lowest order level and uncovered a simple static
non-$\Phi$-derivable $U$-flow scheme (StUF) that was not noticed
in Ref.~\onlinecite{Ren15}.

We tested the different approximation schemes by computing typical
observables of the equilibrium Anderson impurity model.  Compared to
1PI fRG approximations the results are in general rather poor.  For
cfRG and PUF they reflect the kinship with FLEX, a method that is
known to be of limited usefulness for the model at hand.\cite{Whi92}
Based on an analytic prediction by Hamann~\cite{Ham69} for a similar
approximation we expected an $\exp(cU^2)$-behavior of the approximate
effective mass.  This is not confirmed by the data, not even for the
very approximation analyzed by Hamann.  We consider it improbable that
errors in our numerics are the reason for the discrepancy. Our FLEX
data coincide with published ones,\cite{Whi92} and data for Hamann's
approximation can be generated by changing only a few prefactors in
the code.  Further work is required to understand the discrepancy.

Concerning the modified $U$-flow, the artifacts
introduced by the Hartree-Fock initial condition proved to constitute a major
obstacle for the flow.  Neither the spin symmetry breaking of the
unrestricted nor the negative spin susceptibility of the restricted
Hartree-Fock starting point were overcome by the RG flow. In contrast,
they impeded the convergence of the numerical flow.

Comparing the numerical errors of the different $U$-flow schemes to
those of self-consistent perturbation theory (which corresponds to
straightforwardly truncated $C$-flow) we considered it conceivable
that a combination of both methods might improve the approximation
quality.  Therefore we devised the CUF$_\Lambda$ approximation, where
the parameter $\Lambda$ allows for a smooth interpolation between the
PUF approximation and an appropriate (non-$\Phi$-derivable) $C$-flow
truncation.  We found that the range of validity of the approximation
can indeed be extended for all discussed observables, however only to
still moderate $U/\Gamma = 2 \dots 3$.

In contrast to the more elaborate schemes, the simple static variant
StUF performed remarkably well, in particular in regard of the
marginal computational effort it requires.  It describes the linear
conductance as function of the gate voltage better than any 1PI fRG
method that has been applied to the problem. Similar to the static 1PI
fRG without flow of the two-particle vertex, it allows to extract analytically a characteristic scale
$\exp[-U/(\pi \Gamma)]$, where only the prefactor of the exponent
differs from the exact Kondo temperature $\sim\exp[-\pi U/(8
\Gamma)]$.  In view of this success it is an interesting question,
whether the StUF approximation can be extended to higher order
truncations in some other systematic way than studied here.


\section*{Acknowledgements}
We thank D. Manske for helpful discussions on the FLEX approximation, S. Andergassen for her comments on a preprint version of the paper, and N. Dupuis for stimulating comments regarding the spin symmetry breaking in the MUF scheme. We acknowledge support by the Deutsche Forschungsgemeinschaft via the Research Training Group 1995 ``Quantum many-body
methods in condensed matter systems''.


\appendix


\section{Derivation of implementable equations}
\label{sec:Der_impl_eqs}

In this appendix, we present more details on how to derive the
self-consistency or flow equations in a form that exploits the
symmetries and conservation laws specific for the Anderson impurity
model. These equations can then serve as a starting point for the
numerical implementation.


\subsection{Reducing the number of indices}
\label{sec:Reducing_number_indices}

Like the interaction $U^{(2)}$ (cf.~Sec.~\ref{sec:Ham_and_action_SIAM}), all four-point functions $B=\Pi,\Upsilon,\ldots$ turn out to be sparse for the Anderson model. Moreover, they have components which are connected by symmetry. We want to refer to the components in a suitably reduced form. For this purpose, we group the four $y=(c,\sigma)$-indices together in pairs (the first two and the latter two). Such a pair may never take the index combination ${}^{cc}_{\sigma \sigma}$ because this would correspond to a double creation/annihilation of a spin-$\sigma$ electron on the dot (this statement does not hold for $B^\R$). This leaves the following set of index combinations which are allowed for the pairs:
\begin{equation}
\mathbb{I}= \left\{ {}^{++}_{\uparrow \downarrow}, {}^{--}_{\uparrow \downarrow}, {}^{+-}_{\uparrow \uparrow}, {}^{+-}_{\downarrow \downarrow}, {}^{+-}_{\uparrow \downarrow}, {}^{+-}_{\downarrow \uparrow}, {}^{++}_{\downarrow \uparrow}, {}^{--}_{\downarrow \uparrow}, {}^{-+}_{\uparrow \uparrow}, {}^{-+}_{\downarrow \downarrow}, {}^{-+}_{\downarrow \uparrow}, {}^{-+}_{\uparrow \downarrow}\right\} .
\end{equation}
Let $s_1,s_2$ be such indices $\in \mathbb{I}$. Then, we can refer to all non-zero components of $B$ via $B_{s_1 s_2, n_1 n_1^\prime n_2 n_2^\prime}$. As a side-note, the symmetry $B_{s_1 s_2,n_1 n_1^\prime n_2 n_2^\prime}=B_{s_2 s_1,n_2 n_2^\prime n_1 n_1^\prime}$ holds if $B=B^\ix{T}$. However, this is not always the case (e.g. $\Upsilon\neq \Upsilon^\ix{T}$). Thus, we do not use this property in the following considerations. We define an operation \,$\widetilde{}$\, on $s$ via
\begin{equation}
 \widetilde{s}=\widetilde{\left({}^{c_1 c_2}_{\sigma_1 \sigma_2}\right)}=\left({}^{c_2 c_1}_{\sigma_2 \sigma_1}\right).
\end{equation}
Then, the relation $B_{s_1 s_2,n_1 n_1^\prime n_2 n_2^\prime}=-B_{\widetilde{s}_1 s_2,n_1^\prime n_1 n_2 n_2^\prime}$ holds in general [cf.~Eq.~\eqref{eq:symm_interchange_alphas}]. The first six indices and the last six indices in the set $\mathbb{I}$ are connected via the $\widetilde{\phantom{s}}$-operation: Let $\mathbb{S}=\big\{ {}^{++}_{\uparrow \downarrow}, {}^{--}_{\uparrow \downarrow}, {}^{+-}_{\uparrow \uparrow}, {}^{+-}_{\downarrow \downarrow}, {}^{+-}_{\uparrow \downarrow}, {}^{+-}_{\downarrow \uparrow}\big\}$, then $\mathbb{I}=\mathbb{S} \cup \widetilde{\mathbb{S}}$. If we base a $12 \times 12$-matrix notation of $B$ on the order of indices as chosen above, each matrix $B$ can be written in terms of a $6 \times 6$-matrix $\underline{B}$:
\begin{equation}
B_{s_1 s_2,n_1 n_1^\prime n_2 n_2^\prime}=
\left(\begin{array}{cc}
{\underline{B}_{n_1 n_1^\prime n_2 n_2^\prime}} & -{\underline{B}_{n_1 n_1^\prime n_2^\prime n_2}} \\
-{\underline{B}_{n_1^\prime n_1 n_2 n_2^\prime}} & {\underline{B}_{n_1^\prime n_1 n_2^\prime n_2}}
\end{array} \right)_{s_1 s_2}
\end{equation}
Obviously, it is sufficient to work with the underlined matrices. For example, a contraction of $s$-indices $\in \mathbb{I}$ is the same as a contraction of $S$-indices $\in \mathbb{S}$ taking into account a factor of $\frac{1}{2}$, i.e. $\frac{1}{2} \sum_s \leftrightarrow \sum_S$. As an example for an underlined matrix, we provide the interaction in this notation:
\begin{multline}
\label{eq:Matrix_U}
\underline{U}_{S_1 S_2,n_1 n_1^\prime n_2 n_2^\prime}=\beta \delta_{n_1+n_1^\prime+n_2+n_2^\prime,0}
\\ \times \left(\begin{array}{cccccc}
0 & -U & 0 & 0 & 0 & 0 \\
-U & 0 & 0 & 0 & 0 & 0 \\
0 & 0 & 0 & U & 0 & 0 \\
0 & 0 & U & 0 & 0 & 0 \\
0 & 0 & 0 & 0 & 0 & -U \\
0 & 0 & 0 & 0 & -U & 0 
\end{array} \right)_{S_1 S_2}.
\end{multline}
This matrix is block-diagonal consisting of three $2 \times 2$-matrices. It can be shown that this is a general feature for all underlined matrices $\underline{B}$. This is a consequence of particle-number conservation and spin conservation which is fulfilled by propagation and interaction in the Anderson model. Particle-number conservation implies that the number of $c$'s equal to $+$ must be even and so must the number of $c=-$. It does, however, not imply that the sum of all $c$'s must be $0$. This holds only for vertex-like (e.g. $U$) or propagator-like (e.g. $\Pi$) quantities. Combinations of these quantities which are $I$-like (e.g. $\Upsilon$) may have all $c$'s equal to $+$ or equal to $-$ [see e.g. Eq.~\eqref{eq:Psi_++++_component}]. Spin-conservation implies that $c_1 \sigma_1 + c_1^\prime \sigma_1^\prime$ must be equal to $\pm (c_2 \sigma_2 +c_2^\prime \sigma_2^\prime)$. The sign depends on whether the quantity is vertex-, propagator- or $I$-like. For illustration let us consider two examples. A ($S_1 S_2= {}^{++}_{\uparrow \downarrow} {}^{+-}_{\uparrow \uparrow}$)-component would violate particle-number conservation. A ($S_1 S_2= {}^{+-}_{\uparrow \uparrow} {}^{+-}_{\uparrow \downarrow}$)-component would violate spin conservation.


\subsection{Calculating $\overline{\Upsilon}$ for the Anderson model}
\label{sec:calc_Ups_for_SIAM}

In this section, we calculate $\overline{\Upsilon}$ for the Anderson model. We will need this quantity in (almost) all methods.

We use Eq.~\eqref{eq:def-pair-prop} to find
\begin{multline}
 \overline{\Pi}{}^{y_1 y_1^\prime y_2 y_2^\prime}_{n_1 n_1^\prime n_2 n_2^\prime}=\beta^2 \delta_{n_1^\prime+n_2,0} \delta_{n_1+n_2^\prime,0} \overline{G}^{y_1^\prime y_2}_{n_1^\prime} \overline{G}^{y_1 y_2^\prime}_{n_1}
 \\ -\beta^2 \delta_{n_1+n_2,0} \delta_{n_1^\prime+n_2^\prime,0} \overline{G}^{y_1 y_2}_{n_1} \overline{G}^{y_1^\prime y_2^\prime}_{n_1^\prime} .
\end{multline}
With
\begin{equation}
U^{y_1 y_1^\prime y_2 y_2^\prime}_{n_1 n_1^\prime n_2 n_2^\prime}=\beta \delta_{n_1+n_1^\prime+n_2+n_2^\prime,0} U^{y_1 y_1^\prime y_2 y_2^\prime},
\end{equation}
we then find
\begin{equation}
 \left(U \cdot \overline{\Pi}\right)^{y_1 y_1^\prime y_2 y_2^\prime}_{n_1 n_1^\prime n_2 n_2^\prime}= \beta \delta_{n_1+n_1^\prime-n_2-n_2^\prime,0} \overline{\Psi}{}^{y_1 y_1^\prime y_2 y_2^\prime}_{-n_2,-n_2^\prime}
\end{equation}
in which
\begin{equation}
 \label{eq:Def_Psi}
 \overline{\Psi}{}^{y_1 y_1^\prime y_2 y_2^\prime}_{-n_2,-n_2^\prime}= - \sum_{y_3 y_3^\prime} U^{y_1 y_1^\prime y_3 y_3^\prime} \overline{G}^{y_3 y_2}_{-n_2} \overline{G}^{y_3^\prime y_2^\prime}_{-n_2^\prime}.
\end{equation}
We then prove
\begin{align}
 &\left[\left(-U \cdot \overline{\Pi}\right) \cdot \left(-U \cdot \overline{\Pi}\right)\right]^{y_1 y_1^\prime y_2 y_2^\prime}_{n_1 n_1^\prime n_2 n_2^\prime}
\\ \notag &= \beta \delta_{n_1+n_1^\prime-n_2-n_2^\prime,0} \frac{1}{2\beta } \sum_{y_3 y_3^\prime n_3} \overline{\Psi}{}^{y_1 y_1^\prime y_3 y_3^\prime}_{-n_3,n_3-n_2-n_2^\prime} 
\overline{\Psi}{}^{y_3 y_3^\prime y_2 y_2^\prime}_{-n_2,-n_2^\prime}.
\end{align}

Let us now introduce some notations concerning the space of $y$-indices only: $\mathds{1}^{y_1 y_1^\prime y_2 y_2^\prime}=\delta_{y_1 y_2} \delta_{y_1^\prime y_2^\prime} - \delta_{y_1 y_2^\prime} \delta_{y_1^\prime y_2}$ and $(A \circ B)^{y_1 y_1^\prime y_2 y_2^\prime}=\frac{1}{2} \sum_{y_3 y_3^\prime} A^{y_1 y_1^\prime y_3 y_3^\prime} B^{y_3 y_3^\prime y_2 y_2^\prime}$. Then, we show by induction that
\begin{align}
  & \left[\left(-U \cdot \overline{\Pi}\right)^k \cdot \left(-U \cdot \overline{\Pi}\right)\right]^{y_1 y_1^\prime y_2 y_2^\prime}_{n_1 n_1^\prime n_2 n_2^\prime}
\\ \notag =& -\beta \delta_{n_1+n_1^\prime-n_2-n_2^\prime,0} 
\\ \notag &\times \left\{\left[-\frac{1}{\beta}\sum_n \overline{\Psi}_{-n,n-n_2-n_2^\prime} \right]^{\circ k} \circ
\overline{\Psi}_{-n_2,-n_2^\prime} \right\}^{y_1 y_1^\prime y_2 y_2^\prime} .
\end{align}
On the right-hand side, we introduced $^{\circ k}$ to refer to the $k$-fold $\circ$ operation. We define the abbreviation
\begin{eqnarray}
 \label{eq:Def_Psi_one_freq}
 \widetilde{\Psi}_{m=n_2+n_2^\prime}=\frac{1}{\beta}\sum_n \overline{\Psi}_{-n,n-n_2-n_2^\prime}.
\end{eqnarray}
Making use of the geometric series, we now find
\begin{align}
\label{eq:Ups_four_freq}
\overline{\Upsilon}{}^{y_1 y_1^\prime y_2 y_2^\prime}_{n_1 n_1^\prime n_2 n_2^\prime}
=& \beta \delta_{n_1+n_1^\prime-n_2-n_2^\prime,0} 
\\ \notag
&\times \left\{\left[\mathds{1}+\widetilde{\Psi}_{n_2+n_2^\prime} \right]^{\circ(-1)} \!\circ
\overline{\Psi}_{-n_2,-n_2^\prime} \right\}^{y_1 y_1^\prime y_2 y_2^\prime} \!\!\!.
\end{align}
Here, the inverse $^{\circ(-1)}$ is to be understood with respect to the $\circ$ operation in the space of $y$-indices. For this inversion, we resort to the matrix notation introduced in Sec.~\ref{sec:Reducing_number_indices}.

Before performing this inversion, we reduce the frequency structure:
In Eq.~\eqref{eq:Ups_four_freq}, the fourth index $n_2^\prime$ is
determined by the $\delta$-function. Thus, $\overline{\Upsilon}{}^{y_1
  y_1^\prime y_2 y_2^\prime}_{n_1 n_1^\prime n_2 n_2^\prime}$ actually
depends on three frequency indices only. Conveniently,
$\overline{\Upsilon}$ turns out to be needed only in a form in which
the third index is always summed over independently. Thus, this summed
$\overline{\Upsilon}$ depends only on the first and second indices. In
fact, it turns out to only depend on the sum of the two indices and we
define:
\begin{align}
\label{eq:Def_Upsilon_tilde}
 \widetilde{\Upsilon}^{y_1 y_1^\prime y_2 y_2^\prime}_{n_1+n_1^\prime}&=\frac{1}{\beta} \sum_{n_2} \overline{\Upsilon}{}^{y_1 y_1^\prime y_2 y_2^\prime}_{n_1,n_1^\prime,n_2}
\\ \notag
&= \left\{\left[\mathds{1}+\widetilde{\Psi}_{n_1+n_1^\prime} \right]^{\circ(-1)} \circ
\widetilde{\Psi}_{n_1+n_1^\prime} \right\}^{y_1 y_1^\prime y_2 y_2^\prime} .
\end{align}

Now, let us turn to the inversion of $[\mathds{1}+\widetilde{\Psi} {}_{n_1+n_1^\prime}]$. Using Eqs.~\eqref{eq:Def_Psi} and~\eqref{eq:Def_Psi_one_freq}, we determine $\widetilde{\Psi}^{S_1 S_2}_{m}$ for indices $S_1,S_2 \in \mathbb{S}=\{ {}^{++}_{\uparrow \downarrow}, {}^{--}_{\uparrow \downarrow}, {}^{+-}_{\uparrow \uparrow}, {}^{+-}_{\downarrow \downarrow}, {}^{+-}_{\uparrow \downarrow}, {}^{+-}_{\downarrow \uparrow}\}$:
\begin{equation}
 \widetilde{\Psi}^{S_1 S_2}_{m}
= \left(
\begin{array}{cccccc}
 \widetilde{\Psi}^\ix{p}_{m} & 0 & 0 & 0 & 0 & 0 \\
 0 & \widetilde{\Psi}^{\ix{p}\ast}_{m} & 0 & 0 & 0 & 0 \\
 0 & 0 & 0 & \widetilde{\Psi}^{\ix{d}\uparrow}_{m} & 0 & 0 \\
 0 & 0 & \widetilde{\Psi}^{\ix{d}\downarrow}_{m} & 0 & 0 & 0 \\
 0 & 0 & 0 & 0 & \widetilde{\Psi}^{\ix{x}\uparrow}_{m} & 0 \\
 0 & 0 & 0 & 0 & 0 & \widetilde{\Psi}^{\ix{x}\downarrow}_{m}
\end{array}
\right)_{\!\!S_1 S_2} \!.
\end{equation}
Here, we used the abbreviations
\begin{align}
 \label{eq:Psi_++++_component}
 \widetilde{\Psi}^\ix{p}_{m} &=\widetilde{\Psi}^{++++}_{\uparrow \downarrow \uparrow \downarrow, m} = U \frac{1}{\beta} \sum_n \overline{G}^{-}_{\uparrow,-n} \overline{G}^{-}_{\downarrow,n-m} ,
\\ \widetilde{\Psi}^{\ix{d}\uparrow}_{m} &=\widetilde{\Psi}^{+-+-}_{\uparrow \uparrow \downarrow \downarrow, m}=-U \frac{1}{\beta} \sum_n \overline{G}^{-}_{\downarrow,n} \overline{G}^{-}_{\downarrow,n+m} \in \mathbb{R} ,
\\  \widetilde{\Psi}^{\ix{d}\downarrow}_{m} &=\widetilde{\Psi}^{+-+-}_{\downarrow \downarrow \uparrow \uparrow, m}=-U \frac{1}{\beta} \sum_n \overline{G}^{-}_{\uparrow,n} \overline{G}^{-}_{\uparrow,n+m} \in \mathbb{R} ,
\\ \widetilde{\Psi}^{\ix{x}\uparrow}_{m} &=\widetilde{\Psi}^{+-+-}_{\uparrow \downarrow \uparrow \downarrow, m}=U \frac{1}{\beta} \sum_n \overline{G}^{-}_{\uparrow,n} \overline{G}^{-}_{\downarrow,n+m} ,
\\ \widetilde{\Psi}^{\ix{x}\downarrow}_{m} &=\widetilde{\Psi}^{+-+-}_{\downarrow \uparrow \downarrow \uparrow, m}=\widetilde{\Psi}^{\ix{x}\uparrow \ast}_{m}.
\end{align}
The labeling of these abbreviations is inspired by the role of the corresponding components in, for example, the FLEX ladder summations. There are particle-particle and direct particle-hole as well as exchange particle-hole contributions. The extra-labeling with $\uparrow$ or $\downarrow$ refers to which $\sigma$-component of the self-energy is affected by the contribution.

Because of its block-diagonal structure, the inverse of $[\mathds{1}+\widetilde{\Psi}_m]$ is easily computed. Multiplying the result with $\widetilde{\Psi}_{m}$ according to Eq.~\eqref{eq:Def_Upsilon_tilde} yields
\begin{equation}
 \widetilde{\Upsilon}^{S_1 S_2}_{m} = \left(
\begin{array}{cccccc}
 \widetilde{\Upsilon}^\ix{p}_{m} & 0 & 0 & 0 & 0 & 0 \\
 0 & \widetilde{\Upsilon}^{\ix{p}\ast}_{m} & 0 & 0 & 0 & 0 \\
 0 & 0 & \widetilde{\Upsilon}^{\overline{\ix{d}}}_{m} & \widetilde{\Upsilon}^{\ix{d}\uparrow}_{m} & 0 & 0 \\
 0 & 0 & \widetilde{\Upsilon}^{\ix{d}\downarrow}_{m} & \widetilde{\Upsilon}^{\overline{\ix{d}}}_{m} & 0 & 0 \\
 0 & 0 & 0 & 0 & \widetilde{\Upsilon}^{\ix{x}\uparrow}_{m} & 0 \\
 0 & 0 & 0 & 0 & 0 & \widetilde{\Upsilon}^{\ix{x}\downarrow}_{m}
\end{array}
\right)_{S_1 S_2} \!\!\!\!.
\end{equation}
Here, we used the following abbreviations in which again the labels are inspired by the role of the component in the calculation of the self-energy:
\begin{align}
 \widetilde{\Upsilon}^\ix{p}_{m} &=\widetilde{\Upsilon}^{++++}_{\uparrow \downarrow \uparrow \downarrow,m}=\widetilde{\Psi}^\ix{p}_{m} \left[1+\widetilde{\Psi}^\ix{p}_{m} \right]^{-1} ,
\\ \widetilde{\Upsilon}^{\overline{\ix{d}}}_{m} &=\widetilde{\Upsilon}^{+-+-}_{\uparrow \uparrow \uparrow \uparrow,m}=-\widetilde{\Psi}^{\ix{d}\uparrow}_{m} \widetilde{\Psi}^{\ix{d}\downarrow}_{m} \left[1-\widetilde{\Psi}^{\ix{d}\uparrow}_{m} \widetilde{\Psi}^{\ix{d}\downarrow}_{m} \right]^{-1} \in \mathbb{R} ,
\\ \widetilde{\Upsilon}^{\ix{d}\uparrow}_{m} &=\widetilde{\Upsilon}^{+-+-}_{\uparrow \uparrow \downarrow \downarrow,m}=\widetilde{\Psi}^{\ix{d}\uparrow}_{m} \left[1-\widetilde{\Psi}^{\ix{d}\uparrow}_{m} \widetilde{\Psi}^{\ix{d}\downarrow}_{m} \right]^{-1} \in \mathbb{R} ,
\\ \widetilde{\Upsilon}^{\ix{d}\downarrow}_{m} &=\widetilde{\Upsilon}^{+-+-}_{\downarrow \downarrow \uparrow \uparrow,m}=\widetilde{\Psi}^{\ix{d}\downarrow}_{m} \left[1-\widetilde{\Psi}^{\ix{d}\uparrow}_{m} \widetilde{\Psi}^{\ix{d}\downarrow}_{m} \right]^{-1} \in \mathbb{R} ,
\\ \widetilde{\Upsilon}^{\ix{x}\uparrow}_{m} &=\widetilde{\Upsilon}^{+-+-}_{\uparrow \downarrow \uparrow \downarrow,m}=\widetilde{\Psi}^{\ix{x}\uparrow}_{m} \left[1+\widetilde{\Psi}^{\ix{x}\uparrow}_{m} \right]^{-1} .
\\ \widetilde{\Upsilon}^{\ix{x}\downarrow}_{m} &=\widetilde{\Upsilon}^{+-+-}_{\downarrow \uparrow \downarrow \uparrow,m}=\widetilde{\Upsilon}^{\ix{x}\uparrow\ast}_{m}.
\end{align}

Note that the complex conjugation relation $\overline{\Sigma}{}^{c \ast}_{\sigma,n}=\overline{\Sigma}{}^c_{\sigma,-n}$ holds. For its proof, consider the following reasoning: In the derivation of the flow (or self-consistency) equations it was assumed to hold. These equations are found to not lead to a violation of the relation. Furthermore, we start with initial conditions (or guesses) which do not violate the relation. Consequently, the relation is self-consistently fulfilled. The relation also implies $\widetilde{\Psi}{}^i_{m}=\widetilde{\Psi}{}^{i \ast}_{-m}$ and $\widetilde{\Upsilon}{}^i_{m}=\widetilde{\Upsilon}{}^{i \ast}_{-m}$.


\subsection{Self-consistency equations for the conserving schemes}
\label{sec:impl_flow_eq_for_c_schemes}

The goal of this section is to provide self-consistency equations for
the FLEX and the cfRG approximation that can be used for the numerical
implementation.  For FLEX, we insert $\alpha=(c,\sigma,\nu_n)$ in
Eq.~\eqref{eq:FLEX_after_redefinition} and exploit the symmetries
and the sparseness of components to find
\begin{align}
 \notag
 \overline{\Sigma}_{\sigma,n}^{\ix{FLEX}+}=& \frac{U}{\beta} \sum_m \left\{ \left[\frac{2}{3} \widetilde{\Psi}^{\ix{p}}_m-\widetilde{\Upsilon}^{\ix{p}}_m  \right]\overline{G}^{-}_{\bar{\sigma},n-m}\right.
 \\ \notag
 & \qquad \qquad \left.+\left[\widetilde{\Upsilon}^{\ix{d}\sigma}_m - \frac{2}{3}\widetilde{\Psi}^{\ix{d}\sigma}_m \right]\overline{G}^{-}_{\sigma,m-n}\right.
 \\ \notag
 & \qquad \qquad \left. +\left[\frac{2}{3} \widetilde{\Psi}^{\ix{x}\sigma}_m -\widetilde{\Upsilon}^{\ix{x}\sigma}_m \right]\overline{G}^{-}_{\bar{\sigma},m-n}\right\}
 \\ &+ \frac{U}{\beta} \sum_{n^\prime} \overline{G}^{-}_{\bar{\sigma},n^\prime} e^{-i \nu_{n^\prime} 0^+}- \frac{U}{2}  
 \label{eq:FLEX_Sigma_flow_eq_SIAM}
\end{align}
The convergence factor in the last line is necessary and a consequence of correct imaginary time ordering. It can be ``canceled'' with the addend $-\frac{U}{2}$. Remember that the second order diagram contribution
\begin{equation}
 -\left(\frac{U}{\beta}\right)^2 \sum_{m,l} \overline{G}^-_{\sigma,-l} \overline{G}^-_{\bar{\sigma},l-m} \overline{G}^-_{\bar{\sigma},n-m}
\end{equation}
is included correctly in FLEX. In
Eq.~\eqref{eq:FLEX_Sigma_flow_eq_SIAM}, it can be seen that the
three channels contribute each $\frac{1}{3}$ of this
contribution. This arises naturally from the charge index notation. We
now perform the $T=0$ limit ($\frac{1}{\beta}\sum_n \to
\int_{-\infty}^\infty \frac{d\nu}{2\pi}$). Furthermore, we drop the
charge index $+$ on the self-energy and the $-$ on the propagator and
find Eq.~\eqref{eq:cons_Sigma_flow_eq}.  In this equation, we
introduced constants $\kappa_i$ which must be chosen as
$\kappa_0=\frac{2}{3}$ and
$\kappa_\ix{p}=\kappa_\ix{d}=\allowbreak\kappa_\ix{x}=1$ for FLEX. The
cfRG leads to the same form as shown in
Eq.~\eqref{eq:cons_Sigma_flow_eq}. We must then choose
$\kappa_0=0$ and
$\kappa_\ix{p}=\kappa_\ix{d}=\allowbreak\kappa_\ix{x}=\frac{1}{3}$.


\subsection{Flow equations for the non-conserving schemes}
\label{sec:impl_flow_eq_for_nc_schemes}

In this section, we specify the relevant equations for the
implementation of the non-conserving schemes. We start out by
discussing the PUF, StUF and MUF approximations and proceed then with
the CUF and CF approximations.

For the PUF approximation, we find Eq.~\eqref{eq:PUF_fl_eq_Sigma_T=0}.
The last term does not need a convergence factor because
$\dot{\overline{G}}{}^{\lambda}_{\sigma}(\nu)$ goes as $\sim
1/\nu^2$. The initial condition is
$\overline{\Sigma}{}^{\ix{PUF}}_{\sigma,\lambda_\ix{i}}(\nu)=0$ (and
$U_{\lambda_\ix{i}}=0$).

The flow equation for the StUF approximation reads as
\begin{equation}
\label{eq:flow_eq_StUF}
\dot{\overline{\Sigma}}{}^{\ix{StUF}}_{\sigma,\lambda}=\dot{U}_\lambda \int_{-\infty}^\infty \frac{d\nu^\prime}{\pi} \ix{Re}\left[\overline{G}{}^{\lambda}_{\bar{\sigma}}(\nu^\prime) \right]
\end{equation}
with initial condition $\overline{\Sigma}{}^{\ix{StUF}}_{\sigma,\lambda_\ix{i}}=0$. As no frequency dependence is acquired in this scheme, we directly perform the frequency integral analytically and find Eq.~\eqref{eq:flow_eq_StUF_RHS_integrated}.

The equation for the MUF approximation is Eq.~\eqref{eq:MUF_fl_eq_Sigma_T=0}.
The non-zero initial condition is
$\overline{\Sigma}{}^{\ix{MUF}}_{\sigma,\lambda_\ix{i}}(\nu)=\overline{\Sigma}{}^\ix{HF}_\sigma=-\frac{U}{\pi}
\ix{atan}[(\epsilon_{\bar{\sigma}}+\overline{\Sigma}{}^\ix{HF}_{\bar{\sigma}})/\Gamma]$.
For the PUF, StUF and MUF schemes, we take $U_\lambda=\lambda U$,
$\lambda=0\ldots 1$.

Now, we specify the flow equations for the CUF scheme. In this
case all propagators depend explicitly on $\lambda$ [in addition to
their implicit dependence due
to~$\overline{\Sigma}{}^\lambda_\sigma(\nu)$]:
\begin{align}
\notag
\overline{G}^{-}_{\sigma}(\nu)&=-\frac{1}{i \nu + \epsilon_\sigma + i \ix{sgn}(\nu)\Gamma + \overline{\Sigma}{}^{\lambda}_{\sigma}(\nu)^\ast} 
\\ \label{eq:Def_G_Theta}
&\to -\frac{\Theta(|\nu|-\lambda)}{i \nu + \epsilon_\sigma + i \ix{sgn}(\nu)\Gamma + \overline{\Sigma}{}^{\lambda}_{\sigma}(\nu)^\ast}.
\end{align} 
Because of the sharp frequency cut-off we can replace $\overline{\Pi}\cdot \dot{C}^{-1}$ by $\overline{G}|_{\Theta \to \delta}$ when calculating the extra addend from Eq.~\eqref{eq:Def_dot_Delta_Sigma}. This $\delta$-function cancels the frequency integral and we find
\begin{multline}
\label{eq:CF2O_DeltaSigmaUp}
 -\frac{U_\lambda}{2\pi} \sum_{x=\pm\lambda} \left\{ \left[\widetilde{\Psi}^{\ix{x}\sigma}_\lambda(\nu\!+\!x) -\widetilde{\Psi}^\ix{p}_\lambda(\nu\!-\!x)\right] \overline{G}{}^{\lambda}_{\bar{\sigma}}\!(x) \right.
 \\ \left.+ \widetilde{\Psi}^{\ix{d}\sigma}_\lambda(\nu\!+\!x) \overline{G}{}^{\lambda}_\sigma\!(x) \right\} = \!\mathop{:} \Delta^{\lambda}_\sigma (\nu).
\end{multline}
In total, we obtain the flow equation~\eqref{eq:CUF_fl_eq_Sigma_T=0}.
Note that $\dot{\overline{\Sigma}}{}^\ix{PUF}_{\sigma} (\nu)$
represents symbolically what is specified in
Eq.~\eqref{eq:PUF_fl_eq_Sigma_T=0}. However, we must now use the
explicitly $\lambda$-dependent $\overline{G}{}^\lambda_\sigma(\nu)$
from Eq.~\eqref{eq:Def_G_Theta} and $U_\lambda$ from
Eq.~\eqref{eq:CU-cut-off} in order to calculate the right-hand
side. The initial condition for the CUF scheme is
$\overline{\Sigma}{}^{\ix{CUF}_\Lambda}_{\sigma,\lambda_\ix{i}}(\nu)=0$.

The flow equation for the CF approximation is Eq.~\eqref{eq:CF_fl_eq_Sigma_T=0}.
In the second addend, we must apply the replacement $U_\lambda \to U$ on every level - also in the calculation of the components of $\widetilde{\Psi}_{\lambda}$. The numerical initial condition is $\overline{\Sigma}{}^\ix{CF}_{\lambda_\ix{i}^\ix{num},\sigma} (\nu)=0$.


\section{Details on our numerical implementation}
\label{sec:details_num_impl}

We work with two meshes for the frequencies: a grid of fermionic frequencies $\nu_n$ for the
self-energy and a grid of bosonic frequencies $\omega_m$
for the auxiliary quantities~$\widetilde{\Psi}$
and~$\widetilde{\Upsilon}$. They are given by the geometric formulas
($n,m \geq 0$)
\begin{align}
 \nu_n &= d\nu \frac{(1+f_\nu)^n-1}{f_\nu},
\\ \omega_m &= d\nu \frac{(1+f_\omega)^m-1}{f_\omega}.
\end{align}
Note that $n,m$ now refer to grid points and not to Matsubara frequencies. Setting $\Gamma=1$, the following parameters must be externally specified: $d\nu$, $\nu_\ix{max}$ and $n_\ix{len}$. Then, we set $m_\ix{len}=2 n_\ix{len}$ and $\omega_\ix{max}=\nu_\ix{max}^2$. Thus, the bosonic grid has twice as many points as the fermionic grid but also extends to much larger frequencies. We compute $f_{\nu}$, $f_\omega$ such that $d\nu [(1+f_\nu)^{n_\ix{len}}-1]/f_\nu=\nu_\ix{max}$ and $d\nu [(1+f_\omega)^{m_\ix{len}}-1]/f_\omega=\omega_\ix{max}$. The value of a quantity is determined by cubic interpolation if the frequency is not exactly on one of the grid points. $\overline{G}_\sigma(\nu)$ is an exception, see following.

We restrict the grids to non-negative frequencies.  This is
sufficient, since we can apply complex conjugation relations to
express quantities at negative frequencies through their values at
positive frequencies (cf.~Appendix~\ref{sec:calc_Ups_for_SIAM}).  The
advantages of this procedure are a reduced grid size and an accurate
treatment of discontinuities at zero frequency. A discontinuity of
$\widetilde{\Upsilon}$ at zero frequency results from the appearance
of $\sgn \nu$ in the free propagator, cf.
Eq.~\eqref{eq:reg_FT}. A numerical interpolation close to
discontinuities is difficult and avoided by our approach with a grid
of non-negative frequencies only.

For both conserving and non-conserving schemes, we store $\overline
\Sigma$ by separating the asymptotic value from the rest,
\begin{equation}
 \label{eq:Sigma_num_approx}
 \overline{\Sigma}_\sigma(\nu)=
 \left\{
 \begin{array}{ll}
   \overline{\Sigma}{}^{\ix{C}}_{\sigma} +
   \overline{\Sigma}{}^{\ix{D}}_{\sigma}(\nu) 
   &, 0 \leq \nu<\nu_\ix{max}
   \\ \overline{\Sigma}{}^{\ix{C}}_{\sigma} &, \nu >\nu_\ix{max}
 \end{array}
 \right. ,
\end{equation}
in which $\overline{\Sigma}{}^{\ix{C}}_{\sigma}= \lim_{\nu \to \infty}
\overline{\Sigma}_{\sigma}(\nu)$.  Now, $\overline{G}{}_{\sigma}(\nu)$
can be computed for all $\nu$. If needed,
$\overline{\Sigma}{}^{\ix{D}}_\sigma(\nu)$ is computed by cubic
interpolation. Thus, integrals over integrands consisting purely of
$\overline{G}{}_{\sigma}(\nu)$ can be calculated from $-\infty$ to
$\infty$ (and not only on a finite range). The outer parts can be
calculated analytically because the self-energy is taken as constant
there. We make use of this for the calculation of $\widetilde{\Psi}$.

Splitting $\overline{\Sigma}{}_\sigma(\nu)$ up as in Eq.~\eqref{eq:Sigma_num_approx} constitutes an approximation for finite $\nu_\ix{max}$. This allows to compute
$\overline{G}_\sigma(\nu)$ for arbitrary $\nu$ with an error
of $\mathcal{O}(\frac{1}{\nu_\ix{max}^2})$. Performing $\omega$-integrations which formally go from $-\infty$
to $\infty$ only from $-\omega_\ix{max}$ to $\omega_\ix{max}$ also induces an error. By requiring
$\omega_\ix{max} \sim \nu_\ix{max}^2$ we ensure that this error is as
well of $\mathcal{O}(\frac{1}{\nu_\ix{max}^2})$.

All flow equations in Appendix~\ref{sec:impl_flow_eq_for_nc_schemes}
[except Eq.~\eqref{eq:flow_eq_StUF} for StUF] pose a
self-consistency problem in each step of the flow because the
right-hand side contains
$\dot{\overline{G}}{^\lambda_\sigma}(\nu)$. This problem is easily
solved because $\dot{\overline{G}}{}^\lambda_{\sigma}(\nu)$ always
occurs only in a separate addend contributing to the
frequency independent asymptotic value of the self-energy. By
inserting the separation~\eqref{eq:Sigma_num_approx} into the flow
equation for $\overline{\Sigma}{}^{\lambda}_\sigma(\nu)$, we obtain
separate equations for
$\dot{\overline{\Sigma}}{}^{\lambda,\ix{C}}_\sigma$ and
$\dot{\overline{\Sigma}}{}^{\lambda,\ix{D}}_\sigma(\nu)$.  The
right-hand side of the flow equation for
$\dot{\overline{\Sigma}}{}^{\lambda,\ix{D}}_\sigma(\nu)$ then does not
contain $\dot{\overline{G}}{}^\lambda_{\sigma}(\nu)$. Furthermore, the
equation for $\dot{\overline{\Sigma}}{}^{\lambda,\ix{C}}_\sigma$ has the form
$\dot{\overline{\Sigma}}{}^{\lambda,\ix{C}}_\sigma=A_1^\sigma +
A_2^\sigma \dot{\overline{\Sigma}}{}^{\lambda,\ix{C}}_{\bar{\sigma}}$
in which $A_1^\sigma$ depends on
$\dot{\overline{\Sigma}}{}^{\lambda,\ix{D}}_{\bar{\sigma}}(\nu)$. The
explicit solution of this equation is
\begin{equation}
 \label{eq:expl_sol_SigmaCdot}
  \dot{\overline{\Sigma}}
  {}^{\lambda,\ix{C}}_\sigma=\frac{A_1^\sigma + A_2^\sigma
    A_1^{\bar{\sigma}}}{1-A_2^\sigma A_2^{\bar{\sigma}}}. 
\end{equation} 
In each step of the flow, we thus proceed in the following manner: First, we compute $\dot{\overline{\Sigma}}{}^{\lambda,\ix{D}}_\sigma(\nu)$. Second, we calculate $A_1^\sigma$ and $A_2^\sigma$. Third, we calculate $\dot{\overline{\Sigma}}{}^{\lambda,\ix{C}}_\sigma$ by Eq.~\eqref{eq:expl_sol_SigmaCdot}.

Equation~\eqref{eq:expl_sol_SigmaCdot} is suitable to analyze the divergence that occurs in the
unrestricted MUF for $U$ moderately greater than $\pi\Gamma$, compare
Sec.~\ref{sec:results_PM_UF}. For such $U$, the real part of the self-energy is almost
constant $\ix{Re} \overline{\Sigma} {}^\lambda_\sigma (\nu) \approx
\overline{\Sigma} {}^\lambda_\sigma$. In the unrestricted Hartree-Fock
solution (for $V_\ix{g}=0=B$), the up and down self-energy take
non-zero values of opposite sign, i.e. $\overline{\Sigma}
{}^{\lambda_\ix{i}}_\sigma = \sigma \widetilde{h}$. This
$\widetilde{h}$ can be interpreted as an artificial magnetic field in
a non-interacting model. Performing the flow makes the real parts of
the self-energy move closer to one another, in other words the
artificial magnetic field decreases, i.e. the symmetry breaking is
suppressed. This mechanism does, however, not fully restore spin
symmetry. The reason for this is a divergence on the right-hand side
of the flow equation at a particular value of $\lambda$.  It
originates from the term $A_2^\sigma$ of
Eq.~\eqref{eq:expl_sol_SigmaCdot} which is given by
$-\widetilde{\Psi}{}^{\ix{d}\sigma}(\omega=0)=U \int_{-\infty}^\infty
\tfrac{d\nu}{2\pi} \overline{G}{}^\lambda_{\bar{\sigma}} (\nu)^2$,
where $U$ denotes the bare interaction. If $|A_2^\sigma|=1$, the
denominator in Eq.~\eqref{eq:expl_sol_SigmaCdot} leads to a divergence
which corresponds to the divergence that occurs in the RPA-like first
factor $\left(I+U \cdot \overline \Pi_\lambda \right)^\inv$ discussed
in Sec.~\ref{sec:results_PM_UF}.  For the unrestricted Hartree-Fock solution (i.e. at the
beginning of the flow), one finds that $\widetilde{h}$ is sufficiently
large, as to ensure $|A_2^\sigma|<1$. Now, as the symmetry restoring
effect predicted by Ref.~\onlinecite{Dup14} occurs, $\widetilde{h}$ is
effectively reduced, bringing $|A_2^\sigma|$ closer to its critical
value $1$. When $|A_2^\sigma|=1$ is reached, the flow equation cannot
be integrated any further.  This is what happens for $U\approx
U_\ix{crit}\ldots 8\Gamma$ for the effective mass, the charge
susceptibility and the spin susceptibility. Values can be calculated
for $U/\Gamma >8$ because then the restoring effect does not cause
$|A_2^\sigma|=1$ for any $\lambda \leq 1$.  However, the spurious spin
symmetry breaking is not lifted in this case and the results are not
trustworthy as explained in Sec.~\ref{sec:results_PM_UF}.

Analogous considerations can be applied to the restricted MUF.  The
restricted Hartree-Fock solution does not lead to spin symmetry
breaking, i.e. it corresponds to a vanishing artificial magnetic field
$\widetilde{h}=0$. Thus, $|A_2^\sigma|$ is greater than $1$ at the
beginning of the flow if $U > U_\ix{crit}$. This seems to be dubious
at first sight, because in the RPA-like factor $\left(I+U \cdot
\overline \Pi_\lambda \right)^\inv$ discussed in Sec.~\ref{sec:results_PM_UF} this would
correspond to a series of questionable convergence. However, such a
series also occurs in the Hartree-Fock initial condition. In
order to be able to renormalize this initially present
contribution to the self-energy, the flow equation must contain such a
term. Technically, it does not produce a divergence unless
$|A_2^\sigma|=1$. If $\overline{\Sigma}{}^\ix{C}_{\sigma,\lambda} =
\overline{\Sigma}{}^\ix{C}_{\bar{\sigma},\lambda}$ (and
$A_i^\sigma=A_i^{\bar{\sigma}}$) as it is the case for the restricted
MUF scheme as long as $B=0$, even this point, namely $A_2^\sigma=-1$,
can be crossed in the flow without the occurrence of a divergence.  In
order to do this in a numerically stable way, Eq.~\eqref{eq:expl_sol_SigmaCdot} is rewritten
\begin{equation}
\label{eq:expl_sol_SigmaCdot_simple}
\dot{\overline{\Sigma}}{}^{\lambda,\ix{C}}=\frac{A_1}{1-A_2}.
\end{equation} 
This explains why a numerical solution of the restricted MUF flow
equations remains possible even beyond $U_\ix{crit}$ except in the
presence of an external magnetic field. A small magnetic field is
required for the numerical computation of the spin susceptibility
which is thus not accessible for $U> U_\ix{crit}$.

When working with the CF and CUF schemes, we know that some integrands
will be zero in certain integration regions due to the step function
in the propagator. We take this into account and change integration
limits such that we integrate only over regions where the integrand is
non-zero. The integration limits must be updated in each step of the
flow because the step function in the free propagator directly depends
on the flow parameter.


\section{Details on the calculation of the occupancy from the grand potential}
\label{sec:occupancy_from_Omega}

In this appendix, we provide $\Delta \overline{\Omega}$ for the cfRG
and FLEX methods as well as $\dot{\overline{\Omega}}{}_\lambda$ for
the PUF, StUF and MUF schemes.

\paragraph*{The conserving case.} We find
\begin{align}
  \Delta \overline{\Omega}&=\overline{\Omega}-\left.\overline{\Omega}\right|_{U^{(1)}=0=U^{(2)}}
 \\ \notag
 &=\frac{1}{2\beta} \ix{tr} \ix{ln} \left(\overline{G} C^{-1}\right) -\frac{1}{2\beta} \ix{tr} \left[ \left(\overline{\Sigma} + U^{(1)} \right) \overline{G} \right] + \frac{1}{\beta}\overline{\Phi}
\end{align}
in which $\overline{\Phi}$ is given by
\begin{multline}
 \overline{\Phi}=\eta_0 \textrm{Tr}\left(U \cdot \overline{\Pi}\right) +\eta_1 \textrm{Tr}\left[U \cdot \overline{\Pi} \cdot U \cdot \overline{\Pi} \right] 
\\+ \eta_2 \textrm{Tr} \textrm{Ln} \left(I+U \cdot \overline{\Pi}\right) .
\end{multline}
Here, we defined some prefactors which must be chosen as $\eta_0=-\frac{1}{4}$, $\eta_1=\frac{1}{6}$ and $\eta_2=\frac{1}{2}$ for FLEX and as $\eta_0=\frac{1}{12}$, $\eta_1=0$ and $\eta_2=\frac{1}{6}$ for cfRG. The logarithmic expressions are defined via their series expansions
\begin{align}
  \notag
  \ix{tr} \ix{ln} \left(\overline{G} C^{-1}\right)&=-\ix{tr} \ix{ln} \left[\left(C^{-1}-\overline{\Sigma}\right) C\right]
  \\ &=-\textrm{tr} \sum_{k=1}^\infty \frac{(-1)^{k+1}}{k} \left(-\overline{\Sigma}C\right)^{k}
\end{align}
\begin{equation}
\textrm{Tr} \textrm{Ln} \left(I+U \cdot \overline{\Pi}\right)= \sum_{k=1}^\infty\frac{(-1)^{k+1}}{k} \textrm{Tr}\left[ \left(U \cdot \Pi\right)^k \right] .
\end{equation}
In each series, a convergence factor must be taken into account in the lowest contribution. Overall one finds
\begin{align}
  &\Delta \overline{\Omega}=
 \\ \notag
 &\sum_\sigma \int_0^{\infty} \frac{d\nu}{\pi} \left\{\ix{ln}\left| \overline{G}{}_\sigma(\nu) C^{-1}_\sigma(-\nu)\right| -\ix{Re}\left[\overline{G}_\sigma (\nu) \overline{\Sigma} {}^\ix{D}_\sigma (\nu)^\ast\right]\right\}
 \\ \notag
 &+\frac{1}{2}\sum_\sigma \overline{\Sigma}{}^\ix{C}_\sigma -\sum_\sigma \left(\overline{\Sigma}{}^\ix{C}_\sigma + \frac{U}{2}\right) n_\sigma + \left(\eta_0+\eta_2\right)  4 U n_\uparrow n_\downarrow
 \\  \notag & - \eta_2 4 U \left(n_\uparrow -\frac{1}{2}\right)\left(n_\downarrow-\frac{1}{2}\right)
 \\  \notag &+ 2\eta_1 \int_0^{\infty} \frac{d\omega}{\pi} \ix{Re}\left\{ \widetilde{\Psi}^\ix{p}(\omega)^2 + \widetilde{\Psi}^{\ix{d}\downarrow}(\omega) \widetilde{\Psi}^{\ix{d}\uparrow}(\omega)+\widetilde{\Psi}^{\ix{x}\uparrow}(\omega)^2 \right\}
 \\  \notag & + 2\eta_2 \!\int_0^\infty \!\!\frac{d\omega}{\pi} \left\{ \vphantom{\frac{1}{2}}\ix{ln} \left| 1\!+\! \widetilde{\Psi}^\ix{p}\!(\omega)\right|+\ix{ln} \left| 1\!+\! \widetilde{\Psi}^{\ix{x}\uparrow}\!(\omega)\right| \right.
 \\ \notag
 & \qquad \qquad \qquad \qquad \left. +\frac{1}{2}\ix{ln}\left(1\!-\!\widetilde{\Psi}^{\ix{d}\uparrow}(\omega)\widetilde{\Psi}^{\ix{d}\downarrow}(\omega)\right)\right\} .
\end{align}

\paragraph*{The non-conserving case.}  In Ref.~\onlinecite{Ren15}, it was shown
that
$\dot{\overline{\Omega}}_\lambda=\overline{\dot{\Gamma}}_\lambda/\beta$.
This still holds.  However, this is not equal to
$\overline{\dot{\Phi}}_\lambda /\beta$ any more in the PUF and StUF
schemes because $U^{(1)}$ in
Eq.~\eqref{eq:formula_free_eff_action} acquires a
$\lambda$-dependence. We thus have
\begin{align}
  \dot{\overline{\Omega}}{}_\lambda^\ix{PUF}&= 
 \frac{1}{\beta} \overline{\dot{\Phi}}{}^\ix{PUF}_\lambda 
 + \frac{1}{\beta} \dot{U}{}^{(1)}_\lambda \! \cdot \overline{G}_\lambda 
 \\ \notag
 &= \frac{1}{3!\beta} \ix{Tr}\left(\dot{U}_\lambda^{(2)} \!\cdot\! \left[ \frac{3}{2} \overline{\Pi}_\lambda - \overline{\Pi}_\lambda \!\cdot\! \overline{\Upsilon}_\lambda \right]\right)
 + \frac{1}{\beta} \dot{U}{}^{(1)}_\lambda \!\cdot \overline{G}_\lambda
\end{align}
for the PUF approximation. Here, we used Eq.~(106)
of Ref.~\onlinecite{Ren15}. Explicitly, this means
\begin{align}
  &\dot{\overline{\Omega}}{}_\lambda^\ix{PUF}=
 \\ \notag
 &\dot{U}_\lambda \int_0^\infty \frac{d\nu}{\pi} \ix{Re}\left[\overline{G}_\uparrow(\nu)\right] \int_0^\infty \frac{d\nu}{\pi} \ix{Re}\left[\overline{G}_\downarrow(\nu)\right] -\frac{\dot{U}_\lambda}{4}
 \\  \notag &-\frac{1}{3!} \int_{-\infty}^{\infty} \frac{d\omega}{2\pi} \left\{ 2\ix{Re}\left[\widetilde{\Psi}{}_{\lambda \bullet}^\ix{p}(\omega)\widetilde{\Upsilon}{}_{\lambda}^\ix{p}(\omega)+\widetilde{\Psi}{}_{\lambda \bullet}^{\ix{x}\uparrow}(\omega)\widetilde{\Upsilon}{}_{\lambda}^{\ix{x}\uparrow}(\omega)\right] \right.
 \\  \notag
 & \qquad \qquad \qquad \qquad \left. +\widetilde{\Psi}{}_{\lambda \bullet}^{\ix{d}\uparrow}(\omega)\widetilde{\Upsilon}{}_{\lambda}^{\ix{d}\downarrow}(\omega)+\widetilde{\Psi}{}_{\lambda \bullet}^{\ix{d}\downarrow}(\omega)\widetilde{\Upsilon}{}_{\lambda}^{\ix{d}\uparrow}(\omega)\right\}.
\end{align}
Here, we introduced $\widetilde{\Psi}{}^i_{\lambda \bullet}=\widetilde{\Psi}{}^i_{\lambda}|_{U_\lambda \to \dot{U}{}_\lambda}$. Similarly, we obtain
\begin{equation}
 \dot{\overline{\Omega}}{}_\lambda^\ix{StUF}=\dot{U}_\lambda \!\int_0^\infty \!\frac{d\nu}{\pi} \ix{Re}\left[\overline{G}_\uparrow(\nu)\right] \int_0^\infty \!\frac{d\nu}{\pi} \ix{Re}\left[\overline{G}_\downarrow(\nu)\right] -\frac{\dot{U}_\lambda}{4}.
\end{equation}
For the MUF approximation we do not have an additional term and we can
simply use Eq.~(108) of Ref.~\onlinecite{Ren15} to find
\begin{align}
  &\dot{\overline{\Omega}}{}_\lambda^\ix{MUF}=- \frac{1}{3!\beta} \ix{Tr}\left(\dot{U}_\lambda^{(2)} \cdot \overline{\Pi}_\lambda \cdot \overline{\Upsilon}_\lambda \right)
 \\  \notag &=-\frac{1}{3!} \int_{-\infty}^{\infty} \frac{d\omega}{2\pi} \left\{ 2\ix{Re}\left[\widetilde{\Psi}{}_{\lambda \bullet}^\ix{p}(\omega)\widetilde{\Upsilon}{}_{\lambda}^\ix{p}(\omega)+\widetilde{\Psi}{}_{\lambda \bullet}^{\ix{x}\uparrow}(\omega)\widetilde{\Upsilon}{}_{\lambda}^{\ix{x}\uparrow}(\omega)\right] \right.
 \\  \notag & \qquad \qquad \qquad \qquad \left. +\widetilde{\Psi}{}_{\lambda \bullet}^{\ix{d}\uparrow}(\omega)\widetilde{\Upsilon}{}_{\lambda}^{\ix{d}\downarrow}(\omega)+\widetilde{\Psi}{}_{\lambda \bullet}^{\ix{d}\downarrow}(\omega)\widetilde{\Upsilon}{}_{\lambda}^{\ix{d}\uparrow}(\omega)\right\} .
\end{align}


\bibliographystyle{apsrev}
\bibliography{paper}

\end{document}